\documentclass{article}

\usepackage{arxiv}

\usepackage[utf8]{inputenc} 
\usepackage[T1]{fontenc}    
\usepackage{hyperref}       
\usepackage{url}            
\usepackage{booktabs}       
\usepackage{amsfonts}       
\usepackage{nicefrac}       
\usepackage{microtype}      
\usepackage{lipsum}
\usepackage{graphicx}
\graphicspath{ {./images/} }

\usepackage{natbib}
\usepackage{wrapfig}   
\usepackage{rotating}
\usepackage{bbm}
\usepackage{bm}
\usepackage{array}  %
\usepackage{multirow}%
\usepackage{amsmath,amssymb}%
\usepackage{mathtools}
\usepackage{amsthm}%
\usepackage{mathrsfs}%
\usepackage[title]{appendix}%
\usepackage{xcolor}%
\usepackage{textcomp}%
\usepackage{manyfoot}%
\usepackage{lscape}%
\usepackage{pdfpages}%
\usepackage{algorithm}%
\usepackage{algorithmicx}%
\usepackage{algpseudocode}%
\usepackage{listings}%
\usepackage{placeins} 

\usepackage{amsthm}
\newtheorem{lemma}{Lemma}
\theoremstyle{definition}
\newtheorem*{remark}{Remark}

\title{Bayesian Modal Regression for Forecast Combinations}

\author{
  Henry D. van Eijk\thanks{Corresponding author. Email: \texttt{hdvaneij@ncsu.edu}} \\
  Department of Statistics\\
  North Carolina State University\\
  Raleigh, NC 27695 \\
  \texttt{hdvaneij@ncsu.edu}
  \And
  Sujit K. Ghosh \\
  Department of Statistics\\
  North Carolina State University\\
  Raleigh, NC 27695 \\
  \texttt{sujit.ghosh@ncsu.edu}
}

\begin{document}
\maketitle
\begin{abstract}
Forecast combination methods have traditionally emphasized symmetric loss functions, particularly squared error loss, with equally weighted combinations often justified as a robust approach under such criteria. However, these justifications do not extend to asymmetric loss functions, where optimally weighted combinations may provide superior predictive performance. This study introduces a novel contribution by incorporating modal regression into forecast combinations, offering a Bayesian hierarchical framework that models the conditional mode of the response through combinations of time-varying parameters and exponential discounting. The proposed approach utilizes error distributions characterized by asymmetry and heavy tails, specifically the asymmetric Laplace, asymmetric normal, and reverse Gumbel distributions. Simulated data validate the parameter estimation for the modal regression models, confirming the robustness of the proposed methodology. Application of these methodologies to a real-world analyst forecast dataset shows that modal regression with asymmetric Laplace errors outperforms mean regression based on two key performance metrics: the hit rate, which measures the accuracy of classifying the sign of revenue surprises, and the win rate, which assesses the proportion of forecasts surpassing the equally weighted consensus. These results underscore the presence of skewness and fat-tailed behavior in forecast combination errors for revenue forecasting, highlighting the advantages of modal regression in financial applications.
\end{abstract}

\keywords{asymmetric loss functions \and Bayesian methods \and modal regression \and revenue forecasts \and rolling window forecasts}

\section{Introduction}\label{sec1}
In the literature on forecast combinations, \cite{bib1-} and \cite{bib25-} note that squared error loss is the most widely used loss function. Minimizing squared error loss, or more generally, mean squared error (MSE), is equivalent to minimizing the negative log-likelihood with a normality assumption for the error distribution. However, this assumption of normality for errors is not always supported in financial applications \citep[p.~14]{bib26-}, which is not surprising since financial data typically exhibit both asymmetry and fat tails; see \cite{bib23-}, \cite{bib20-}, \cite{bib24-}, and \cite{bib22-}. When the normality assumption is violated, even after transformations such as the logarithmic transformation, an asymmetric error distribution may help to capture the asymmetry. \cite{bib27-} note several asymmetric loss functions that are useful in forecast combinations: (i) lin-lin, (ii) asymmetric quadratic, and (iii) linex. Furthermore, \cite{bib27-} support the idea that, with respect to MSE, the equally weighted combination may lead to better performance than the optimally weighted combination; however, they overturn these results with asymmetric loss functions for skewed error distributions. 

In order to find the optimally weighted combination with an asymmetric loss function, one could use direct optimization. However, this approach lacks measures of uncertainty. To overcome this limitation, consider a Bayesian hierarchical framework. This framework does not easily support setting a loss function as an optimization problem; thus, one must find an equivalent distribution for the errors. We connect the lin-lin, asymmetric quadratic, and linex loss functions to the asymmetric Laplace, asymmetric normal (a reparameterized split/two-piece normal distribution) and the reverse Gumbel distribution, respectively. Consequently, we show how these three error distributions relate to the quantile, expectile, and extreme value regressions. For each pair of the loss function and error distribution described above, we show that by fixing the value of the asymmetry parameter, minimizing the negative log-likelihood with respect to the location parameter is equivalent to minimizing the corresponding loss function. We present the derivations of these equivalences in Section~\ref{sec2}.

It is well-known that minimizing mean absolute error (MAE) is equivalent to minimizing the negative log-likelihood under the assumption that the error distribution follows a Laplace distribution. The asymmetric Laplace distribution is an extension of the Laplace that can capture asymmetry. Since the asymmetric Laplace distribution can handle fat tails and asymmetry, \cite{bib21-} note that it applies to financial data (e.g., currency exchange rates). Next, \cite{bib19-} mentions that the financial industry has utilized the split/two-piece normal (the asymmetric normal in our case) since the 1990s. Specifically, the Bank of England used this distribution to release probabilistic forecasts of future inflation through its fan chart. Lastly, the reverse Gumbel distribution is not as well known as the Gumbel distribution. Both Gumbel distributions are advantageous in forecasting extreme events. \cite{bib32-} suggest that the Gumbel distribution is advantageous in option pricing. \cite{bib18-} extended the Gumbel distribution by introducing the flexible Gumbel to handle both the maxima (Gumbel) and minima (reverse Gumbel) cases. As shown in Section~\ref{sec2}, regression assuming that the error distribution follows an asymmetric Laplace, asymmetric normal, or reverse Gumbel is essentially a quantile, expectile, or extreme value regression, respectively. However, this result is only valid for a fixed value of the asymmetry parameter.

What happens if we do not fix a value of the asymmetry parameter? By treating this parameter as a random variable (r.v.) in our Bayesian hierarchical framework and placing a prior distribution on it, the problems of quantile, expectile, and extreme value regression translate to modal regression since the interest is now to forecast the conditional mode of the response $Y$ conditioned on $X$. This result is because these distributions all have their location parameter as the mode.

\begin{figure}[ht] 
  \centering
  \includegraphics[width=0.8\textwidth]{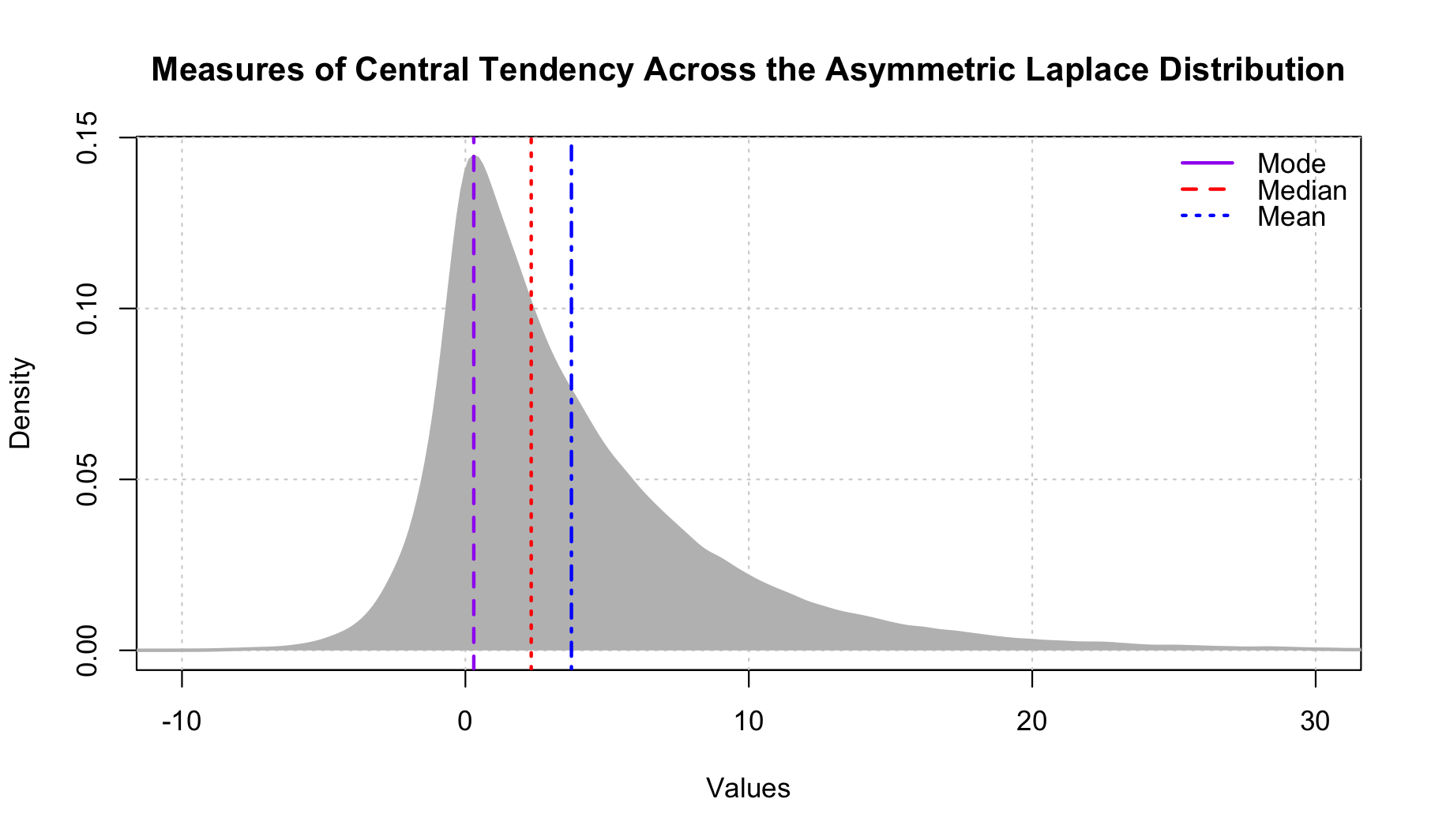} 
  \caption{Comparison of the mode, median, and mean for the asymmetric Laplace distribution density plot with $\mu=0$, $\sigma=1$, and $\tau=0.2$. See \eqref{eq:ald} for the pdf definition.}
  \label{fig:mode_median_mean}
\end{figure}

Our interest lies in modeling \textit{central tendency}. Thus, we present why modal regression is advantageous over mean and median regression; see \cite{bib17-} as an example of modal regression outperforming both mean and median regression for COVID-19 data. For a simple illustration, consider Figure~\ref{fig:mode_median_mean} that displays how the mode, median, and mean differ across an asymmetric distribution, specifically the asymmetric Laplace distribution. Notice that the mode is more robust as it is not pulled away from the center, unlike the mean and median when the distribution exhibits asymmetry. Another interesting takeaway is that the likelihood of the mode is approximately 0.15; meanwhile, the median is approximately 0.10, and the mean is approximately 0.075. This result indicates that the mode is approximately two times as likely as the mean and about one and a half times as likely as the median. \cite{bib6-} mention three benefits of modal regression over mean and median regression: (i) more robust to outliers, (ii) narrower prediction intervals, and (iii) better interpretability. Result (i) is beneficial in financial forecasting as outliers are common. For example, consider forecasting quarterly revenue—outliers could include revenues during COVID-19 or other periods of market uncertainty that present large deviations or asymmetry in the error distributions. Removing these outliers would be subjective and risks discarding valuable information. Next, modal regression exhibits narrower predictive intervals than mean and median regression when the error distribution is unimodal and skewed; see \cite{bib28-} and \cite{bib29-}. Lastly, the interpretability of modal regression may be ideal over mean or median regression since we can predict the \textit{most likely} or \textit{most probable} value instead of the mean or median value. 

To summarize, the conditional mode, as opposed to the mean or median, provides several advantages:
\begin{itemize}
\item 
{\bf Robustness to skewed and heavy-tailed distributions:} In financial forecasting, error distributions frequently deviate from normality, exhibiting skewness and excess kurtosis. The conditional mean can be unduly influenced by extreme values, leading to biased predictions, while the median only accounts for central tendency without fully capturing the dominant forecast. The mode, in contrast, identifies the most probable outcome, making it more robust in such settings.
\item 
{\bf Optimal decision-making in asymmetric loss contexts:} In financial applications, decision-makers often operate under asymmetric loss functions, where overestimation and underestimation carry different penalties. Since the mode represents the most frequently occurring outcome, it aligns with practical forecasting goals in situations where researchers must minimize extreme deviations from the typical forecast.
\item 
{\bf Enhanced predictive accuracy for forecast combinations:} Analyst forecasts often exhibit systematic biases, and their errors follow an asymmetrical distribution. The proposed approach targets the most likely revenue surprise outcome by focusing on the conditional mode rather than the mean or median, improving classification-based forecasting metrics such as the hit and win rates.
\end{itemize}

To the best of our knowledge, this is the first paper on modal regression (Bayesian or frequentist) in forecast combinations. Thus, our main novelty is that we present this original Bayesian hierarchical framework for modal regression that accounts for time-varying parameters through rolling windows, exponential discounting through power likelihoods, and parameter constraints. The rest of the paper is structured as follows. In Section~\ref{sec2.1}, we first present the definitions of the three loss functions from above, and then we derive their equivalent error distribution in Section~\ref{sec2.2}. Next, Section~\ref{sec3} composes our Bayesian modal regression models by specifying these error distributions. We apply our models in Section~\ref{sec4}, where we display Monte Carlo simulation results for using these error distributions for forecast combinations. Section~\ref{sec5} uses these Bayesian modal regression models on a real-life revenue forecasting dataset and compares the results against our previous paper on Bayesian mean regression; see \cite{bib39-}. Lastly, Section~\ref{sec6} presents our concluding remarks. 

\section{Error Distributions for Asymmetric Losses}\label{sec2}
\subsection{Asymmetric Loss Functions}\label{sec2.1}
Let $\epsilon$ denote a generic error variable, which can result by fitting the regression model. For example, if we are predicting $y$ using a regression function $h(x,\beta)$, then $\epsilon=y - h(x,\beta)$ denotes the error of the regression model. Even though we assume that $\mathbb{E}(\epsilon|x)=0$, the conditional distribution of $\epsilon$ given $x$ may not be symmetric around zero. This result leads to asymmetric errors within a regression model. \citet{bib1-} notes that a common way to express an asymmetric loss function is with a threshold, $\kappa$, where forecast errors above this threshold are penalized differently otherwise. Consider the general form with error $\epsilon$ as
\begin{equation*}
    \mathcal{L}(\epsilon) =
    \begin{cases} 
      \mathcal{L}_1(\epsilon), & \text{if } \epsilon \geq \kappa, \\
      \mathcal{L}_2(\epsilon), & \text{if } \epsilon < \kappa,
    \end{cases}
\end{equation*}
where $\mathcal{L}_1(\kappa)=\mathcal{L}_2(\kappa)$ ensures that the loss function is continuous. This result motivates the definition of the lin-lin (i.e., pinball) loss function
\begin{equation}
\label{eq:linlin}
    \mathcal{L}_\tau(\epsilon) =
    \begin{cases} 
      \tau|\epsilon|, & \text{if } \epsilon \geq 0, \\
      (1-\tau)|\epsilon|, & \text{if } \epsilon < 0,
    \end{cases}
\end{equation}
where $\tau\in(0,1)$ is an asymmetry parameter (e.g., the $\tau$th quantile) \citep{bib7-}. Another representation of lin-lin loss is $\mathcal{L}_\tau(\epsilon)=(\tau-\mathbb{I}(\epsilon<0))\epsilon$ where $\mathbb{I}(\cdot)$ is the indicator function. Note that $\tau=0.5$ denotes equal penalization. However, different choices of $\tau$ penalize $\epsilon$ differently based on where it lies compared to the threshold zero. In addition to the lin-lin loss function, the definition of the asymmetric quadratic loss function is
\begin{equation}
\label{eq:asymquad}
    \mathcal{L}_\tau(\epsilon) =
    \begin{cases} 
      \tau \epsilon^2, & \text{if } \epsilon \geq 0, \\
      (1-\tau) \epsilon^2, & \text{if } \epsilon < 0,
    \end{cases}
\end{equation}
where $\tau\in(0,1)$ is an asymmetry parameter. It can be expressed as $\mathcal{L}_\tau(\epsilon)=|\tau-\mathbb{I}(\epsilon<0)|\epsilon^2$ in indicator function form. Lastly, the definition of the linex loss is
\begin{equation}
\label{eq:linex}
    \mathcal{L}_\tau(\epsilon) = \exp(\tau \epsilon) - \tau \epsilon - 1.
\end{equation}
where $\tau>0$ is the asymmetry parameter.

\subsection{Derivations of Equivalent Error Distributions}\label{sec2.2}
Let $y_t$ denote the true variable to predict and $\bm{x}_t$ denote a m-dimensional vector containing predictions for $m$ total forecasters before time $t$. We define a forecast combination with parameters $\bm{\theta}=(\omega_0, \bm{\omega})$ where $\bm{\omega}=(\omega_1,\ldots,\omega_m)^\prime\in \mathcal{S}_m=\{\bm{\omega}\succeq\bm{0}:\bm{\omega}^\prime\bm{1}_m=1\}$ (i.e., non-negative weights that sum to one) and $\omega_0\in \mathbb{R}$ is the intercept. Thus, an optimally weighted forecast combination is defined as $\hat{y}_t(\bm{\theta}) = \omega_0 + \bm{\omega}^\prime\bm{x}_t$ with forecast errors $e_t(\bm{\theta})=y_t-\hat{y}_t(\bm{\theta})$.

For a one-step-ahead forecasting horizon (e.g., forecasting the next quarter in revenue forecasting), we define the following rolling window cross-validation procedure for folds $f=1,\ldots,F$ and window size $L$: 
\begin{enumerate}
    \item Define the training dataset $\mathcal{D}_f=\{(y_t, \bm{x}_t)\}^L_{t=1}$ for time $t=1,\ldots,L$ that contains the ground truth values and the respective analyst forecasts.
    \item Use model $\mathcal{M}$ with data $\mathcal{D}_f$ to estimate the parameters $\bm{\theta}^{(f)}=(\omega_0^{(f)},\bm{\omega}^{(f)})$.
    \item Produce $\hat{y}_{L+1}^{(f)} = \hat{\omega}_0^{(f)}+(\bm{\hat{\omega}}^{(f)})^\prime\bm{x}^{(f)} _{L+1}$ that is the single out-of-sample optimally weighted forecast for fold $f$.
\end{enumerate}
We repeat this procedure across folds $1, \ldots, F$ resulting in the {\em parameter estimates $\bm{\hat{\omega}}$ and $\hat{\omega}_0$ to change as the training subsets $\mathcal{D}_f$ vary}, which is known as time-varying parameter combinations. For simplicity, we present the rest of this section with a single fold in the rolling window cross-validation and using $n$ as window size $L$. 

Consider performing inference on this collection of parameters $\bm{\theta}$ by minimizing some loss function denoted as $\mathcal{L}(e_t(\bm{\theta}))$. A traditional approach to this problem is to minimize the empirical risk function via optimization, denoted as
\begin{equation*}
    R(\bm{\theta})=\frac{1}{n}\sum_{t=1}^{n} \mathcal{L}(e_t(\bm{\theta}))
\end{equation*}
that leads to the solution $\bm{\hat{\theta}} \in \arg\min_{\bm{\theta}} R(\bm{\theta})$. We posit a probabilistic model to overcome the point uncertainty (i.e., measures of uncertainty are difficult to attain) limitation; a Bayesian inferential approach would help fix the lack of uncertainty. To make the connection between the empirical risk function and likelihood function, typically one assumes the negative log-likelihood as the loss function denoted as $-\log f(\mathcal{D}|\bm{\theta})=\sum_{t=1}^{n} \mathcal{L}(e_t(\bm{\theta}))$ where $f(\mathcal{D}|\bm{\theta})$ is the likelihood function determined by the error distribution.
This is equivalent to $ f(\mathcal{D}|\bm{\theta})\propto\exp[-\sum_{t=1}^{n} \mathcal{L}(e_t(\bm{\theta}))]=\exp(-nR(\bm{\theta}))$ due to the Gibbs posterior formalization; see \cite{bib2-} for more details on Gibbs posteriors. The posterior distribution $p(\bm{\theta}|\mathcal{D})$ is then defined as
\begin{equation*}
    p(\bm{\theta}|\mathcal{D}) = \frac{f(\mathcal{D}|\bm{\theta}) \pi(\bm{\theta})}{\int f(\mathcal{D}|\bm{\theta}) \pi(\bm{\theta}) d\bm{\theta}} \propto f(\mathcal{D}|\bm{\theta}) \pi(\bm{\theta}) \propto  L(\bm{\theta}) \pi(\bm{\theta})
\end{equation*}
where $f(\mathcal{D}|\bm{\theta})$ and $L(\bm{\theta})$ are the likelihood functions, $\int f(\mathcal{D}|\bm{\theta}) \pi(\bm{\theta}) d\bm{\theta}$ is the marginal distribution, and $\pi(\bm{\theta})$ is the prior distribution.

To connect the three asymmetric error distributions with their corresponding loss functions presented in Section~\ref{sec1}, consider the symmetric case as an example with a normality error assumption. We assume that $Y_t|\bm{x_t} \sim \text{N}(\omega_0 + \bm{\omega}^\prime\bm{x}_t, \sigma^2)$ where $\text{N}(\mu, \sigma^2)$ denotes a normal distribution with mean $\mu\in\mathbb{R}$ and variance $\sigma^2>0$. The likelihood function is then
\begin{equation*}
\begin{aligned}
    L(\bm{\theta}) 
    = \prod_{t=1}^n \frac{1}{\sqrt{2\pi\sigma^2}} \exp\left(-\frac{e_t(\bm{\theta})^2}{2\sigma^2}\right)
\end{aligned}
\end{equation*}
and taking the negative log-likelihood results in
\begin{equation}
\label{eq:log_mse_normal}
    -\ell(\bm{\theta}) = \frac{n}{2} \log(2\pi) + n\log(\sigma) + \frac{1}{2\sigma^2} \sum_{t=1}^n e_t(\bm{\theta})^2.
\end{equation}
Thus, minimizing \eqref{eq:log_mse_normal} with respect to $\bm{\theta}$ is equivalent to minimizing the squared error loss because $\frac{n}{2} \log(2\pi\sigma^2) + \frac{1}{2\sigma^2}$ can be ignored for optimization (conditional on $\sigma$). Before we present the equivalences between the three asymmetric error distributions and their loss functions, we present the following results for a generic split distribution. We show that asymmetric Laplace and asymmetric Normal are special cases of this general asymmetric distribution family. \\

\begin{lemma}
    Let $g(u)$ be any symmetric density on $\mathbb{R}$ with mode 0, i.e., $0\leq g(-u)=g(u)\leq g(0)$ for any $u\geq 0$ and $\int_{-\infty}^\infty g(u) du=1$. Then, a class of asymmetric densities with asymmetry parameter $\tau\in (0, 1)$ and scale parameter $\sigma>0$ is given by
    \begin{equation*}
    f(\epsilon;\sigma,\tau) = \frac{2\tau(1-\tau)}{\sigma}
    \begin{cases} 
      g\left(\frac{(1-\tau)\epsilon}{\sigma}\right), & \text{if } \epsilon \leq 0, \\
      g\left(\frac{\tau \epsilon}{\sigma}\right), & \text{if } \epsilon > 0
    \end{cases}
    \end{equation*}
\label{lemma1}    
\end{lemma}

\begin{remark}
Clearly, the mode from the above density is zero for any $\tau\in(0, 1)$ and $\sigma>0$. Notice that the asymmetric Laplace and the asymmetric normal distributions result from choosing $g(u)=\frac{1}{2}\exp(-|u|)$ and $g(u)=\frac{1}{\sqrt{2\pi}}\exp\left(-\frac{u^2}{2}\right)$. However, we will use a slightly different parametrization for asymmetric normal following the standard notations used in the literature.
\end{remark}

We derive the quantiles by computing the cumulative distribution function $F(\epsilon;\sigma,\tau)= \int_{-\infty}^\epsilon f(u;\sigma,\tau) du$ as
 \begin{equation*}
    F(\epsilon;\sigma,\tau) = 
    \begin{cases} 
      2\tau G\left(\frac{(1-\tau)\epsilon}{\sigma}\right), & \text{if } \epsilon \leq 0, \\
      \tau + (1-\tau)\left(2G\left(\frac{\tau\epsilon}{\sigma}\right)-1\right), & \text{if } \epsilon > 0
    \end{cases}
    \end{equation*}
where, $G(u) = \int_{-\infty}^u g(z) dz$ is the cumulative distribution function for the density $g(u)$. Thus, the median of $\epsilon$ is $\frac{\sigma}{1-\tau}G^{-1}(\frac{1}{4\tau})<0$ when $\tau\geq \frac{1}{2}$ and the median is $\frac{\sigma}{\tau}G^{-1}(\frac{3-4\tau}{4-4\tau})>0$ when $\tau\leq\frac{1}{2}$. Next, we provide the first three moments of the above density, assuming that $g(u)$ has the corresponding finite moments.\\

\begin{lemma}
    Let the density $f(\epsilon;\sigma,\tau)$ as defined in Lemma \ref{lemma1} of a generic split distribution. Then, the first three moments are given by the following expressions.
    \begin{enumerate}
        \item Assume that $c_1=\int_{0}^\infty u g(u) du<\infty$. The first moment, $\mathbb{E}(\epsilon|\tau)$ is given by
        \begin{equation*}
        \int_{-\infty}^\infty \epsilon f(\epsilon|\tau) d\epsilon = \frac{2c_1\sigma}{\tau (1-\tau)}(1-2\tau).
        \end{equation*}
        As $c_1>0$, unless $\tau=\frac{1}{2}$, the mean is strictly negative is $\tau>\frac{1}{2}$ and positive is $\tau<\frac{1}{2}$.
        
        \item Assume that $c_2=\int_{0}^\infty u^2 g(u) du<\infty$. The second moment, $\mathbb{E}(\epsilon^2|\tau)$, is given by  
        \begin{equation*}
        \int_{-\infty}^\infty \epsilon^2 f(\epsilon|\tau) d\epsilon 
        = 2c_2\sigma^2\tau(1-\tau)\left[\frac{1}{(1-\tau)^3} + \frac{1}{\tau^3}\right]
        = \frac{2c_2\sigma^2}{\tau^2 (1-\tau)^2} (\tau^3+(1-\tau)^3).
        \end{equation*}
        This result illustrates that the distribution's spread increases as $\tau$ moves away from 0.5.
        
        \item Assume that $c_3=\int_{0}^\infty u^3 g(u) du<\infty$. The third moment, $\mathbb{E}(\epsilon^3|\tau)$, is given by
        \begin{equation*}
        \int_{-\infty}^\infty \epsilon^3 f(\epsilon|\tau) d\epsilon 
        = 2c_3\sigma^3\tau(1-\tau)\left[\frac{1}{\tau^4}-\frac{1}{(1-\tau)^4}\right]
        = \frac{2c_3\sigma^3 (\tau^2+(1-\tau)^2)}{\tau^3 (1-\tau)^3} (1-2\tau).
        \end{equation*}
        As $c_3>0$, it follows that distribution is left skewed if $\tau > \frac{1}{2}$ and right skewed if $\tau < \frac{1}{2}$ and symmetric if $\tau=\frac{1}{2}$.
    \end{enumerate}
    \label{lemma2}
\end{lemma}
The proof of the Lemma \ref{lemma2}, along with a few additional properties of the generalized split distributions, are provided in the appendices.

From the above results it is clear that if we try to estimate the $\bm{\theta}$ based on the model $y_t = \hat{y}_t(\bm{\theta}) +\epsilon_t$ by regressing $y_t$'s on $\bm{x}_t$'s using the ordinary least squares method (which assumes $\mathbb{E}(\epsilon_t)=0$) or by a quantile or expectile method which assumes $\tau$-th quantile or expectile of $\epsilon_t$ is zero, then such methods will necessarily lead to biased estimates, when the value of $\tau$ is not known in advance or assumed known. This result justifies the need to use {\em modal regression} methods which are unaffected by the asymmetry parameter $\tau\in (0, 1)$ (and also the scale parameter $\sigma>0$). Next, we develop methodologies assuming that the conditional mode of $e_t =y_t - \hat{y}(\bm{\theta})$ is zero, but the distribution may depend on the unknown parameters $\tau$ and $\sigma$. Although our methodologies apply to any symmetric density $g(u)$ with mode $0$, we focus on two distinct parametric distribution classes motivated by the aforementioned loss functions. In addition, we apply our modal regression approach to a class of densities that do not necessarily arise from an underlying symmetric density with mode zero.


Let the pdf for the asymmetric Laplace distribution with parameters $\mu\in\mathbb{R}$ as the mode, $\sigma>0$ as the scale parameter, and $\tau \in (0,1)$ as the asymmetry parameter be defined as 
\begin{equation}
\label{eq:ald}
    f_{AL}(\epsilon;\sigma,\tau) = \frac{\tau(1-\tau)}{\sigma}
    \begin{cases} 
      \exp\left(\frac{(1-\tau)\epsilon}{\sigma}\right), & \text{if } \epsilon \leq 0, \\
      \exp\left(-\frac{\tau\epsilon}{\sigma}\right), & \text{if } \epsilon > 0,
    \end{cases}
\end{equation}
with notation $Y\sim\mathcal{AL}(\mu, \sigma, \tau)$ for r.v. $Y$. For our forecast combinations model, we assume that $Y_t|\bm{x_t} \sim \mathcal{AL}(\omega_0 + \bm{\omega}^\prime\bm{x}_t, \sigma, \tau)$ therefore the likelihood function is
\begin{equation*}
\begin{aligned}
    L(\bm{\theta}) 
    = \prod_{t=1}^n \frac{\tau(1-\tau)}{\sigma}
    \begin{cases} 
      \exp\left(\frac{(1-\tau) e_t(\bm{\theta}))}{\sigma}\right), & \text{if } e_t(\bm{\theta}) \leq 0, \\
      \exp\left(-\frac{\tau e_t(\bm{\theta}))}{\sigma}\right), & \text{if } e_t(\bm{\theta}) > 0,
    \end{cases}
\end{aligned}
\end{equation*}
and computing the negative log-likelihood results in
\begin{equation}
\begin{aligned}
\label{eq:nllald}
    -\ell(\bm{\theta}) 
    &= -n\log\left(\frac{\tau(1-\tau)}{\sigma}\right) + \frac{1}{\sigma} \sum_{t=1}^n \mathcal{L}_\tau(e_t(\bm{\theta}))
\end{aligned}
\end{equation}
where $\mathcal{L}_\tau(e_t(\bm{\theta}))$ is lin-lin loss. Thus, for a fixed $\tau\in (0, 1)$, minimizing \eqref{eq:nllald} with respect to $\bm{\theta}$ is equivalent to minimizing lin-lin loss from \eqref{eq:linlin}. This result is quantile regression since one fixes asymmetry parameter $\tau$.

Next, let the pdf for the asymmetric normal distribution, with the same parameters defined in \eqref{eq:ald}, be defined as 
\begin{equation*}
    f_{AN}(\epsilon;\sigma,\tau) = C(\tau)
    \begin{cases} 
      \exp\left(-\frac{(1-\tau)\epsilon^2}{\sigma^2}\right), & \text{if } \epsilon \leq 0, \\
      \exp\left(-\frac{\tau\epsilon^2}{\sigma^2}\right), & \text{if } \epsilon > 0,
    \end{cases}
\end{equation*}
where we define $C(\tau)=\frac{2\sqrt{\tau(1-\tau)}}{\sigma\sqrt{\pi}(\sqrt{\tau}+\sqrt{1-\tau})}$ as a normalizing constant. 

\begin{remark}
Notice that the above parametrization may appear slightly different from the one in Lemma \ref{lemma1}. However, if we replace $\tau$ above by $\tau^2$ and $\sigma$ by $\sqrt{2}\sigma$, then $C(\tau)$ above matches with that defined in Lemma \ref{lemma1}.    
\end{remark}

We define the notation for r.v. $Y$ following an asymmetric normal distribution as $Y\sim\mathcal{AN}(\mu, \sigma, \tau)$. The assumed regression model is then $Y_t|\bm{x_t} \sim \mathcal{AN}(\omega_0 + \bm{\omega}^\prime\bm{x}_t, \sigma, \tau)$ with likelihood function
\begin{equation*}
\begin{aligned}
    L(\bm{\theta}) 
    = \prod_{t=1}^n C(\tau)
    \begin{cases} 
      \exp\left(-\frac{(1-\tau)e_t(\bm{\theta})^2}{\sigma^2}\right), & \text{if } e_t(\bm{\theta}) \leq 0, \\
      \exp\left(-\frac{\tau e_t(\bm{\theta})^2}{\sigma^2}\right), & \text{if } e_t(\bm{\theta}) > 0,
    \end{cases}
\end{aligned}
\end{equation*}
and taking the negative log-likelihood results in
\begin{equation}
\begin{aligned}
\label{eq:log_mse_an}
    -\ell(\bm{\theta}) 
    &= -n\log C(\tau) + \frac{1}{\sigma^2} \sum_{t=1}^n \mathcal{L}_\tau(e_t(\bm{\theta}))
\end{aligned}
\end{equation}
where $\mathcal{L}_\tau(e_t(\bm{\theta}))$ is asymmetric quadratic loss from \eqref{eq:asymquad}. Thus, minimizing \eqref{eq:log_mse_an} with respect to $\bm{\theta}$ is equivalent to minimizing the asymmetric quadratic loss. This result is known as the {\em expectile regression} since one fixes the asymmetry parameter; see \cite{bib9-} and \cite{bib33-} for more details. 

Lastly, the pdf for the reverse Gumbel distribution with parameters $\mu\in\mathbb{R}$ for location and $\beta>0$ for scale is defined as
\begin{equation*}
    f_{RG}(\epsilon;\beta) = \frac{1}{\beta} \exp\left(\frac{\epsilon}{\beta}\right) \exp\left(-\exp\left(\frac{\epsilon}{\beta}\right)\right).
\end{equation*}
We denote r.v. $Y$ following a reverse Gumbel distribution as $Y\sim\mathcal{RG}(\mu, \beta)$. The posited forecast combinations model is then $Y_t|\bm{x_t} \sim \mathcal{RG}(\omega_0 + \bm{\omega}^\prime\bm{x}_t, \sigma)$ with likelihood function 
\begin{equation*}
\begin{aligned}
    L(\bm{\theta}) 
    = \prod_{t=1}^n 
    \frac{1}{\beta} \exp\left(\frac{e_t(\bm{\theta})}{\beta}\right) 
    \exp\left(-\exp\left(\frac{e_t(\bm{\theta})}{\beta}\right)\right)
\end{aligned}
\end{equation*}
and its corresponding negative log-likelihood is 
\begin{equation}
\begin{aligned}
\label{eq:log_gumbel}
    -\ell(\bm{\theta}) &= \sum_{t=1}^n 
    \left(\log \beta + \exp\left(\frac{e_t(\bm{\theta})}{\beta}\right) - \frac{e_t(\bm{\theta})}{\beta}\right) \\
    &= n \log \beta + \sum_{t=1}^n \left(\exp\left(\frac{e_t(\bm{\theta})}{\beta}\right) - \frac{e_t(\bm{\theta})}{\beta}\right) \\
    &= n \log \beta + \sum_{t=1}^n \left(\mathcal{L}_\tau(e_t(\bm{\theta}))+1\right) \\
    &= n \log \beta + n + \sum_{t=1}^n \mathcal{L}_\tau(e_t(\bm{\theta})).
\end{aligned}
\end{equation}
Minimizing $\eqref{eq:log_gumbel}$ is then equivalent to minimizing the linex loss function $\mathcal{L}_\tau(e_t(\bm{\theta}))$ from \eqref{eq:linex} with $\tau=1/\beta$ \citep{bib16-}. The scale parameter is often referred to as an asymmetry parameter since it can be expressed in terms of the asymmetry parameter $\tau$ from \eqref{eq:linex}. The assumed model with a reverse Gumbel error distribution is known as extreme value regression.

We illustrated that minimizing the negative log-likelihood of the asymmetric Laplace, asymmetric normal, and reverse Gumbell distributions with respect to $\bm{\theta}$ are equivalent to minimizing lin-lin, asymmetric quadratic, and linex loss functions. 
\textit{However, this interpretation holds only when the asymmetry parameter $\tau$ remains fixed or arbitrarily chosen by the user}. Since our interest lies in modeling central tendency, we could fix the asymmetry parameter over a discrete grid (i.e., hyperparameter tuning). 

Instead, we treat the asymmetry parameter $\tau$ as parameter within a Bayesian hierarchical framework. This approach has two advantages: (i) the asymmetry parameter can take on a continuous range of values, allowing the data to choose the optimal value, and (ii) it introduces a time-varying parameter aspect to the asymmetry parameter since it changes over each fold in the rolling window cross-validation, unlike hyperparameter tuning. Then, quantile, expectile, and extreme value regression are translated into modal regression since we are modeling the conditional mode of response $Y$ conditioned on $\bm{x}$ denoted as $M(Y|\bm{x})=\omega_0 + \bm{\omega}^\prime\bm{x}$. Next, we present the modal regression models for the three asymmetric error distributions.

\section{Bayesian Modal Regression}\label{sec3}
Although we could derive equivalent error distributions for the lin-lin, asymmetric quadratic, and linex loss functions, performing inference via techniques such as MCMC can still be challenging. For example, popular Bayesian frameworks such as \texttt{JAGS} (Just Another Gibbs Sampler) \citep{bib30-} and \texttt{Stan} \citep{bib31-} do not support the asymmetric Laplace, asymmetric normal, and reverse Gumbel distributions. We resort to using \texttt{Stan} over \texttt{JAGS} for two main reasons: (i) the ability to define custom distributions via the log pdf and (ii) the simplicity of incorporating exponential discounting into our models through the power likelihood. However, we include a stochastic representation of the asymmetric Laplace in \texttt{JAGS}; see Section~\ref{A2} of the appendix for more details.

Rather than relying on frequentist optimization methods that minimize asymmetric loss functions, a Bayesian approach provides a more comprehensive and principled framework for modeling uncertainty and performing predictive inference. The Bayesian inferential framework is particularly well-suited for modal regression and forecast combination due to the following reasons:
\begin{itemize}
\item 
{\bf Quantification of parameter uncertainty:} Bayesian methods offer a probabilistic treatment of parameters, capturing posterior distributions rather than relying on point estimates. This property is critical in modal regression, where uncertainty in parameter estimates can significantly impact predictive inference. Unlike optimization-based methods that yield fixed estimates, Bayesian inference provides a full posterior distribution, allowing for more reliable uncertainty quantification in predictions.
\item
{\bf Flexibility in model specification and hierarchical structuring:} The Bayesian hierarchical framework allows for incorporating time-varying parameter combinations and exponential discounting, accommodating dynamic changes in forecast reliability. This flexibility is difficult to achieve using standard loss-based optimization techniques, which typically assume static parameters.
\item 
{\bf Regularization and shrinkage for improved generalization:} Bayesian priors can impose regularization, preventing overfitting and improving predictive performance, especially in high-dimensional or sparse data settings. This result is beneficial in financial applications where analyst forecasts vary in quality, and an appropriate shrinkage mechanism can help balance the influence of different forecasters.
\item
{\bf Improved predictive inference with full posterior distributions:} Instead of relying solely on point predictions, Bayesian inference enables full posterior predictive distributions, essential in risk-sensitive decision-making. Forecasting in finance often involves assessing the likelihood of extreme events, where understanding the entire predictive distribution is more informative than simply optimizing a loss function.
\item 
{\bf Prior information compensating for limited data:} In financial applications, lack of data is a common problem, especially when working with lower-frequency settings such as quarterly revenue data. A Bayesian framework allows one to incorporate external (prior) knowledge—possibly from a subject matter expert—to compensate for the limited sample size. This benefit is illustrated by \cite{bib42-} that mention “the Bayesian approach to combining forecasts can be adopted to obtain optimal weights even if there is little information available on the performance of the individual forecasts”.
\end{itemize}
By leveraging the conditional mode within a Bayesian inferential framework, the proposed approach provides a more robust, flexible, and uncertainty-aware methodology for combining financial forecasts, outperforming traditional mean- and median-based models.

We present our three Bayesian hierarchical modal regression models for each error distribution. Below, $\text{N}(\mu, \sigma^2)$ denotes a normal distribution with mean $\mu\in\mathbb{R}$ and variance $\sigma^2>0$, $\text{Dir}(\alpha_1, \alpha_2, \ldots, \alpha_m)$ denotes a Dirichlet distribution with parameters $\alpha_1>0, \alpha_2>0, \ldots, \alpha_m>0$, $\text{Half-Cauchy}(x_0, \gamma)$ denotes a Half-Cauchy distribution with location parameter $x_0\in\mathbb{R}$ and scale parameter $\gamma>0$, and $\text{Beta}(\alpha, \beta)$ denotes a Beta distribution with shape parameters $\alpha>0$ and $\beta>0$. See Section~\ref{A1} of the appendix for the corresponding probability density functions. First, we define the Bayesian modal regression model under the assumption that the errors follow an asymmetric Laplace as
\begin{equation}
\begin{aligned}
    &y_t\sim\mathcal{AL}\left(w_0 +\sum_{j=1}^{m} x_{t,j} \omega_j, \sigma, \tau\right) \\
    &\bm{\omega} \sim \text{Dir}(\alpha_1, \alpha_2, \ldots, \alpha_m), \\
    &w_0 \sim \text{N}(0, \sigma_{w_0}^2), \\
    &\sigma \sim \text{Half-Cauchy}(0, \gamma_0), \\
    &\tau \sim \text{Beta}(\alpha_0, \beta_0). \\
\end{aligned}
\label{eq:aldmodelhie}
\end{equation}
The next Bayesian modal regression model, assuming an asymmetric normal error distribution, is defined as 
\begin{equation}
\begin{aligned}
    &y_t\sim\mathcal{AN}\left(w_0 +\sum_{j=1}^{m} x_{t,j} \omega_j, \sigma, \tau \right) \\
    &\bm{\omega} \sim \text{Dir}(\alpha_1, \alpha_2, \ldots, \alpha_m), \\
    &w_0 \sim \text{N}(0, \sigma_{w_0}^2), \\
    &\sigma \sim \text{Half-Cauchy}(0, \gamma_0), \\
    &\tau \sim \text{Beta}(\alpha_0, \beta_0). \\
\end{aligned}
\label{eq:snmodel}
\end{equation}
Lastly, we define the Bayesian framework for a reverse Gumbel error distribution as
\begin{equation}
\begin{aligned}
    &y_t\sim\mathcal{RG}\left(w_0 +\sum_{j=1}^{m} x_{t,j} \omega_j, \beta \right) \\
    &\bm{\omega} \sim \text{Dir}(\alpha_1, \alpha_2, \ldots, \alpha_m), \\
    &w_0 \sim \text{N}(0, \sigma_{w_0}^2), \\
    &\beta \sim \text{Half-Cauchy}(0, \gamma_0). \\
\end{aligned}
\label{eq:gmmodel}
\end{equation}

Next, we present how to incorporate an exponential discounting factor $\lambda$ into the model, allowing one to give more weight to recent data while gradually down-weighting older data based on the choice of $\lambda$. We define the exponential discounting function as
\begin{equation*}
    p_t(\lambda) = \frac{e^{-\lambda(L-t)}}{\sum_{t=1}^{L} e^{-\lambda(L-t)}}=\frac{e^{-\lambda(L-t)}(1-e^{-\lambda})}{1-e^{-\lambda L}}.
\end{equation*}
Ideally, we want to incorporate this function within the variance of our likelihood, i.e., multiply the variance by $\frac{1}{p_t(\lambda)}$ in order to increase the variance as the data get older, such as in weighted least squares. We introduce exponential discounting via a power likelihood that we define as
\begin{equation}
\begin{aligned}
    \prod_{t=1}^n [f(y_t)]^{p_t(\lambda)}. \nonumber
\end{aligned}
\label{eq:powerlike}
\end{equation}
Note that in Section~\ref{sec4} and Section~\ref{sec5} for our simulation and empirical results, respectively, we do not consider exponential discounting due to the weak empirical evidence favoring discounting from our previous paper \cite{bib39-} on revenue forecasting. However, exponential discounting could still be advantageous in other financial applications. 

The \texttt{Stan} software and its corresponding R interface, \texttt{rstan}, employ Hamiltonian Monte Carlo (HMC) methods to generate posterior samples given a specified log-likelihood function and log-prior density; for more details on HMC methods, see \cite{bib40-}. This method contrasts with traditional Metropolis-Hastings (MH) algorithms used in software like \texttt{JAGS}. The workflow of \texttt{Stan} with HMC in \texttt{rstan} is (i) Define the Bayesian Model: The user specifies the likelihood function and prior distributions for parameters in the \texttt{Stan} modeling language; (ii) Automatic Log Posterior Gradient Computation: \texttt{Stan} internally computes the log posterior density by adding the log-likelihood, log-prior, and gradients; (iii) Gradient-Based Sampling with HMC: Rather than proposing random jumps like MH, HMC utilizes Hamiltonian dynamics, leveraging gradients of the log posterior to guide efficient sampling; (iv) Efficient Exploration of High-Dimensional Space: HMC uses momentum variables to simulate a particle moving through the parameter space, reducing random walk behavior and improving convergence speed; and (v) Adaptive Tuning via NUTS (No-U-Turn Sampler) \citep{bib41-}: \texttt{Stan} typically employs the NUTS variant of HMC, which adaptively selects optimal step sizes and trajectory lengths, removing the need for manual tuning. After one specifies a model in \texttt{Stan}, it gets compiled into C++ code which allows for fast computations. Because of these advantages, \texttt{Stan} with HMC is particularly well-suited for Bayesian hierarchical models, high-dimensional parameter spaces, and cases with complex posteriors—making it an excellent choice for modal regression and forecast combination models that require accurate capture of parameter uncertainty.

\section{Simulation Study}\label{sec4}
We perform simulation studies for each of our models posited in \eqref{eq:aldmodelhie}, \eqref{eq:snmodel}, and \eqref{eq:gmmodel}. For each distribution, we generated random data from the following R functions: (i) \texttt{rALD} from \texttt{ald} by \cite{bib35-}, (ii) \texttt{rasynorm} from \texttt{dirttee} by \cite{bib33-}, and (iii) \texttt{rRevGumbel} from \texttt{DescTools} by \cite{bib34-}. Our data generation process was across $N=500$ simulations with a sample size of $n=100$. In order to generate the data, we first set the location parameter as $\mu=w_0 +\sum_{j=1}^{m} x_{t,j} \omega_j$ where $w_0=0$ and $\bm{\omega}=(\frac{1}{4}, \frac{1}{4}, \frac{1}{4}, \frac{1}{4})^\prime$ (implying that $m=4$). The observed data $X$ consists of $n\times m=100\times 4$ observations that are random samples from a standard normal distribution. For both \eqref{eq:aldmodelhie} and \eqref{eq:snmodel}, we fix the scale parameter $\sigma=1$ and run three simulation studies across the asymmetry parameter $\tau\in\{0.25, 0.5, 0.75\}$. Whereas in \eqref{eq:gmmodel}, we ran three simulation studies across the scale/asymmetry parameter $\beta\in\{1, 5, 10\}$. 

We utilized \texttt{Stan} via the R interface \texttt{rstan}. The specified models contained two chains with 5,000 samples for the burn-in period followed by 10,000 samples for a total of $S=$20,000 post burn-in samples. We used the following uninformative priors: (i) $w_0 \sim \text{N}(0, 1000)$, (ii) $\bm{\omega} \sim \text{Dir}(1, 1, 1, 1)$, (iii) $\sigma\sim\text{InvGamma}(2,2)$, (iv) $\tau \sim \text{Beta}(1, 1)$, and (v) $\beta\sim\text{InvGamma}(2,2)$. Initially, we used Inverse Gamma priors for the parameters $\sigma$ and $\beta$ in our simulation study. However, following the recommended practices through documentation, we later switched to Half-Cauchy priors in Section~\ref{sec5} because they are often more well-behaved. Our simulation studies used the built-in \texttt{Stan} functions to check for empirical convergence of the MCMC samples, and the diagnostics showed no issues with numerical convergence. Lastly, the nine specified simulations (three models with three values for the asymmetry parameter) required only 3.66 to 7.56 seconds to run for a given dataset over the  $N=500$ simulations. 

Next, we present the results of the simulation studies. Consider the simulated dataset ${\cal D}_j$ for $j=1,\ldots,N$ (Monte Carlo sample size) with a sample size of $n$. Suppose $\theta_1^{(j)}, \ldots,\theta_S^{(j)}$ are post-burn-in MCMC samples based on the dataset ${\cal D}_j$. We define the following quantities for performance evaluation:
\begin{itemize}
    \item The posterior mean is estimated as $\bar{\theta}^{(j)} = {1\over S}\sum_{s=1}^S \theta_s^{(j)}$. 
    \item The posterior standard deviation is estimated as $\hat{\sigma}_\theta^{(j)} = \sqrt{{1\over S-1}\sum_{s=1}^S(\theta_s^{(j)}-\bar{\theta}^{(j)})^2}$.
    \item The Monte Carlo standard error (MCSE) is $\sqrt{{1\over N(N-1)}\sum_{j=1}^N(\bar{\theta}^{(j)}-\bar{\theta})^2}$, where $\bar{\theta}={1\over N}\sum_{j=1}^N\bar{\theta}^{(j)}$ is the sample average of posterior mean estimates.
    \item The bias as $\text{BIAS}=\bar{\theta} - \theta^{true}$ which we would expect to be nearly zero.
    \item The quantity $\text{AVG.SE}={1\over N}\sum_{j=1}^N \hat{\sigma}_\theta^{(j)}$ as the average standard error of the estimates.
    \item The coverage $\text{COV}=\frac{1}{N}\sum_{j=1}^{N} \mathbb{I}(\theta^{true} \in \{l^{(j)},u^{(j)}\})$ where $l^{(j)}$ and $u^{(j)}$ are the 0.025 and 0.975 quantiles of the posterior samples $\theta_1^{(j)}, \ldots,\theta_S^{(j)}$ providing the coverage of the 95\% credible intervals for $\theta$ based on the dataset ${\cal D}_j$. We should expect this value to be close to 0.95.
\end{itemize}
If $N$ is chosen sufficiently large we should expect $\text{MCSE}\approx {1\over N}\sum_{j=1}^N \hat{\sigma}_\theta^{(j)}=\text{AVG.SE}$ and the bias, $\text{BIAS}=\bar{\theta} - \theta^{true}\approx 0$. The results for our posited models in \eqref{eq:aldmodelhie}, \eqref{eq:snmodel}, and \eqref{eq:gmmodel} are found in Table~\ref{tab:asymetriclaplace_sim}, Table~\ref{tab:asymetricnormal_sim}, and Table~\ref{tab:gumbelmin_sim} respectively. There are no issues with the results of our simulation studies. All the biases were nearly zero (i.e., no significant statistical biases), the average standard errors were close to the MCSE (indicating that we conducted a sufficient number of MC runs), and the coverages were around 0.95, though in a few cases, the COV estimates slightly exceeded the expected values for a given sample size $n$.

\begin{table}[!ht]
\centering
\begin{tabular}{l c r r r r}
\hline
\textbf{Parameter} 
& \boldmath{$\tau$} 
& \textbf{BIAS} 
& \textbf{AVG.SE} 
& \textbf{MCSE} 
& \textbf{COV} \\
\hline
\multirow{3}{*}{$w_{0}$}
  & 0.25 &  0.027 & 0.018 & 0.016 & 0.970 \\
  & 0.75 & -0.044 & 0.018 & 0.018 & 0.970 \\
  & 0.50 &  0.010 & 0.015 & 0.015 & 0.958 \\
\hline
\multirow{3}{*}{$w_{1}$}
  & 0.25 & -0.001 & 0.007 & 0.004 & 0.998 \\
  & 0.75 &  0.001 & 0.007 & 0.004 & 0.998 \\
  & 0.50 & -0.000 & 0.006 & 0.004 & 1.000 \\
\hline
\multirow{3}{*}{$w_{2}$}
  & 0.25 & -0.006 & 0.006 & 0.004 & 0.992 \\
  & 0.75 & -0.006 & 0.006 & 0.004 & 0.998 \\
  & 0.50 & -0.002 & 0.006 & 0.004 & 0.994 \\
\hline
\multirow{3}{*}{$w_{3}$}
  & 0.25 &  0.006 & 0.006 & 0.005 & 1.000 \\
  & 0.75 &  0.002 & 0.006 & 0.004 & 0.998 \\
  & 0.50 &  0.003 & 0.006 & 0.005 & 0.990 \\
\hline
\multirow{3}{*}{$w_{4}$}
  & 0.25 &  0.002 & 0.007 & 0.004 & 0.998 \\
  & 0.75 &  0.003 & 0.006 & 0.004 & 1.000 \\
  & 0.50 & -0.001 & 0.006 & 0.004 & 0.992 \\
\hline
\multirow{3}{*}{$\sigma$}
  & 0.25 &  0.003 & 0.007 & 0.006 & 0.970 \\
  & 0.75 &  0.005 & 0.007 & 0.007 & 0.962 \\
  & 0.50 & -0.011 & 0.005 & 0.005 & 0.942 \\
\hline
\multirow{3}{*}{$\tau$}
  & 0.25 &  0.005 & 0.002 & 0.002 & 0.970 \\
  & 0.75 & -0.007 & 0.002 & 0.002 & 0.960 \\
  & 0.50 &  0.001 & 0.002 & 0.002 & 0.958 \\
\hline
\end{tabular}
\caption{Asymmetric Laplace simulation results (using \texttt{Stan}) across $\tau\in\{0.25, 0.5, 0.75\}$.}
\label{tab:asymetriclaplace_sim}
\end{table}

\begin{table}[!ht]
\centering
\begin{tabular}{l c r r r r}
\hline
\textbf{Parameter} 
& \boldmath{$\tau$} 
& \textbf{BIAS} 
& \textbf{AVG.SE} 
& \textbf{MCSE} 
& \textbf{COV} \\
\hline
\multirow{3}{*}{$w_{0}$}
  & 0.25 &  0.077 & 0.013 & 0.011 & 0.970 \\
  & 0.75 & -0.076 & 0.012 & 0.011 & 0.952 \\
  & 0.50 &  0.002 & 0.011 & 0.010 & 0.972 \\
\hline
\multirow{3}{*}{$w_{1}$}
  & 0.25 & -0.001 & 0.004 & 0.004 & 0.970 \\
  & 0.75 &  -0.001 & 0.004 & 0.004 & 0.974 \\
  & 0.50 & -0.001 & 0.004 & 0.004 & 0.960 \\
\hline
\multirow{3}{*}{$w_{2}$}
  & 0.25 & -0.004 & 0.004 & 0.004 & 0.954 \\
  & 0.75 & -0.004 & 0.004 & 0.004 & 0.952 \\
  & 0.50 & -0.004 & 0.004 & 0.003 & 0.944 \\
\hline
\multirow{3}{*}{$w_{3}$}
  & 0.25 &  0.003 & 0.004 & 0.004 & 0.964 \\
  & 0.75 &  0.002 & 0.004 & 0.004 & 0.964 \\
  & 0.50 &  0.003 & 0.004 & 0.004 & 0.964 \\
\hline
\multirow{3}{*}{$w_{4}$}
  & 0.25 &  0.002 & 0.004 & 0.004 & 0.958 \\
  & 0.75 &  0.003 & 0.004 & 0.004 & 0.966 \\
  & 0.50 & 0.002 & 0.004 & 0.004 & 0.960 \\
\hline
\multirow{3}{*}{$\sigma$}
  & 0.25 &  -0.008 & 0.006 & 0.005 & 0.982 \\
  & 0.75 &  -0.009 & 0.006 & 0.005 & 0.986 \\
  & 0.50 & -0.046 & 0.004 & 0.004 & 0.938 \\
\hline
\multirow{3}{*}{$\tau$}
  & 0.25 &  0.053 & 0.005 & 0.004 & 0.972 \\
  & 0.75 & -0.054 & 0.005 & 0.005 & 0.968 \\
  & 0.50 &  0.000 & 0.006 & 0.005 & 0.966 \\
\hline
\end{tabular}
\caption{Asymmetric normal simulation results (using \texttt{Stan}) across $\tau\in\{0.25, 0.5, 0.75\}$.}
\label{tab:asymetricnormal_sim}
\end{table}

\begin{table}[ht!]
\centering
\begin{tabular}{l c r r r r}
\hline
\textbf{Parameter} 
& \boldmath{$\sigma$}
& \textbf{BIAS} 
& \textbf{AVG.SE} 
& \textbf{MCSE} 
& \textbf{COV} \\
\hline
\multirow{3}{*}{$w_{0}$}
 & 1   & -0.003 & 0.005 & 0.005 & 0.958 \\
 & 5   &  0.001 & 0.024 & 0.023 & 0.954 \\
 & 10  &  0.014 & 0.047 & 0.046 & 0.954 \\
\hline
\multirow{3}{*}{$w_{1}$}
 & 1   &  0.005 & 0.004 & 0.004 & 0.968 \\
 & 5   &  0.004 & 0.008 & 0.003 & 1.000 \\
 & 10  &  0.002 & 0.008 & 0.002 & 1.000 \\
\hline
\multirow{3}{*}{$w_{2}$}
 & 1   &  0.002 & 0.004 & 0.003 & 0.950 \\
 & 5   &  0.004 & 0.008 & 0.004 & 1.000 \\
 & 10  &  0.003 & 0.008 & 0.002 & 1.000 \\
\hline
\multirow{3}{*}{$w_{3}$}
 & 1   & -0.003 & 0.004 & 0.004 & 0.952 \\
 & 5   & -0.003 & 0.008 & 0.004 & 1.000 \\
 & 10  & -0.002 & 0.008 & 0.002 & 1.000 \\
\hline
\multirow{3}{*}{$w_{4}$}
 & 1   & -0.003 & 0.004 & 0.003 & 0.940 \\
 & 5   & -0.005 & 0.008 & 0.003 & 1.000 \\
 & 10  & -0.002 & 0.008 & 0.002 & 1.000 \\
\hline
\multirow{3}{*}{$\beta$}
 & 1   &  0.009 & 0.004 & 0.003 & 0.950 \\
 & 5   & -0.010 & 0.018 & 0.017 & 0.944 \\
 & 10  & -0.014 & 0.035 & 0.034 & 0.944 \\
\hline
\end{tabular}
\caption{Reverse Gumbel simulation results (using \texttt{Stan}) across $\beta\in\{1, 5, 10\}$.}
\label{tab:gumbelmin_sim}
\end{table}

\section{Application to Real Datasets}\label{sec5}
We produced our empirical results with the exact out-of-sample procedure as our previous paper to compare to Bayesian mean regression with normally distributed errors. For all the details and justifications, see \cite{bib39-}; however, we briefly recap the overall procedure here. We utilize quarterly revenue data from the Institutional Brokers' Estimate System (I/B/E/S) within the Wharton Research Data Services (WRDS). 

Our sample contains 23 technology companies across the years 2015 through 2023. The dataset includes the ground truth quarterly revenues with corresponding one-quarter-ahead analyst forecasts. Since an analyst can update their forecast for a given quarter, we only include an analyst's most recent forecast, i.e., the forecast closest to the ground truth release date. We obtained our empirical results with a rolling window cross-validation procedure with 24 folds and a window length of 12. For each fold in the rolling window cross-validation, the goal is to predict the $y_{t+1}$ value given the analyst forecasts $\bm{x}_{t+1}$ (that is available before $y_{t+1}$ is released) and the training data $\mathcal{D}_f$. See Section~\ref{sec2.2} for a recap of the rolling window cross-validation procedure.

For all models shown in \eqref{eq:aldmodelhie}, \eqref{eq:snmodel}, and \eqref{eq:gmmodel}, we define the prior distributions for the parameters as
\begin{equation}
\begin{aligned}
    &\bm{\omega} \sim \text{Dir}(1, 1, \ldots, 1), \\
    &w_0 \sim \text{N}(0, 1), \\
    &\sigma, \beta \sim \text{Half-Cauchy}(0, 1), \\
    &\tau \sim \text{Beta}(2,2). \\
\end{aligned}
\label{eq:priors_specified}
\end{equation}
See Section~\ref{A1} of the appendix for definitions of the corresponding probability density functions. Setting all of the hyperparameters $\alpha_1=\alpha_2=\ldots=\alpha_m=1$ for $\bm{\omega}$ indicates a uniform distribution across all weights $\omega_1, \omega_2, \ldots, \omega_m$ because the $\text{Dir}(1,1,\ldots,1)$ distribution is uniform supported on the simplex ${\cal S}_m$. Next, based on our previous parameter estimates from our previous paper, we had a general idea of the range of the parameter estimates for the intercept $w_0$ and scale parameters and note that these are still somewhat large ranges as our data is on the logarithmic scale. Lastly, we chose a $\text{Beta}(2,2)$ for $\tau$ since there was limited asymmetry in the residual plots from our previous paper. This choice gives slightly more weight when $\tau$ is closer to 0.5. Our Bayesian models utilized four chains with a burn-in of 5,000 samples followed by 10,000 samples for a total of 20,000 MCMC samples. Similar to Section~\ref{sec4}, we used the built-in functions to check for numerical convergence and encountered no problems. Lastly, if an analyst $j$ was missing a forecast for quarter $t$, we used the following missing data model: $X_{t,j} \sim \text{N}(\hat{\mu}_t, \hat{\sigma}^2_t)$ where $\hat{\mu}_t$ and $\hat{\sigma}^2_t$ are the mean and variance across all analysts for a given quarter $t$. Using the cross-sectional estimates in revenue forecasting aligns with \cite{bib25-} that imputed missing values with the mean across all analysts for a given quarter. Our missing data model allows us to utilize the posterior predictive distribution (PPD) to impute any missing values given the observed data through Monte Carlo sampling.

\begin{table}[ht]
\centering
\begin{tabular}{lccc}
\toprule
\textbf{Ticker} 
& \textbf{Asymmetric Laplace} 
& \textbf{Asymmetric Normal} 
& \textbf{Reverse Gumbel} \\
\midrule
AAPL & 83.3 & 83.3 & 83.3 \\
ACN  & 87.5 & 87.5 & 87.5 \\
ADBE & 91.7 & 91.7 & 91.7 \\
ADI  & 87.5 & 87.5 & 87.5 \\
ADP  & 70.8 & 54.2 & 70.8 \\
AMAT & 87.5 & 87.5 & 87.5 \\
AMD  & 79.2 & 79.2 & 79.2 \\
ANET & 95.8 & 100.0 & 100.0 \\
AVGO & 83.3 & 79.2 & 79.2 \\
CRM  & 100.0 & 100.0 & 100.0 \\
CSCO & 87.5 & 87.5 & 87.5 \\
IBM  & 45.8 & 37.5 & 66.7 \\
INTC & 83.3 & 83.3 & 83.3 \\
INTU & 87.5 & 83.3 & 79.2 \\
KLAC & 100.0 & 100.0 & 100.0 \\
MSFT & 87.5 & 87.5 & 87.5 \\
MU   & 83.3 & 83.3 & 83.3 \\
NOW  & 87.5 & 87.5 & 87.5 \\
NVDA & 87.5 & 87.5 & 87.5 \\
ORCL & 54.2 & 50.0 & 58.3 \\
PANW & 91.7 & 91.7 & 91.7 \\
QCOM & 75.0 & 75.0 & 75.0 \\
TXN  & 75.0 & 75.0 & 75.0 \\
\midrule
\textbf{Mean} & 83.2 & 81.7 & 83.9 \\
\bottomrule
\end{tabular}
\caption{Comparison of hit rates (in \%) across three Bayesian modal regression models with asymmetric Laplace, asymmetric normal, and reverse Gumbel error distributions.}
\label{tab:hr_comparison}
\end{table}

\begin{table}[ht]
\centering
\begin{tabular}{lccc}
\toprule
\textbf{Ticker} 
& \textbf{Asymmetric Laplace} 
& \textbf{Asymmetric Normal} 
& \textbf{Reverse Gumbel} \\
\midrule
AAPL & 66.7 & 62.5 & 41.7 \\
ACN  & 66.7 & 58.3 & 45.8 \\
ADBE & 66.7 & 62.5 & 54.2 \\
ADI  & 79.2 & 70.8 & 54.2 \\
ADP  & 62.5 & 41.7 & 54.2 \\
AMAT & 79.2 & 79.2 & 79.2 \\
AMD  & 41.7 & 41.7 & 37.5 \\
ANET & 70.8 & 70.8 & 54.2 \\
AVGO & 66.7 & 54.2 & 50.0 \\
CRM  & 79.2 & 79.2 & 62.5 \\
CSCO & 79.2 & 70.8 & 66.7 \\
IBM  & 33.3 & 29.2 & 50.0 \\
INTC & 62.5 & 66.7 & 58.3 \\
INTU & 75.0 & 70.8 & 54.2 \\
KLAC & 83.3 & 83.3 & 79.2 \\
MSFT & 75.0 & 79.2 & 66.7 \\
MU   & 58.3 & 62.5 & 54.2 \\
NOW  & 66.7 & 62.5 & 62.5 \\
NVDA & 75.0 & 70.8 & 58.3 \\
ORCL & 41.7 & 37.5 & 33.3 \\
PANW & 62.5 & 66.7 & 62.5 \\
QCOM & 66.7 & 66.7 & 54.2 \\
TXN  & 58.3 & 52.5 & 50.0 \\
\midrule
\textbf{Mean} & 66.0 & 63.0 & 55.8 \\
\bottomrule
\end{tabular}
\caption{Comparison of win rates (in \%) across three Bayesian modal regression models with asymmetric Laplace, asymmetric normal, and reverse Gumbel error distributions.}
\label{tab:wr_comparison}
\end{table} 

We present our results in terms of two relevant metrics in revenue forecasting. To justify why the hit and win rates are relevant in revenue forecasting, refer to \cite{bib39-}. First, we define the hit rate as the proportion of times our model correctly classifies the sign of the revenue surprise. Next, we define the win rate as the proportion of times our model's forecast is closer to the actual value than the equally weighted consensus. We proceed to compare our three Bayesian modal regression models defined in \eqref{eq:aldmodelhie}, \eqref{eq:snmodel}, and \eqref{eq:gmmodel} against our Bayesian mean regression with normally distributed errors from \cite{bib39-}. Note that our previous mean regression model achieved a hit rate of 82.4\% and a win rate of 62.7\%. In contrast, our asymmetric Laplace, asymmetric normal, and reverse Gumbel models achieved 83.2\%, 81.7\%, and 83.9\% hit rates and 66.0\%, 63.0\%, and 55.8\% win rates, respectively. 

\textit{The asymmetric Laplace was the only approach that achieved a higher average hit and win rate than mean regression with normally distributed errors}. One possible reason is that our rolling window cross-validation was performed over COVID—a time of market uncertainty—meaning it could have handled outliers better. After all, it is a fat-tailed distribution, unlike the asymmetric normal, normal, and reverse Gumbel distributions. This result suggests that some degree of asymmetry and fat tails may be present in the error distributions, which aligns with \cite{bib21-}, who note that the asymmetric Laplace distribution is highly applicable to financial data because it can accommodate both fat tails and asymmetry. Next, the asymmetric normal achieved almost identical results to the symmetric normal case, suggesting that the degree of asymmetry is mild—or possibly the fat tails dominated, suppressing the amount of asymmetry captured. Lastly, the reverse Gumbel achieved a high hit rate but a low win rate, possibly because it can only handle a single skew direction, unlike the asymmetric Laplace and asymmetric normal distributions. Due to this requirement, we did not expect the reverse Gumbel to perform well.

\begin{figure}[ht]
    \centering
    \includegraphics[width=0.48\textwidth]{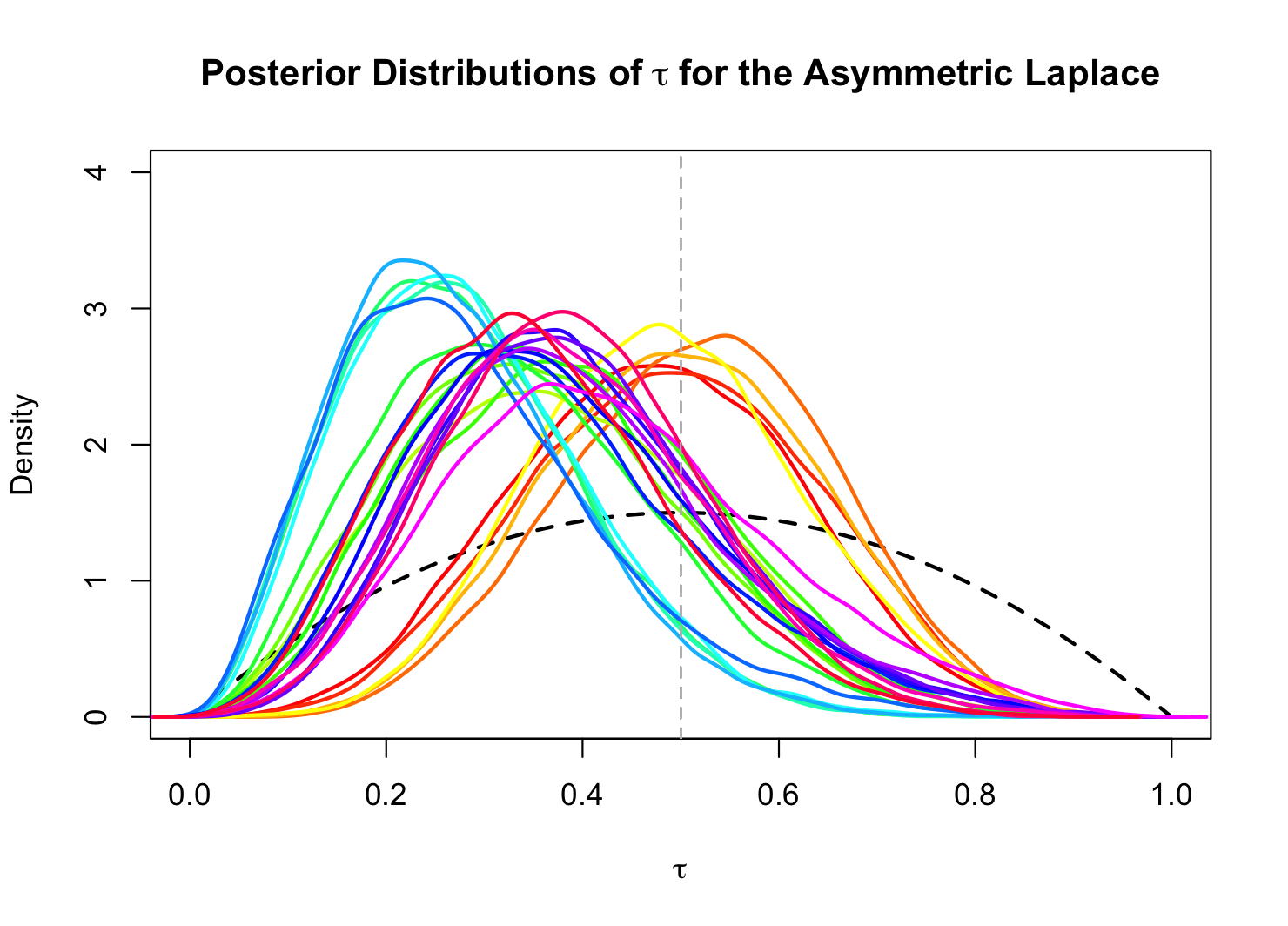}
    \includegraphics[width=0.48\textwidth]{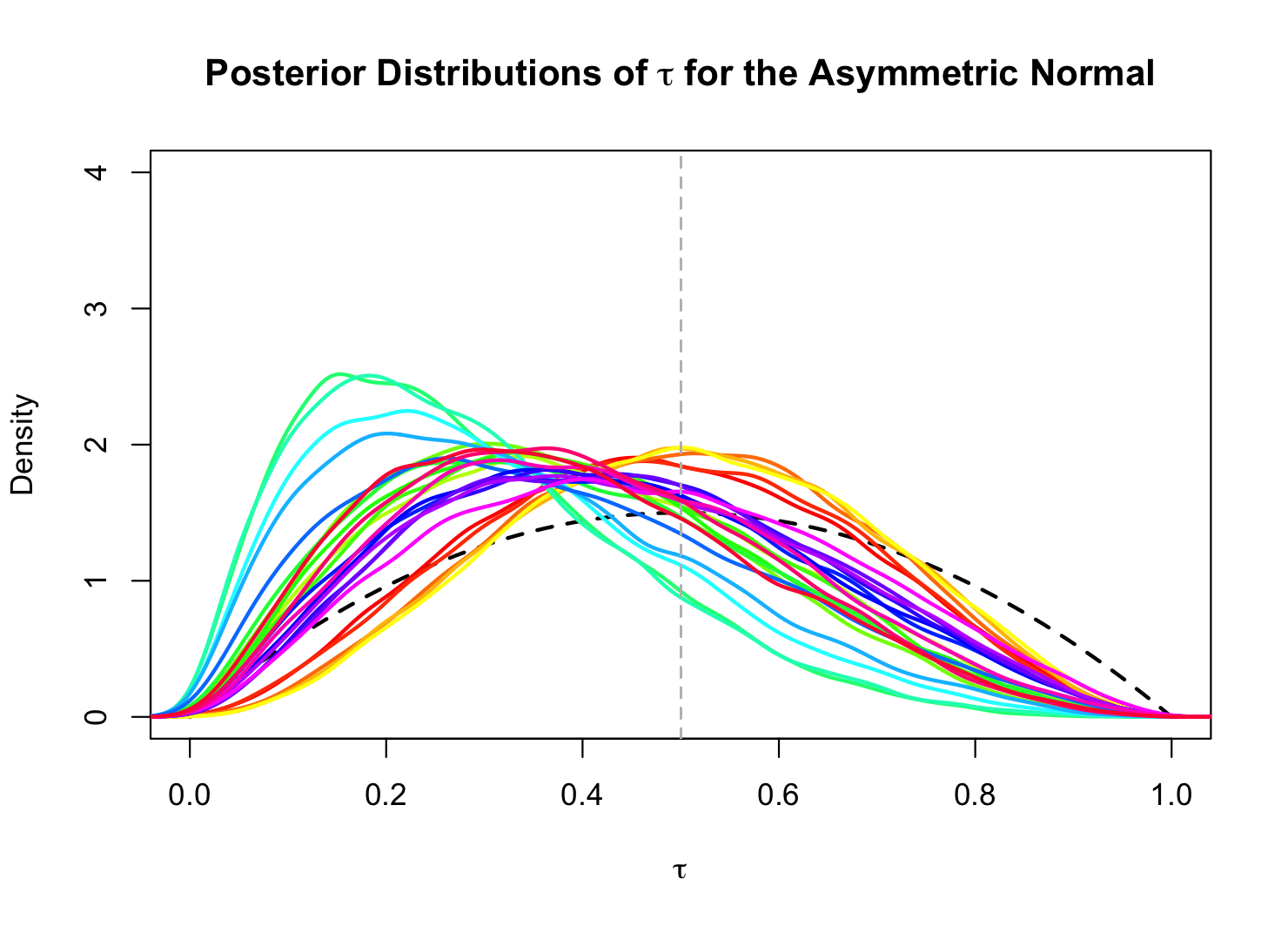}
    \caption{Posterior distributions of $\tau$ for AAPL across all 24 folds. The black dashed line represents the $\text{Beta}(2,2)$ prior distribution.}
    \label{fig:Tau}
\end{figure}

One interesting takeaway from our approach is that since no hyperparameter tuning is needed, each model's asymmetry parameter can vary across a continuous range in the rolling window cross-validation. To illustrate this time-varying parameter concept for the asymmetry parameter, we present an example for the ticker AAPL. Figure~\ref{fig:Tau} depicts the posterior distributions of the asymmetry parameter $\tau$ across all 24 folds for the asymmetric normal and asymmetric Laplace models along with an overlayed $\tau \sim \text{Beta}(2,2)$ prior. This example provides evidence of time-varying parameter combinations and asymmetric error distributions, as the densities change over time and $\tau$ does not consistently center at 0.5 (i.e., the symmetric case). For a more holistic view of time-varying parameter parameters in the models, we included posterior predictive distribution (PPD) plots across all 24 folds and tickers; see Section~\ref{A4} of the appendix. We center the PPD by taking the residuals defined as $y_t - \hat{y}_t$, where $y_t$ denotes the actual value and $\hat{y}_t$ denotes the predictive value from the PPD. Clearly, the PPDs for specific tickers fluctuate over time and contain asymmetry. 

We compare our work to two other studies, as seen in our previous paper. Regarding hit rate, we came across the real-world revenue nowcasting company AKAnomics, and \cite{bib37-} disclosed an average hit rate of 66\%. Note that their modeling approaches are inaccessible to the public. This result is shy of our 83.2\%, 81.7\%, and 83.9\% hit rates for the models posited in \eqref{eq:aldmodelhie}, \eqref{eq:snmodel}, and \eqref{eq:gmmodel} respectively. Next, we compare our results against prior research from \cite{bib38-} that reported an average win rate of 57.2\% across 34 companies from 2015 to 2019. Their single classical linear systems model had 306 out-of-sample predictions, whereas each of our Bayesian modal regression models had 552 out-of-sample predictions. We note that their modeling approach only considered alternative data, specifically, a credit card transactions dataset to forecast companies' quarterly revenue, which could be why our models performed better. Comparing our work to \cite{bib37-} and \cite{bib38-} is difficult since both did not disclose which companies make up their hit and win rates, respectively. 

\section{Discussion}\label{sec6}
While forecast combinations primarily have been centered around symmetric losses, we introduce a framework to use asymmetric losses through Bayesian modal regression. This approach is unique since some asymmetric error distributions can also handle the symmetric case. For example, the asymmetric Laplace and normal distributions reduce to the Laplace and normal distributions when the asymmetry parameter $\tau=0.5$. Thus, the Laplace and normal distributions are special cases of their asymmetric counterparts. 

Our empirical study only focused on quarterly revenue forecasting. However, there is an extensive history of asymmetry and fat tails present in financial data. Researchers have previously utilized the three error distributions we considered. For the asymmetric Laplace, see \cite{bib21-} for currency exchange rates, \cite{bib3-} for option pricing (jump models), and \cite{bib4-} for portfolio selection. Next, refer to \cite{bib19-} for inflation forecasting with split/two-piece normal (the asymmetric normal in our case). Lastly, see \cite{bib32-} for option pricing via the Gumbel distribution. Although we did not use exponential discounting in our out-of-sample results due to low-frequency quarterly data, we are optimistic that this feature can be highly advantageous in higher-frequency financial applications (e.g., volatility forecasting and FX trading). Our code for the three Bayesian modal regression models will be available on GitHub after our work is published.

\bibliographystyle{unsrtnat}  
\bibliography{references}
\newpage

\appendix 

\section{List of Distributions}\label{A1}
\begin{table}[ht]
\centering
\label{tab:distPDF}
\begin{tabular}{lll}
\hline
\textbf{Distribution} & \textbf{Notation} & \textbf{PDF} \\
\hline

Asymmetric Laplace 
& $\mathcal{AL}(\mu,\sigma,\tau)$
& $\frac{\tau(1-\tau)}{\sigma}
    \begin{cases} 
      \exp\left(\frac{(1-\tau)(x-\mu)}{\sigma}\right), & \text{if } x \leq \mu, \\
      \exp\left(-\frac{\tau(x-\mu)}{\sigma}\right), & \text{if } x > \mu,
    \end{cases}$ \\[2.5em]

Asymmetric Normal 
& $\mathcal{AN}(\mu,\sigma,\tau)$
& $\frac{2 \sqrt{\tau (1 - \tau)}}{\sigma \sqrt{\pi} (\sqrt{\tau} + \sqrt{1 - \tau})}
    \begin{cases} 
      \exp\left(-\frac{(1-\tau)(x-\mu)^2}{\sigma^2}\right), & \text{if } x \leq \mu, \\
      \exp\left(-\frac{\tau(x-\mu)^2}{\sigma^2}\right), & \text{if } x > \mu,
    \end{cases}$ \\[2.5em]

Beta 
& $\text{Beta}(\alpha,\beta)$
& $\frac{\Gamma(\alpha+\beta)}{\Gamma(\alpha) \Gamma(\beta)} x^{\alpha - 1}(1 - x)^{\beta - 1}$ \\[2.5em]

Dirichlet 
& $\text{Dir}(\alpha_1,\dots,\alpha_m)$
& $\frac{\Gamma(\alpha_1 + \dots + \alpha_m)}{\Gamma(\alpha_1)\cdots \Gamma(\alpha_m)}
  \prod_{i=1}^m x_i^{\alpha_i - 1}$ \\[2.5em]

Half-Cauchy 
& $\text{Half-Cauchy}(x_0,\gamma)$
& $\begin{cases}
    \frac{2}{\pi \gamma}\ \frac{1}{1+(x - x_0)^2 / \gamma^2}, & \text{if } x \geq x_0,\\
    0, & \text{if } x < x_0,
  \end{cases}$ \\[2.5em]

Inverse Gamma 
& $\text{InvGamma}(\alpha,\beta)$
& $\frac{\beta^\alpha}{\Gamma(\alpha)}x^{-\alpha - 1}\exp\left(-\frac{\beta}{x}\right)$ \\[2.5em]

Normal 
& $\text{N}(\mu,\sigma^2)$
& $\frac{1}{\sqrt{2\pi\sigma^2}}\exp\left(-\frac{(x-\mu)^2}{2 \sigma^2}\right)$ \\[2.5em]

Reverse Gumbel 
& $\mathcal{RG}(\mu,\beta)$
& $\frac{1}{\beta} \exp\left(\frac{x-\mu}{\beta}\right) \exp\left(-\exp\left(\frac{x-\mu}{\beta}\right)\right)$ \\[1em]

\hline
\end{tabular}
\caption{List of distributions and their corresponding notation and PDF.}
\end{table}

\FloatBarrier

\section{Stochastic Representation of ALD}\label{A2}
We came across one stochastic representation for the asymmetric Laplace; thus, we consider using data augmentation by introducing latent (i.e., unobservable) variables to transform the target distribution into simpler, well-known distributions. This procedure leads to closed-form, full-conditional distributions for the model parameters and the latent variables, allowing for the use of Gibbs sampling. 
\vspace{1em}
\begin{lemma}
    Let $Y|V \sim N(\mu + \beta \left(\frac{1}{\kappa} - \kappa\right)V, 2\beta^2 V)$ with unknown parameter $V \sim Exp(1)$. Then the result of compounding $f_{Y|V}$ with $f_{V}$ is
    \begin{equation*}
    \begin{aligned}
        f_Y(y) = \int f_{Y|V}(y|v) f_V(v) dv,
    \end{aligned}
    \end{equation*}
    where the resulting marginal distribution $f_{Y}$ is the asymmetric Laplace distribution, i.e., $Y\sim\mathcal{AL}^*(\mu, \beta, \kappa)$ with pdf
    \begin{equation}
    \label{eq:ald_different}
        f_{AL}(y;\mu,\beta,\kappa) = \frac{1}{\beta} \frac{\kappa}{1+\kappa^2}
        \begin{cases} 
          \exp(-\frac{\kappa}{\beta}(y-\mu)), & \text{if } y < \mu, \\
          \exp(-\frac{1}{\kappa\beta}(\mu-y)), & \text{if } y \geq \mu.
        \end{cases}
    \end{equation}
    \cite{bib5-} note this stochastic representation; however, they did not include a formal proof. We include the proof of this Lemma in Section~\ref{lemma_proof_3} of the appendix.
\end{lemma}
\vspace{1em}
Using this stochastic representation, we can use a forecast combinations framework to express our Bayesian modal regression model. We define the Bayesian hierarchical model as
\begin{equation}
\label{eq:latentmodel}
\begin{aligned}
    &v_t \sim \text{Exp}(1) \\
    &y_t|v_t\sim \text{N}\left(w_0 +\sum_{j=1}^{m} x_{t,j} \omega_j + \beta \left(\frac{1}{\kappa} - \kappa\right)v_t, 2\beta^2v_t\right), \\
    &\bm{\omega} \sim \text{Dir}(\alpha_1, \alpha_2, \ldots, \alpha_m), \\
    &w_0 \sim \text{N}(0, \tau_{w_0}^2), \\
    &\kappa \sim \text{U}(c_0, d_0), \\
    &\beta \sim \text{InvGamma}(a_0, b_0),
\end{aligned}
\end{equation}
with the goal of forecasting $y_{n+1}|v_{n+1}$. In \eqref{eq:latentmodel}, $\text{Exp}(\lambda)$ denotes an exponential distribution with rate parameter $\lambda>0$, $\text{N}(\mu, \sigma^2)$ denotes a normal distribution with mean $\mu\in\mathbb{R}$ and variance $\sigma^2>0$, $\text{Dir}(\alpha_1, \alpha_2, \ldots, \alpha_m)$ denotes a Dirichlet distribution with parameters $\alpha_1>0, \alpha_2>0, \ldots, \alpha_m>0$, $\text{U}(c_0, d_0)$ denotes a uniform distribution with lower bound $c_0>0$ and upper bound $d_0>c_0$, and $\text{InvGamma}(a_0, b_0)$ denotes Inverse Gamma distribution with shape parameter $a_0>0$ and scale parameter $b_0>0$. 

\begin{table}[!ht]
\centering
\begin{tabular}{l c r r r r}
\hline
\textbf{Parameter} 
& \boldmath{$\kappa$} 
& \textbf{BIAS} 
& \textbf{AVG.SE} 
& \textbf{MCSE} 
& \textbf{COV} \\
\hline
\multirow{3}{*}{$\kappa$}
 & 0.5 &  0.005 & 0.003 & 0.003 & 0.956 \\
 & 1.0 &  0.025 & 0.005 & 0.005 & 0.954 \\
 & 2.0 &  0.187 & 0.016 & 0.016 & 0.932 \\
\hline
\multirow{3}{*}{$\beta$}
 & 0.5 & -0.010 & 0.006 & 0.006 & 0.952 \\
 & 1.0 & -0.007 & 0.005 & 0.005 & 0.952 \\
 & 2.0 & -0.040 & 0.006 & 0.006 & 0.934 \\
\hline
\multirow{3}{*}{$w_{0}$}
 & 0.5 &  0.016 & 0.009 & 0.009 & 0.974 \\
 & 1.0 &  0.011 & 0.008 & 0.008 & 0.966 \\
 & 2.0 &  0.029 & 0.008 & 0.008 & 0.956 \\
\hline
\multirow{3}{*}{$w_{1}$}
 & 0.5 &  0.004 & 0.004 & 0.004 & 0.966 \\
 & 1.0 &  0.003 & 0.004 & 0.004 & 0.954 \\
 & 2.0 &  0.000 & 0.004 & 0.003 & 0.960 \\
\hline
\multirow{3}{*}{$w_{2}$}
 & 0.5 & -0.001 & 0.004 & 0.004 & 0.966 \\
 & 1.0 & -0.005 & 0.004 & 0.004 & 0.946 \\
 & 2.0 & -0.001 & 0.004 & 0.004 & 0.942 \\
\hline
\multirow{3}{*}{$w_{3}$}
 & 0.5 &  0.006 & 0.004 & 0.004 & 0.964 \\
 & 1.0 &  0.010 & 0.004 & 0.004 & 0.974 \\
 & 2.0 &  0.006 & 0.004 & 0.004 & 0.960 \\
\hline
\multirow{3}{*}{$w_{4}$}
 & 0.5 & -0.009 & 0.004 & 0.004 & 0.966 \\
 & 1.0 & -0.008 & 0.004 & 0.004 & 0.968 \\
 & 2.0 & -0.005 & 0.004 & 0.004 & 0.952 \\
\hline
\end{tabular}
\caption{Asymmetric Laplace simulation results (in \texttt{JAGS}) across $\kappa\in\{0.5, 1, 2\}$.}
\label{tab:asymetriclaplace_sim_jags}
\end{table}

Next, we present a simulation study for the model in \eqref{eq:latentmodel} via the Bayesian software package \texttt{JAGS}. Using a MCMC sample size of $N=500$, $n=100$ data points were sampled from $Y_t|V_t\sim \text{N}\left(w_0 +\sum_{j=1}^{m} x_{t,j} \omega_j + \beta \left(\frac{1}{\kappa} - \kappa\right)V_t, 2\beta^2V_t\right)$ where $V_t \sim \text{Exp}(1)$. We used the \texttt{rnorm} and \texttt{rexp} functions from R to sample from these distributions. We set the scale parameter $\beta=1$ then performed three simulation studies across the asymmetry parameter $\kappa\in\{0.5, 1, 2\}$. The uninformative priors specified were $\kappa \sim \text{U}(0.001, 4)$ and $\beta \sim \text{Gamma}(2, 2)$. Figure 1 from \cite{bib5-} shows that the parameter $\kappa$ asymptotically falls within the range zero to four, which motivates us to use this range for the prior.

The model in \texttt{JAGS} contained 10,000 samples for the burn-in followed by 20,000 samples. The trace plots produced by JAGS assessed convergence and no issues were present. Table~\ref{tab:asymetriclaplace_sim_jags} presents the results of our simulation study across $\kappa\in\{0.5, 1, 2\}$. One concern with this model is the slightly elevated bias when $\kappa = 2$, which further supports our decision to use \texttt{Stan}.

\FloatBarrier

\section{Proof of Lemma 2}
\label{lemma_proof_2}
Without any loss of generality, we assume that $\sigma=1$, because $\mathbb{E}(\epsilon^k)=\sigma^k\mathbb{E}(\tilde{\epsilon})$ for $k=1,2,3$, where $\tilde{\epsilon}$ can be assumed to have a density with $\sigma=1$. Consider the density $f(\epsilon;\sigma,\tau)$ as defined in Lemma \ref{lemma1} of a generic split distribution. Then, the first three moments are given by
\begin{enumerate}
    \item 
        Assume that $c_1=\int_{0}^\infty u g(u) du<\infty$. The first moment, $\mathbb{E}(\epsilon|\tau)$ is given by
        \begin{equation*}
        \begin{aligned}
            \int_{-\infty}^\infty \epsilon f(\epsilon|\tau) d\epsilon 
            &= 2\tau(1-\tau)\left[\int_{-\infty}^0 \epsilon g((1-\tau)\epsilon) d\epsilon + \int_{0}^\infty \epsilon g(\tau\epsilon) d\epsilon\right] \\
            &= 2\tau(1-\tau)\left[\int_{-\infty}^0 \left(\frac{u}{1-\tau}\right) g(u) \frac{1}{1-\tau} du + \int_{0}^\infty \left(\frac{u}{\tau}\right) g(u) \frac{1}{\tau} du\right] \\
            &= 2\tau(1-\tau)\left[-\frac{1}{(1-\tau)^2} \int_{0}^\infty u g(u) du + \frac{1}{\tau^2} \int_{0}^\infty u g(u) du\right] \\
            &= 2\tau(1-\tau)\left[-\frac{c_1}{(1-\tau)^2} + \frac{c_1}{\tau^2}\right] \\
            &= -\frac{2\tau}{1-\tau}c_1 + \frac{2\tau(1-\tau)}{\tau^2}c_1 \\
            &= 2c_1 \left[-\frac{\tau}{1-\tau} + \frac{(1-\tau)}{\tau}\right] \\
            &= 2c_1 \left[\frac{-\tau^2+(1-\tau)^2}{\tau(1-\tau)}\right] \\
            &= \frac{2c_1}{\tau (1-\tau)}(1-2\tau).
        \end{aligned}
        \end{equation*}
    \item 
        Assume that $c_2=\int_{0}^\infty u^2 g(u) du<\infty$. The second moment, $\mathbb{E}(\epsilon^2|\tau)$, is given by
        \begin{equation*}
        \begin{aligned}
            \int_{-\infty}^\infty \epsilon^2 f(\epsilon|\tau) d\epsilon  
            &= 2\tau(1-\tau)\left[\int_{-\infty}^0 \epsilon^2 g((1-\tau)\epsilon) d\epsilon + \int_{0}^\infty \epsilon^2 g(\tau\epsilon) d\epsilon\right] \\
            &= 2\tau(1-\tau)\left[\int_{-\infty}^0 \left(\frac{u}{1-\tau}\right)^2 g(u) \frac{1}{1-\tau} du + \int_{0}^\infty \left(\frac{u}{\tau}\right)^2 g(u) \frac{1}{\tau}du\right] \\
            &= 2\tau(1-\tau)\left[\frac{1}{(1-\tau)^3}\int_{-\infty}^0 u^2 g(u) du + \frac{1}{\tau^3}\int_{0}^\infty u^2 g(u) du\right] \\
            &= 2\tau(1-\tau)\left[\frac{c_2}{(1-\tau)^3} + \frac{c_2}{\tau^3}\right] \\
            &= \frac{2\tau}{(1-\tau)^2}c_2 + \frac{2\tau(1-\tau)}{\tau^3}c_2 \\
            &= 2c_2 \left[\frac{\tau}{(1-\tau)^2} + \frac{1-\tau}{\tau^2}\right] \\
            &= 2c_2 \left[\frac{\tau^3 + (1-\tau)^3}{\tau^2 (1-\tau)^2}\right] \\
            &= \frac{2c_2}{\tau^2 (1-\tau)^2} (\tau^3 + (1-\tau)^3).
        \end{aligned}
        \end{equation*}
    \item 
        Assume that $c_3=\int_{0}^\infty u^3 g(u) du<\infty$. The third moment, $\mathbb{E}(\epsilon^3|\tau)$, is given by
        \begin{equation*}
        \begin{aligned}
            \int_{-\infty}^\infty \epsilon^3 f(\epsilon|\tau) d\epsilon 
            &= 2\tau(1-\tau)\left[\int_{-\infty}^0 \epsilon^3 g((1-\tau)\epsilon) d\epsilon + \int_{0}^\infty \epsilon^3 g(\tau\epsilon) d\epsilon\right] \\
            &= 2\tau(1-\tau)\left[\int_{-\infty}^0 \left(\frac{u}{1-\tau}\right)^3 g(u) \frac{1}{1-\tau} du + \int_{0}^\infty \left(\frac{u}{\tau}\right)^3 g(u) \frac{1}{\tau} du\right] \\
            &= 2\tau(1-\tau)\left[\frac{1}{(1-\tau)^4}\int_{-\infty}^0 u^3 g(u) du + \frac{1}{\tau^4}\int_{0}^\infty u^3 g(u) du\right] \\
            &= 2\tau(1-\tau)\left[- \frac{c_3}{(1-\tau)^4} + \frac{c_3}{\tau^4}\right] \\
            &= -\frac{2\tau}{(1-\tau)^3} c_3 + \frac{2\tau(1-\tau)}{\tau^4} c_3 \\
            &= 2c_3 \left[-\frac{\tau}{(1-\tau)^3} + \frac{1-\tau}{\tau^3}\right] \\
            &= 2c_3 \left[\frac{-\,\tau^4 + (1-\tau)^4}{\tau^3 (1-\tau)^3}\right] \\
            &= \frac{2c_3(\tau^2 + (1-\tau)^2)}{\tau^3 (1-\tau)^3} (1 - 2\tau).
        \end{aligned}
        \end{equation*}
\end{enumerate}

More generally, assuming that $c_k=\int_{0}^\infty u^k g(u) du<\infty$ for any given integer $k\geq 1$, we can derive the $k$-th moment $\mathbb{E}(\epsilon^k|\tau)$ given by
        \begin{equation*}
        \begin{aligned}
            \int_{-\infty}^\infty \epsilon^k f(\epsilon|\tau) d\epsilon 
            &= 2\tau(1-\tau)\left[\int_{-\infty}^0 \epsilon^k g((1-\tau)\epsilon) d\epsilon + \int_{0}^\infty \epsilon^k g(\tau\epsilon) d\epsilon\right] \\
            &= 2\tau(1-\tau)\left[\int_{-\infty}^0 \left(\frac{u}{1-\tau}\right)^k g(u) \frac{1}{1-\tau} du + \int_{0}^\infty \left(\frac{u}{\tau}\right)^k g(u) \frac{1}{\tau} du\right] \\
            &= 2\tau(1-\tau)\left[\frac{1}{(1-\tau)^{k+1}}\int_{-\infty}^0 u^k g(u) du + \frac{1}{\tau^{k+1}}\int_{0}^\infty u^k g(u) du\right] \\
            &= 2\tau(1-\tau)\left[\frac{(-1)^k c_k}{(1-\tau)^{k+1}} + \frac{c_k}{\tau^{k+1}}\right] \\
            &= \frac{2c_k}{\tau^k(1-\tau)^k}\left[(1-\tau)^{k+1} + (-1)^k\tau^{k+1}\right]
        \end{aligned}
        \end{equation*}

\section{Proof of Lemma 3}
\label{lemma_proof_3}
Let $Y\sim\mathcal{AL^*}(\theta,\sigma,\kappa)$ with pdf
\begin{equation*}
    f_{\theta,\sigma,\kappa}(y) = \frac{\sqrt{2}}{\sigma} \frac{\kappa}{1 + \kappa^2} 
    \begin{cases} 
    \exp \left( -\frac{\sqrt{2} \kappa}{\sigma} |y - \theta| \right), & \text{if } y \geq \theta, \\
    \exp \left( -\frac{\sqrt{2}}{\sigma \kappa} |y - \theta| \right), & \text{if } y < \theta.
    \end{cases}
\end{equation*}
then its moment generating function (mgf) is
\begin{equation*}
\begin{aligned}
    M_{\theta,\sigma,\kappa}(t) = E[e^{tY}] = \frac{\exp(\theta t)}{1 - \frac{1}{2} \sigma^2 t^2 - \frac{\sigma}{\sqrt{2}} \left( \frac{1}{\kappa} - \kappa \right) t}, \quad -\frac{\sqrt{2}}{\sigma \kappa} < t < \frac{\sqrt{2} \kappa}{\sigma}
\end{aligned}
\end{equation*}
as shown in \cite{bib11-}. This parametrization is equivalent to \eqref{eq:ald_different} where $\mu=\theta$, $\kappa=\kappa$, and $\beta=\frac{\sigma}{\sqrt{2}}$. The reparameterized mgf according to \eqref{eq:ald_different} is therefore
\begin{equation*}
\begin{aligned}
    M_{\mu, \kappa, \beta}(t) = E[e^{tY}] = \frac{\exp(\mu t)}{1 - \beta^2 t^2 - \beta \left( \frac{1}{\kappa} - \kappa \right) t}, \quad -\frac{1}{\beta \kappa} < t < \frac{\kappa}{\beta}.
\end{aligned}
\end{equation*}
Let $Y|V \sim N(\mu + \beta \left(\frac{1}{\kappa} - \kappa\right)V, 2\beta^2 V)$ with latent variable $V \sim Exp(1)$. The mgf of r.v. $Y$ given $V=v$ is
\begin{equation*}
\begin{aligned}
    M_{Y|V=v}(t) = \mathbb{E}[e^{tY} | V = v] &= \exp(\mu t + \sigma^2 t^2/2) \\
    &= \exp([\mu + \beta(\frac{1}{\kappa} - \kappa)v] t + [2\beta^2v] t^2/2) \\
    &= \exp(t\mu + t\beta(\frac{1}{\kappa} - \kappa)v + t^2\beta^2v) \\
\end{aligned}
\end{equation*}
because the mgf of a normal r.v. with mean $\mu$ and variance $\sigma^2$ is $\exp(\mu t+\sigma^2 t^2/2)$. Using the law of total expectation, the mgf of r.v. $Y$ is denoted as 
\begin{equation*}
\begin{aligned}
    M_Y(t) = \mathbb{E}[e^{tY}] &= \int_{0}^{\infty} M_{Y|V=v}(t)f_V(v)dv \\
    &= \int_{0}^{\infty} \exp(t\mu + t\beta(\frac{1}{\kappa} - \kappa)v + t^2\beta^2v) \exp(-v) dv \\
    &= \exp(t\mu) \int_{0}^{\infty} \exp(t\beta(\frac{1}{\kappa} - \kappa)v + t^2\beta^2v - v) dv \\
    &= \exp(t\mu) \int_{0}^{\infty} \exp(-v[1-\beta(\frac{1}{\kappa} - \kappa)t - \beta^2t^2]) dv \\
    &= \frac{\exp(\mu t)}{1-\beta(\frac{1}{\kappa} - \kappa)t - \beta^2t^2} \\
\end{aligned}
\end{equation*}
due to the Laplace transformation of a constant $c=1$ where
\begin{equation*}
\begin{aligned}
    \int_{0}^{\infty} c\exp(-s v)dv = \left[ \frac{\exp(-s v)}{-s} \right]_{v=0}^{v=\infty}= \left(0 - \frac{1}{-s}\right) = \frac{1}{s}.
\end{aligned}
\end{equation*}

\FloatBarrier

\section{Additional Properties of Generalized Split Distributions}
\label{addl_properties}
 Let $g(u)$ be any symmetric density on $\mathbb{R}$ with mode $0$, i.e., $0\leq g(-u)=g(u)\leq g(\mu)$ for any $u\geq 0$ and $\int_{-\infty}^\infty g(u) du=1$. Consider the class of general asymmetric split densities with mode $\mu\in\mathbb{R}$, asymmetry parameter $\tau\in (0, 1)$ and scale parameter $\sigma>0$ given by
\begin{equation*}
    f(y;\mu,\sigma,\tau) = \frac{2\tau(1-\tau)}{\sigma}
    \begin{cases} 
      g\left(\frac{(1-\tau)(y-\mu)}{\sigma}\right), & \text{if }\;\; y \leq \mu, \\
      g\left(\frac{\tau (y-\mu)}{\sigma}\right), & \text{if }\;\; y \geq\mu
    \end{cases}
\end{equation*}

It is easy to see that $f(y;\mu,\sigma,\tau)\leq f(\mu;\mu,\sigma,\tau)$ for any $\sigma>0$ and $\tau\in (0,1)$, establishing that $\mu$ is the mode of distribution regardless of the scale and asymmetry parameters. Moreover, it is easy to derive the cumulative distribution function and the quantile function, which we provide below without proof:
\begin{enumerate}
    \item {\em Cumulative distribution function} is  given by 
 \begin{equation*}
    F(y;\mu,\sigma,\tau) = \int_{-\infty}^y f(x;\mu\sigma,\tau)\,dx =
    \begin{cases} 
      2\tau G\left((1-\tau)\frac{(y-\mu)}{\sigma}\right), & \text{if }\;\; y \leq \mu, \\
      \tau + 2(1-\tau) \left[G\left(\tau\frac{(y-\mu)}{\sigma}\right)-\frac{1}{2}\right], & \text{if }\;\; y \geq\mu
    \end{cases}
\end{equation*}  
where $G(x) = \int_{-\infty}^x g(u)\,du$ denotes the cumulative distribution function corresponding to the density $g(u)$. Notice that $F(\mu;\mu,\sigma,\tau)=\tau$ and $G(-x)=1-G(x)$ for any $x\geq 0$.

 \item {\em Quantile function} is given by 
 \begin{equation*}
    Q(q;\mu,\sigma,\tau) = F^{-1}(q;\mu\sigma,\tau) =
    \begin{cases} 
      \mu + \frac{\sigma}{1-\tau}G^{-1}\left(\frac{q}{2\tau}\right), & \text{if }\;\; 0< q\leq \tau, \\
      \mu+\frac{\sigma}{\tau}G^{-1}\left(\frac{1+q-2\tau}{2-2\tau}\right), & \text{if }\;\; \tau\leq q<1
    \end{cases}
\end{equation*}  
where $G^{-1}(q)$ denotes the quantile function corresponding to the density $g(u)$. Notice that $Q(\tau;\mu,\sigma,\tau)=\tau$ and $G^{-1}(q)+G^{-1}(1-q)=0$ for $q\in (0,1)$.
\end{enumerate}
\FloatBarrier

\section{Posterior Predictive Distribution Plots}\label{A4}

\begin{figure}[ht]
    \centering
    \includegraphics[width=0.33\textwidth]{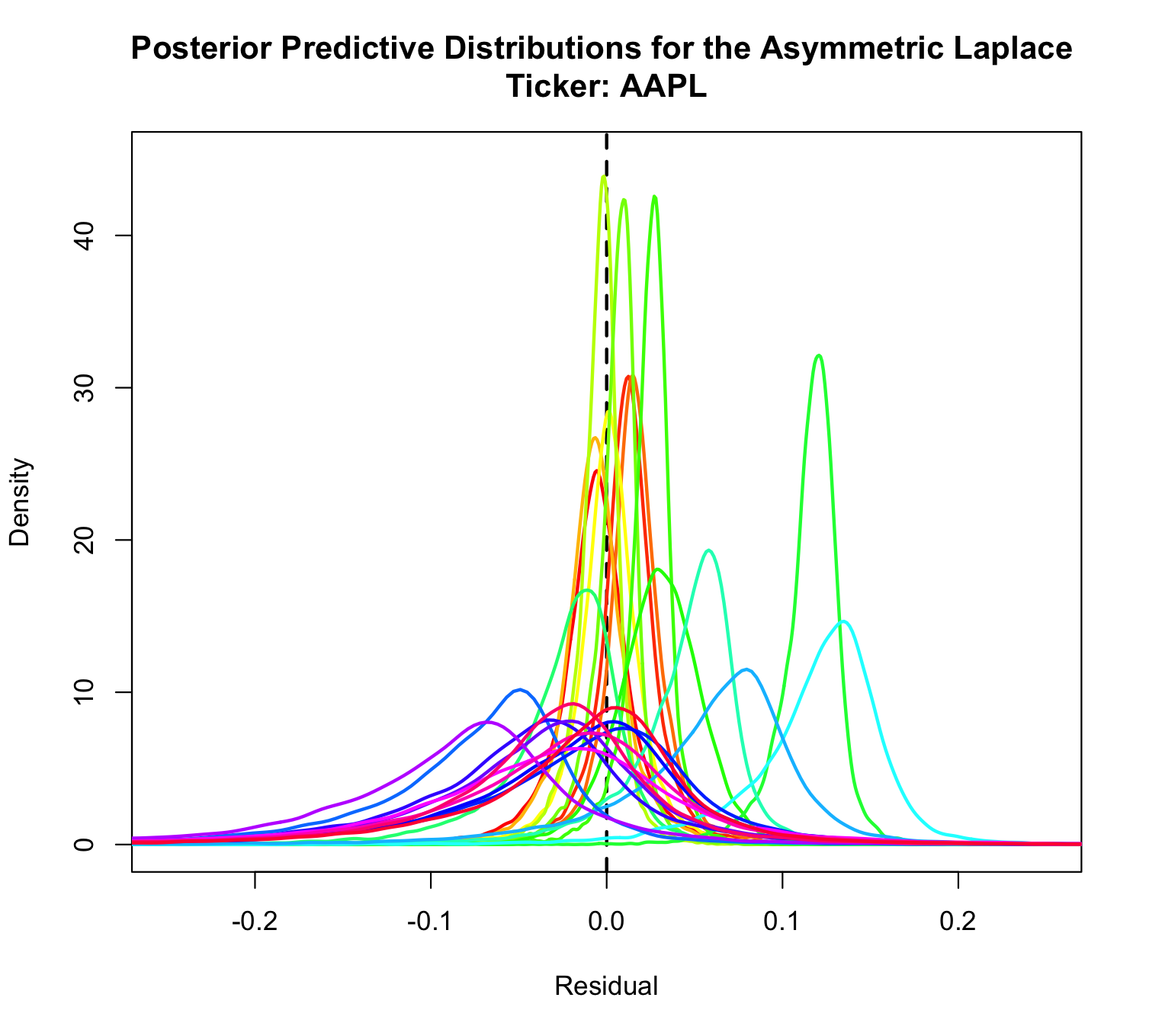}
    \includegraphics[width=0.33\textwidth]{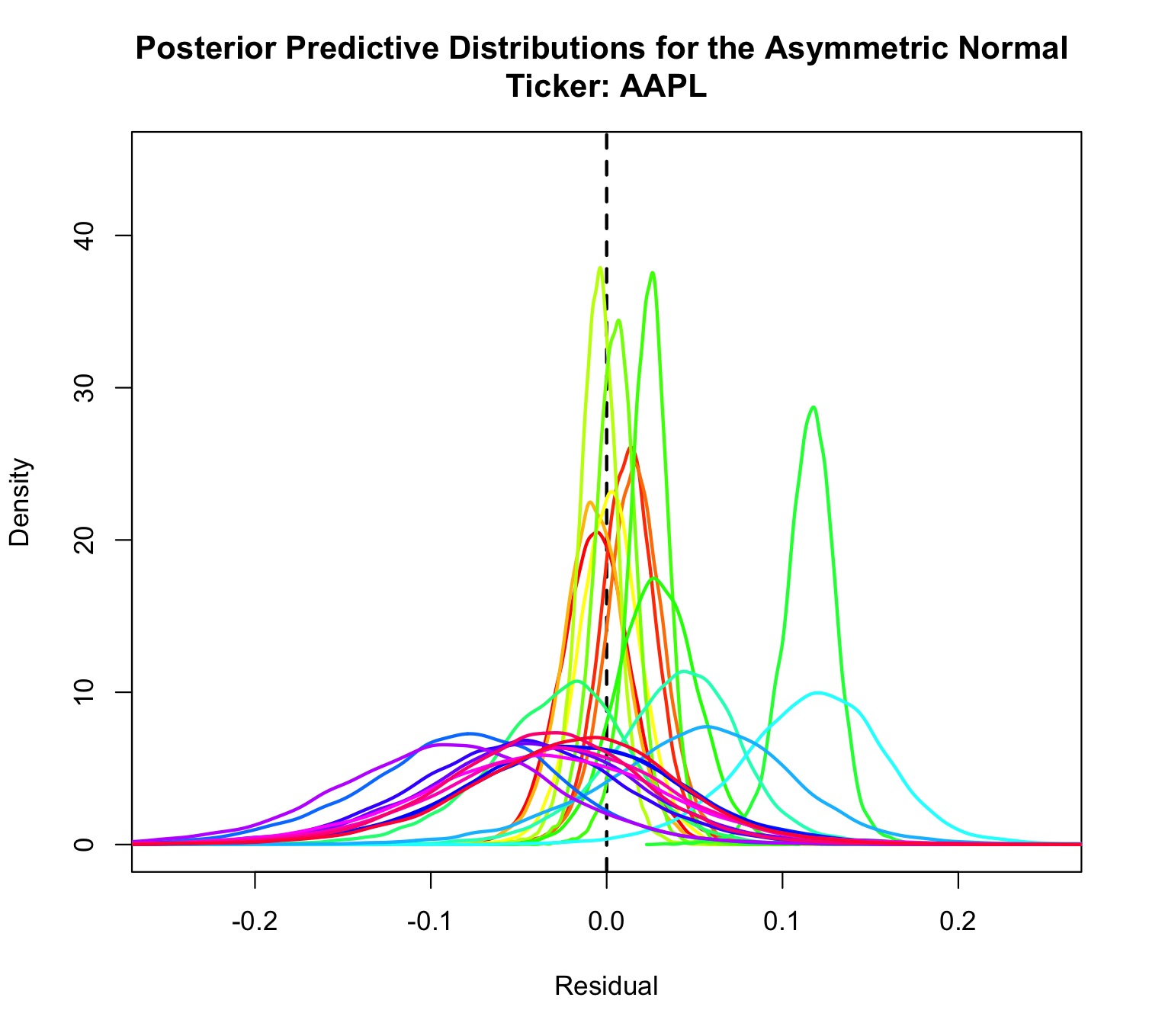}
    \includegraphics[width=0.33\textwidth]{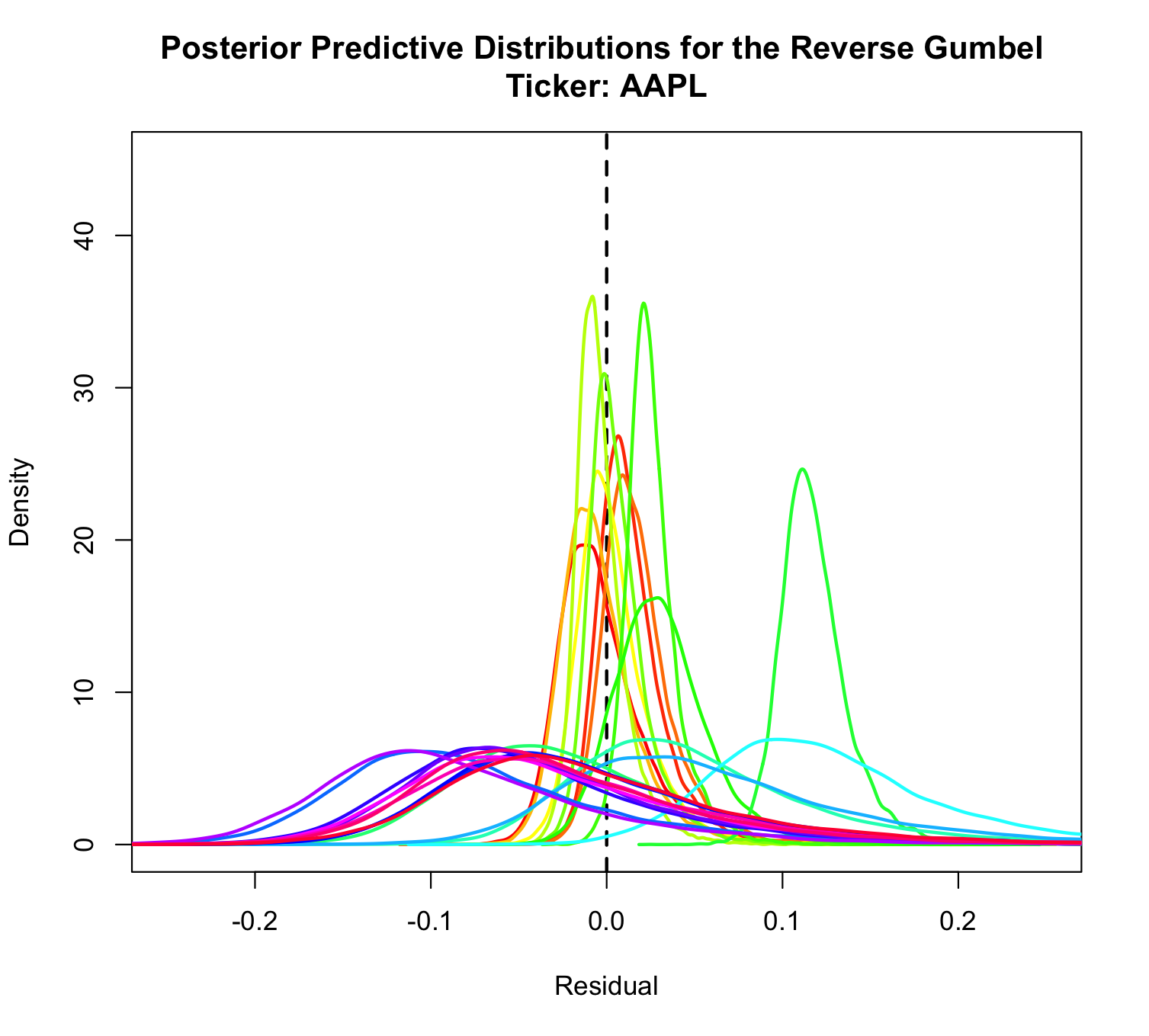}
    \caption{Posterior predictive distributions for AAPL.}
    \label{fig:AAPL}
\end{figure}

\begin{figure}[ht]
    \centering
    \includegraphics[width=0.33\textwidth]{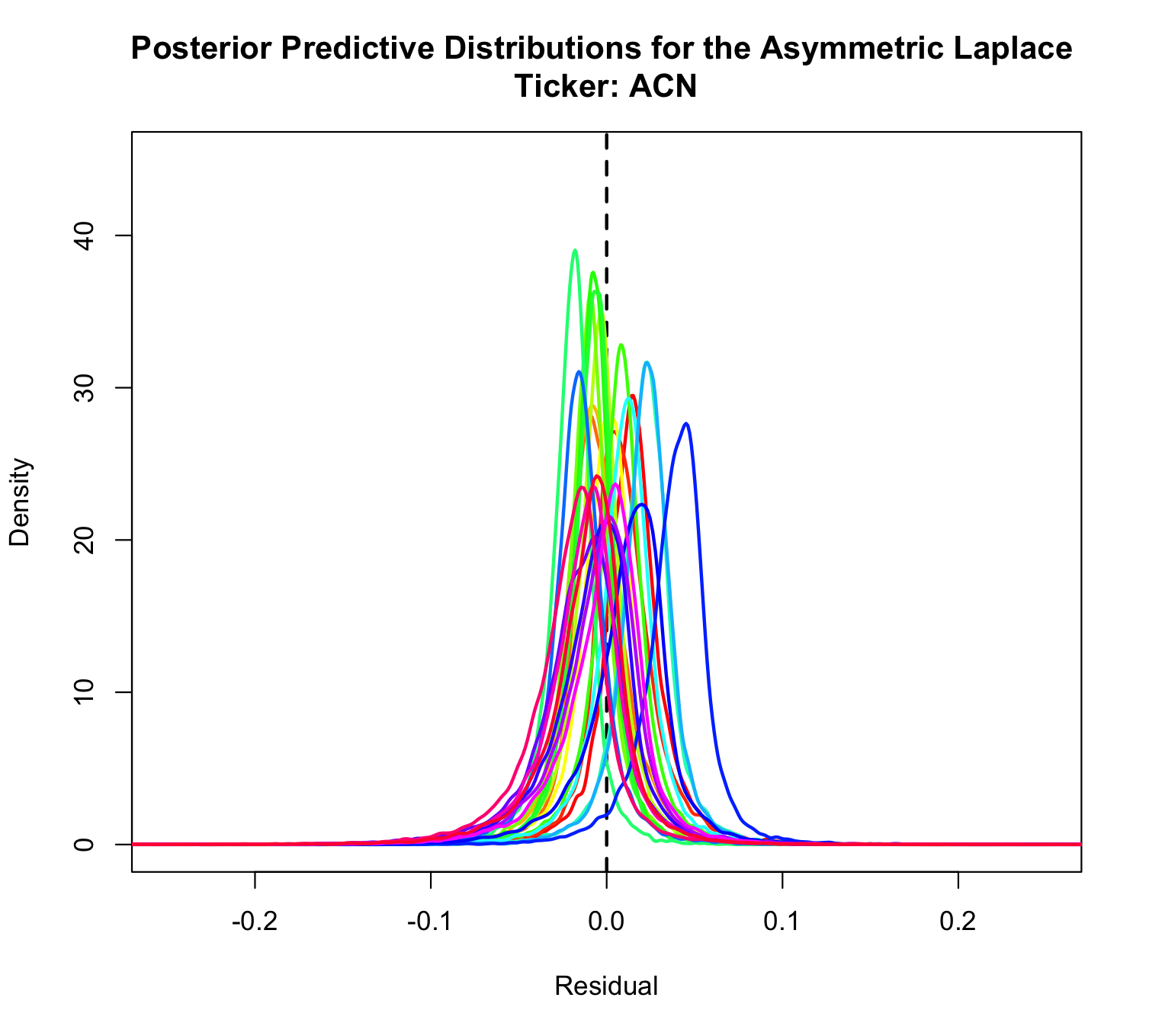}
    \includegraphics[width=0.33\textwidth]{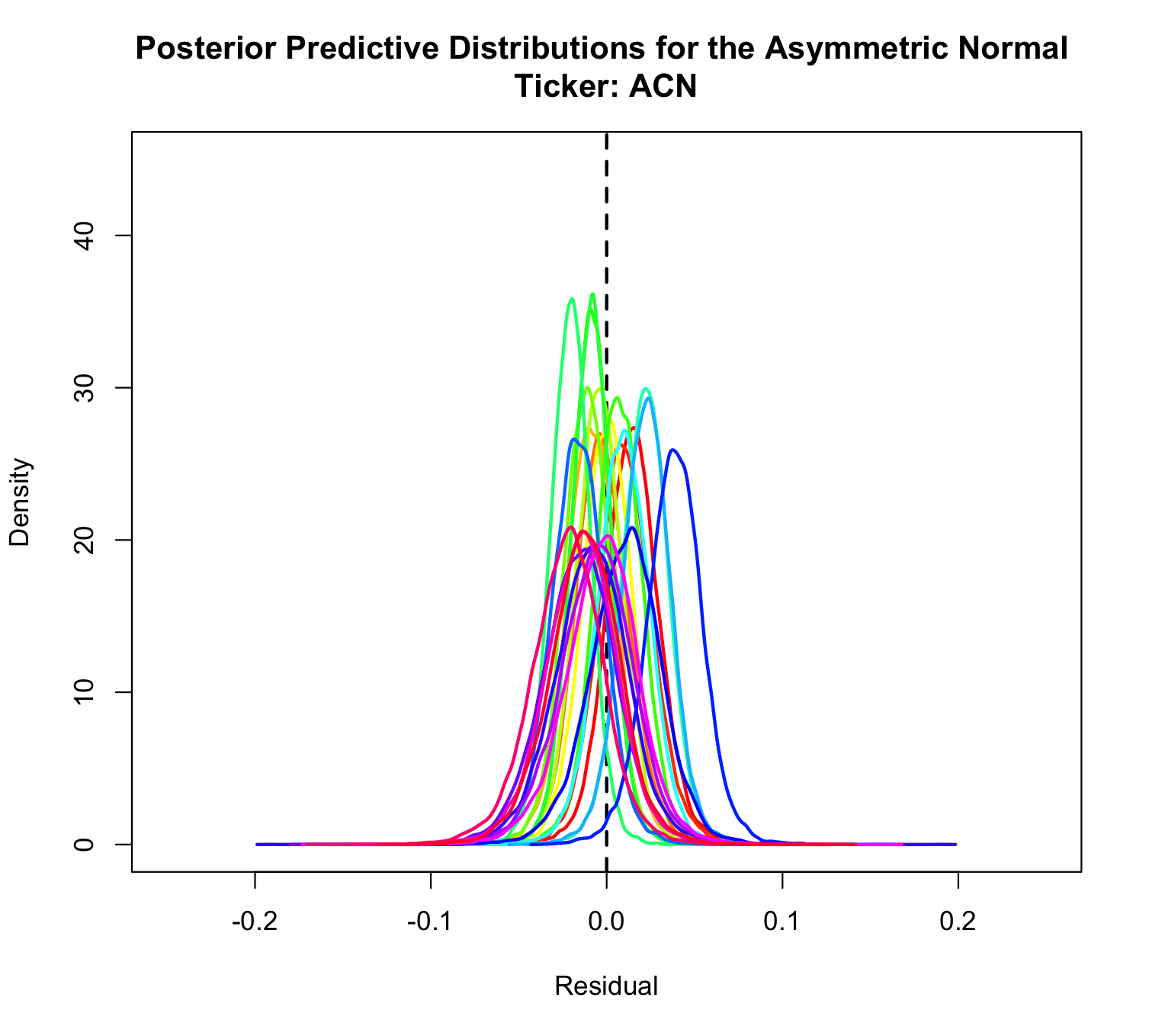}
    \includegraphics[width=0.33\textwidth]{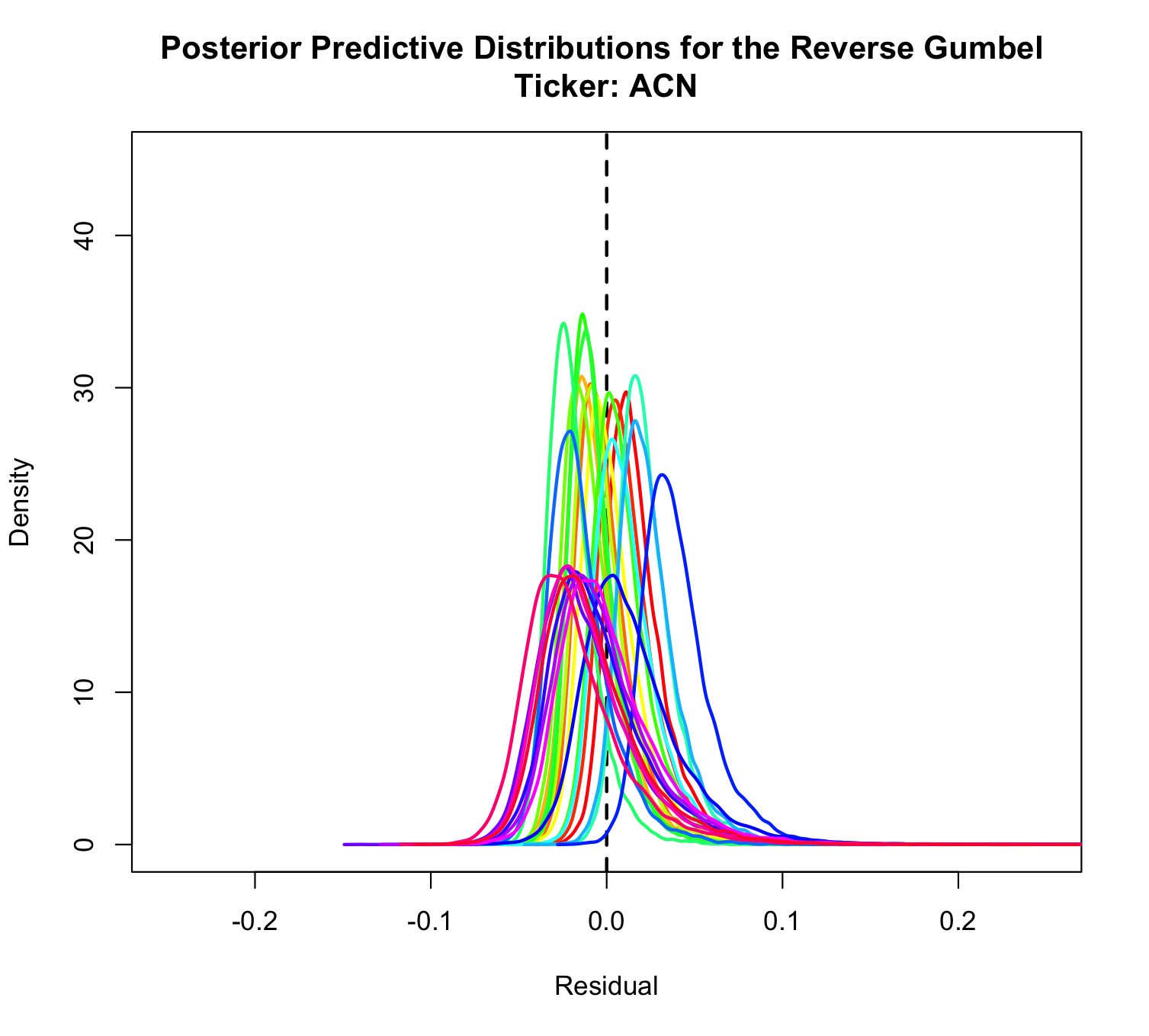}
    \caption{Posterior predictive distributions for ACN.}
    \label{fig:ACN}
\end{figure}

\begin{figure}[ht]
    \centering
    \includegraphics[width=0.33\textwidth]{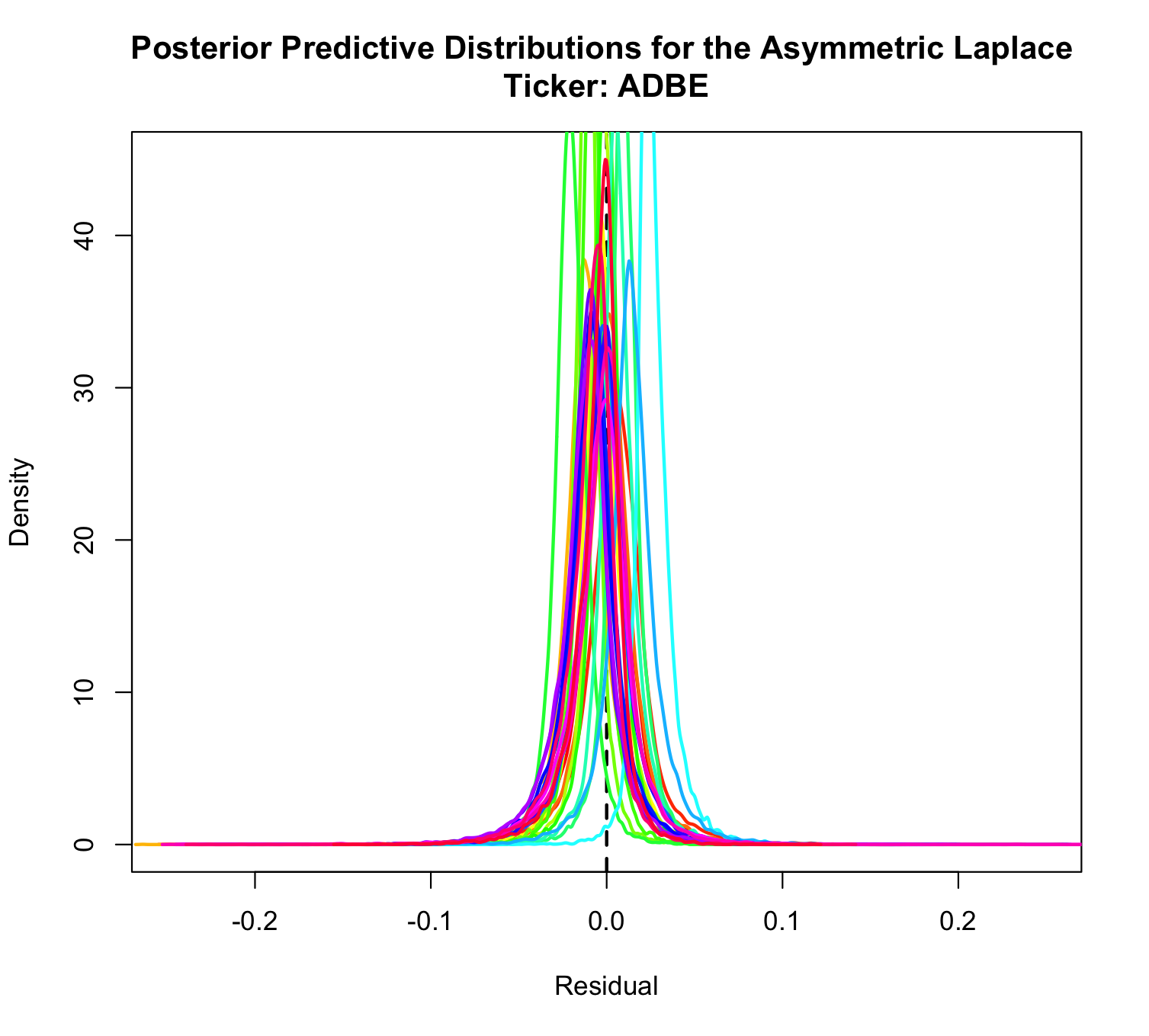}
    \includegraphics[width=0.33\textwidth]{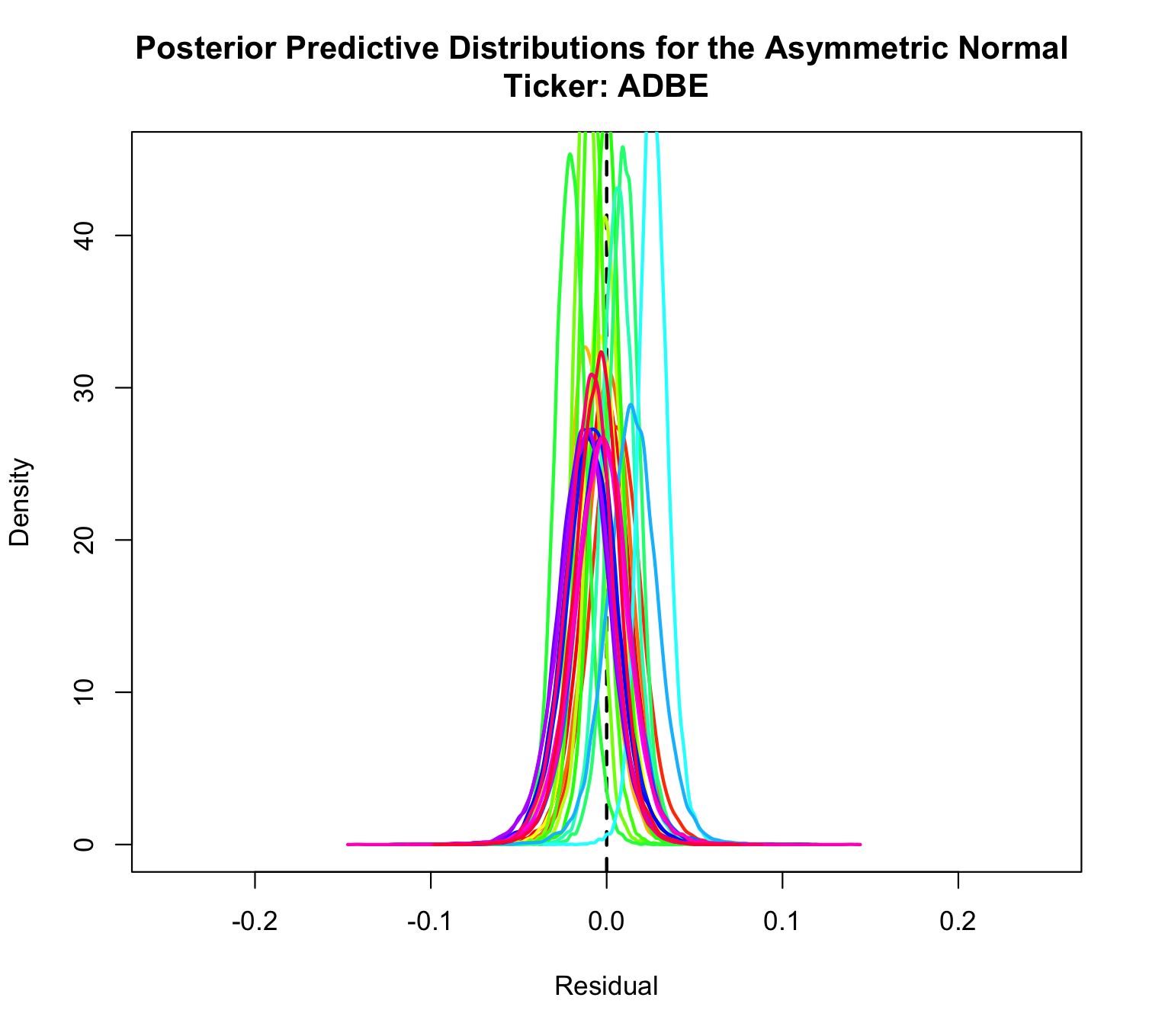}
    \includegraphics[width=0.33\textwidth]{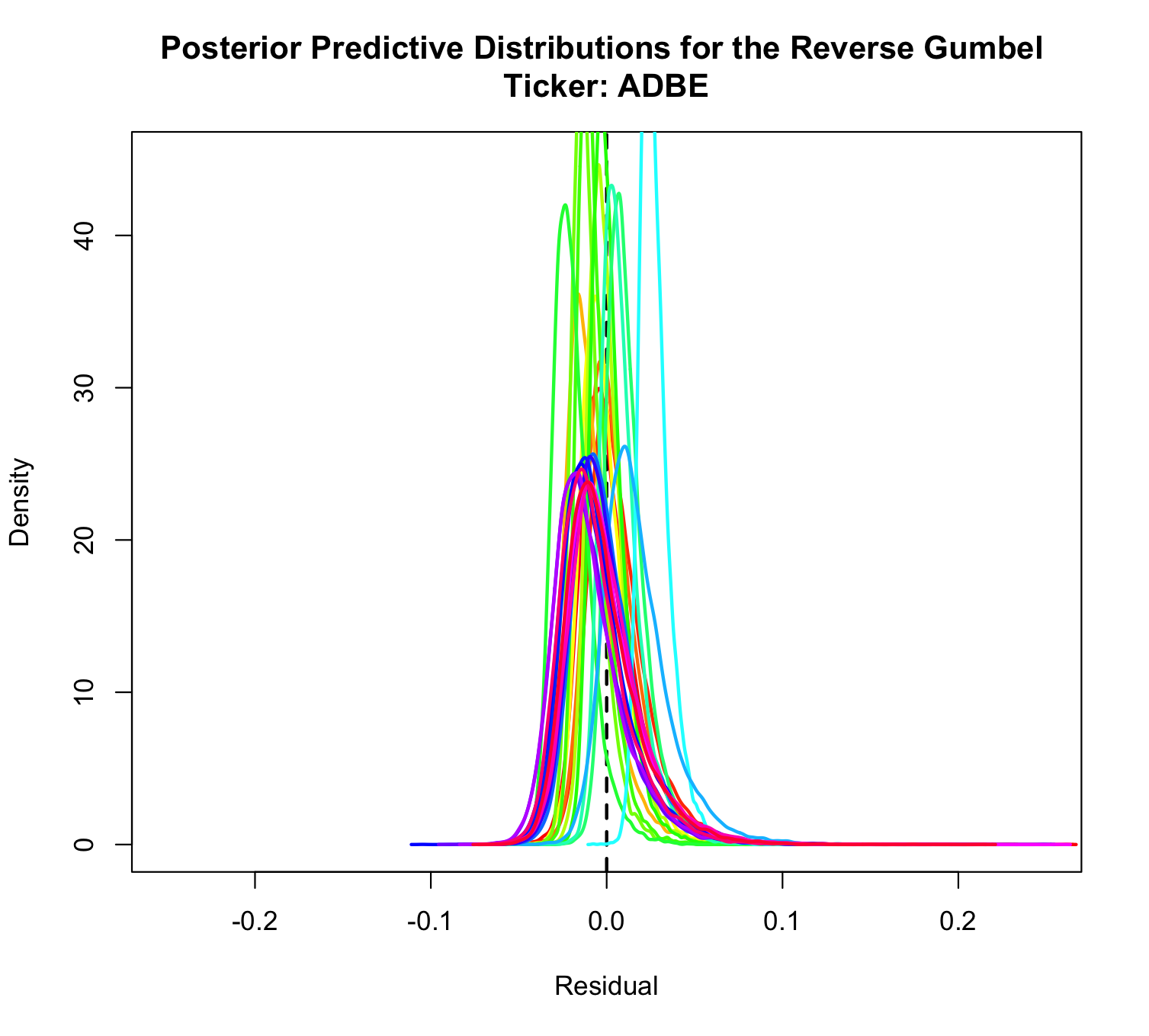}
    \caption{Posterior predictive distributions for ADBE.}
    \label{fig:ADBE}
\end{figure}

\begin{figure}[ht]
    \centering
    \includegraphics[width=0.33\textwidth]{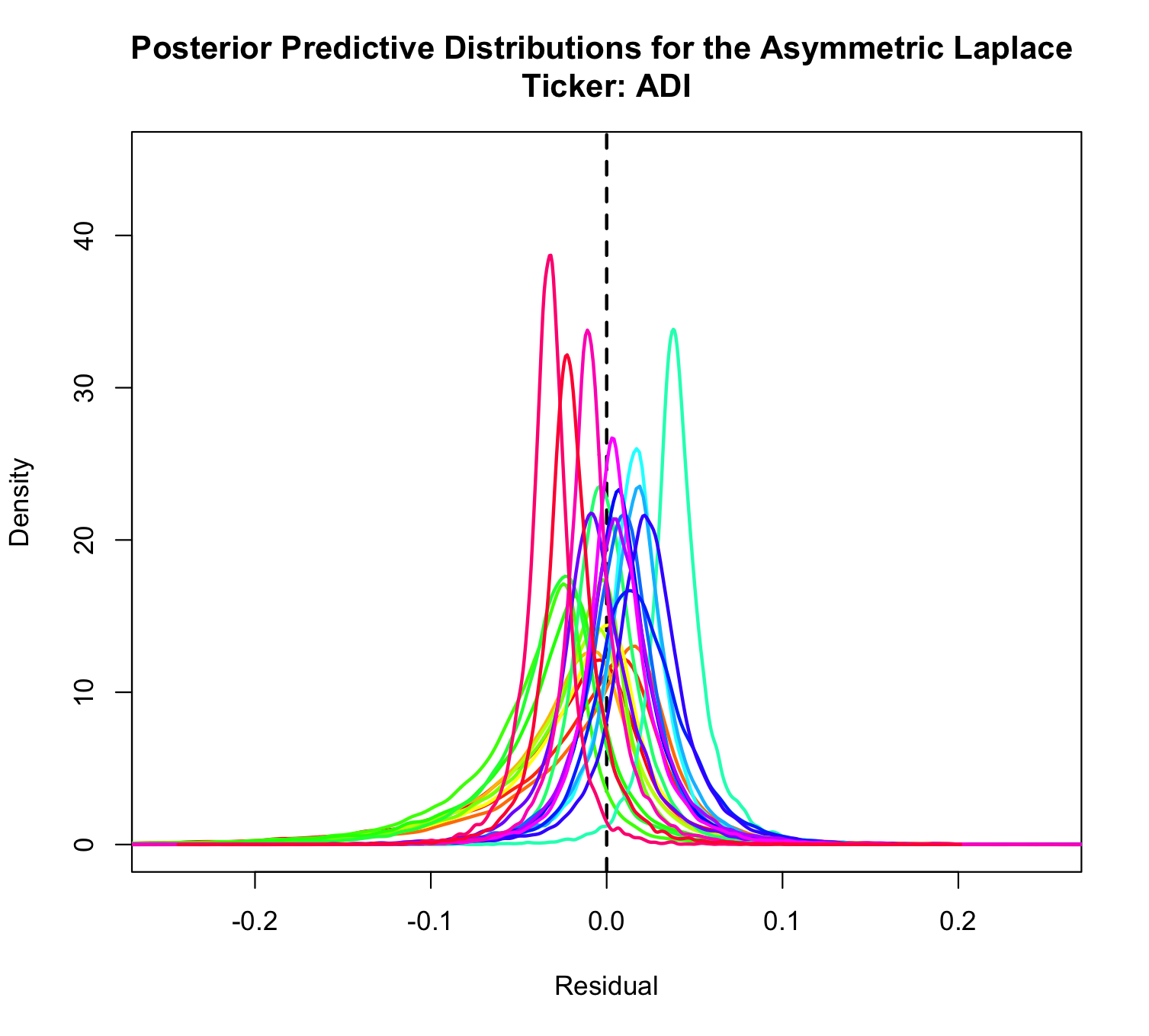}
    \includegraphics[width=0.33\textwidth]{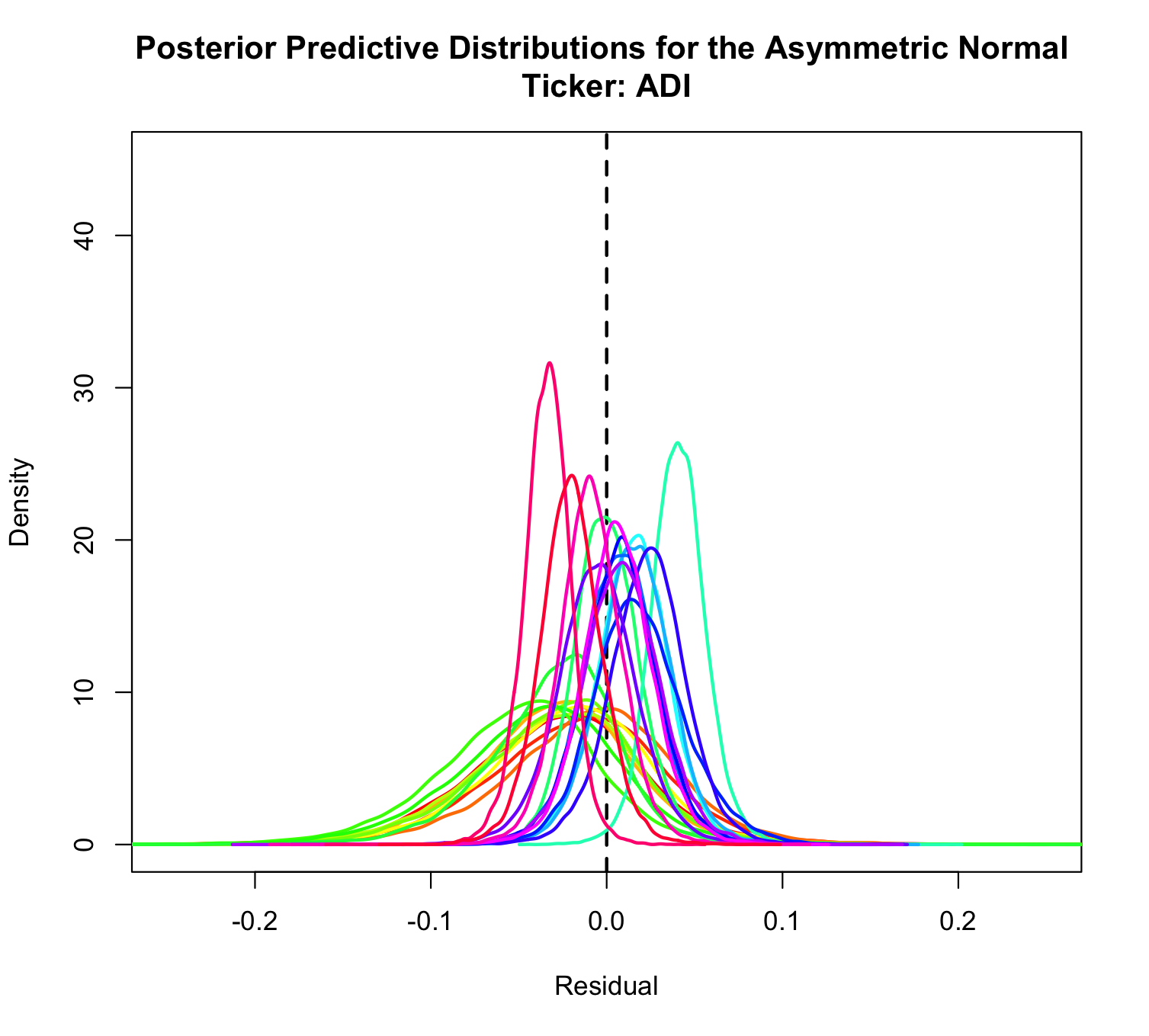}
    \includegraphics[width=0.33\textwidth]{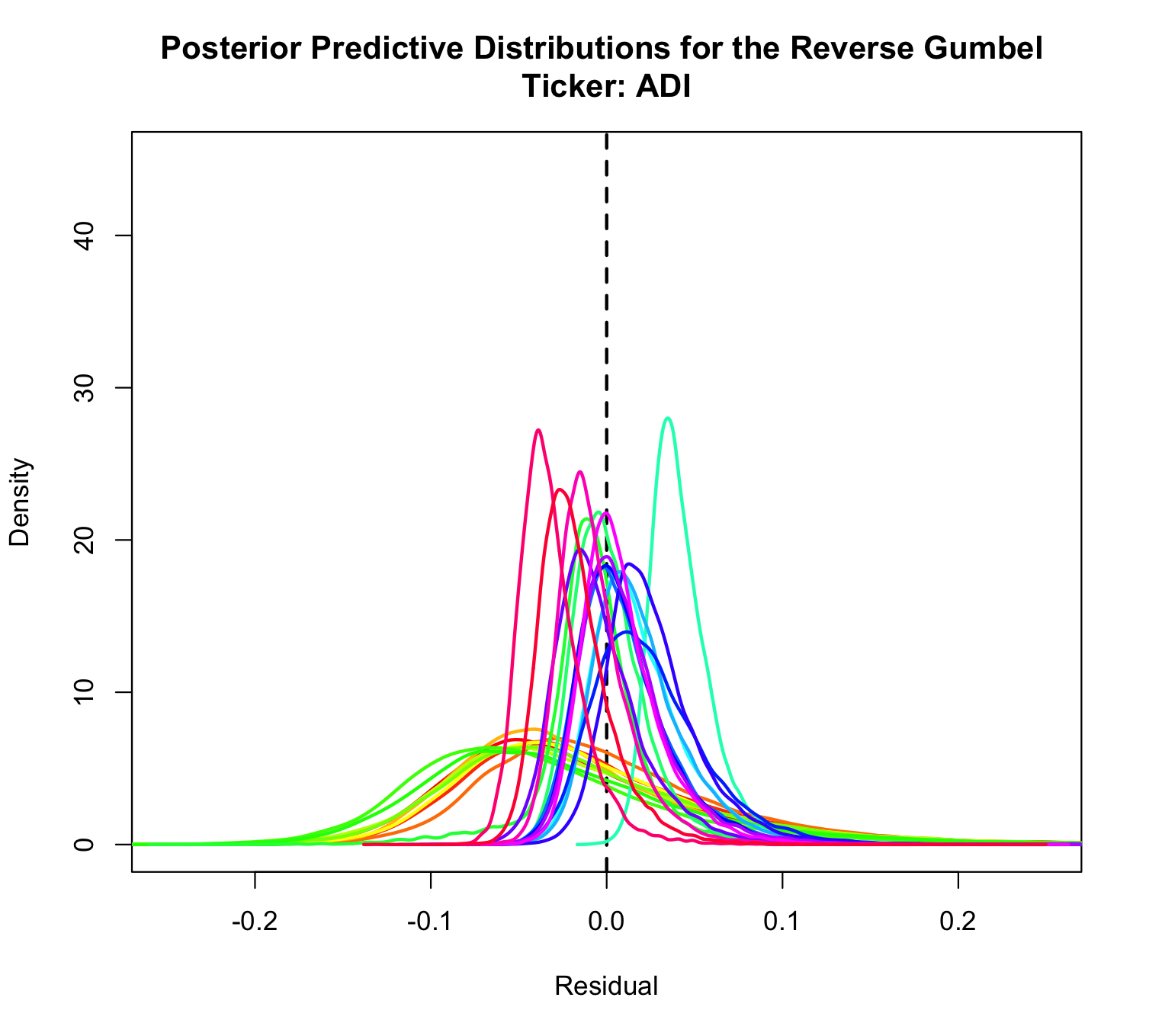}
    \caption{Posterior predictive distributions for ADI.}
    \label{fig:ADI}
\end{figure}

\begin{figure}[ht]
    \centering
    \includegraphics[width=0.33\textwidth]{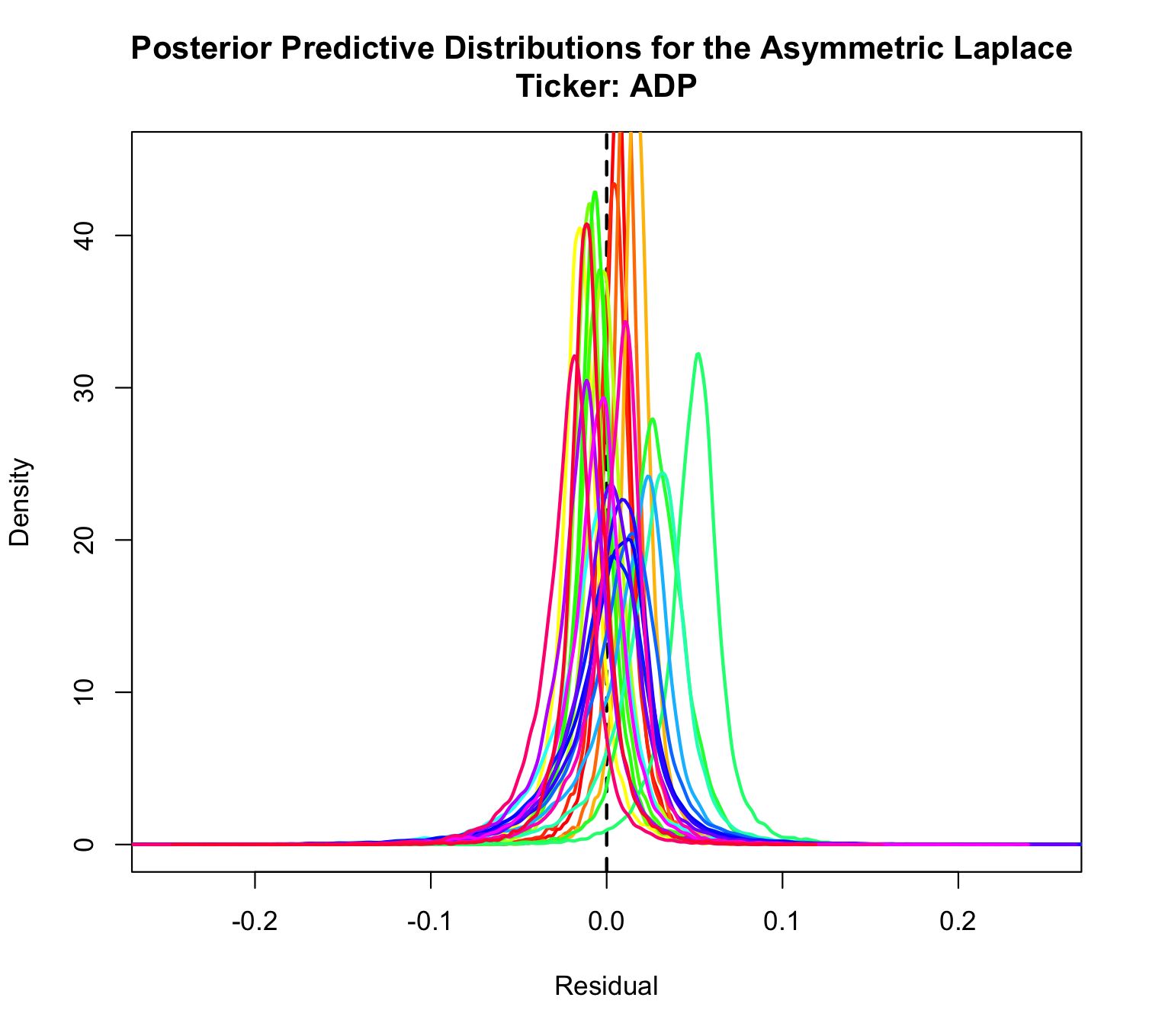}
    \includegraphics[width=0.33\textwidth]{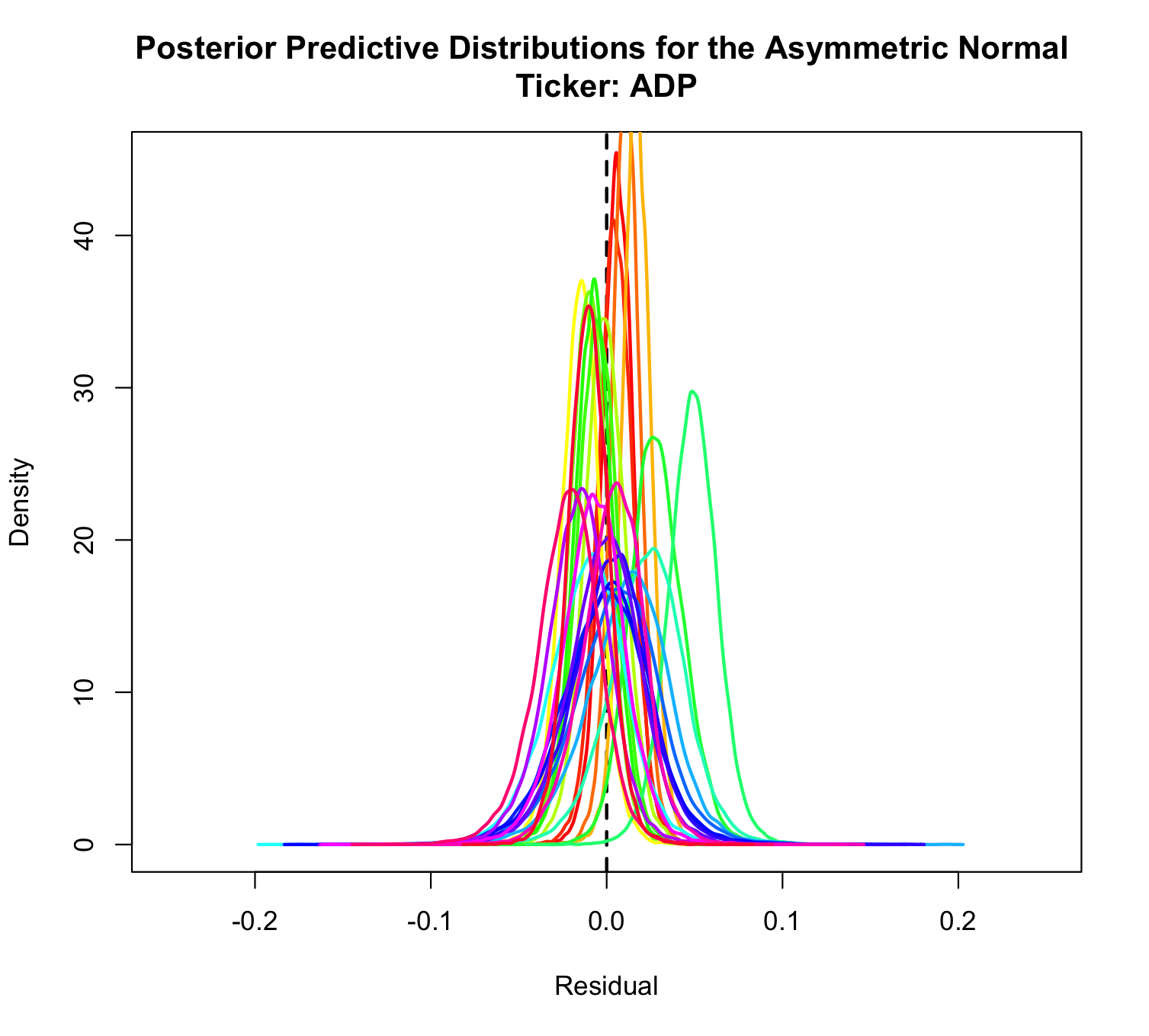}
    \includegraphics[width=0.33\textwidth]{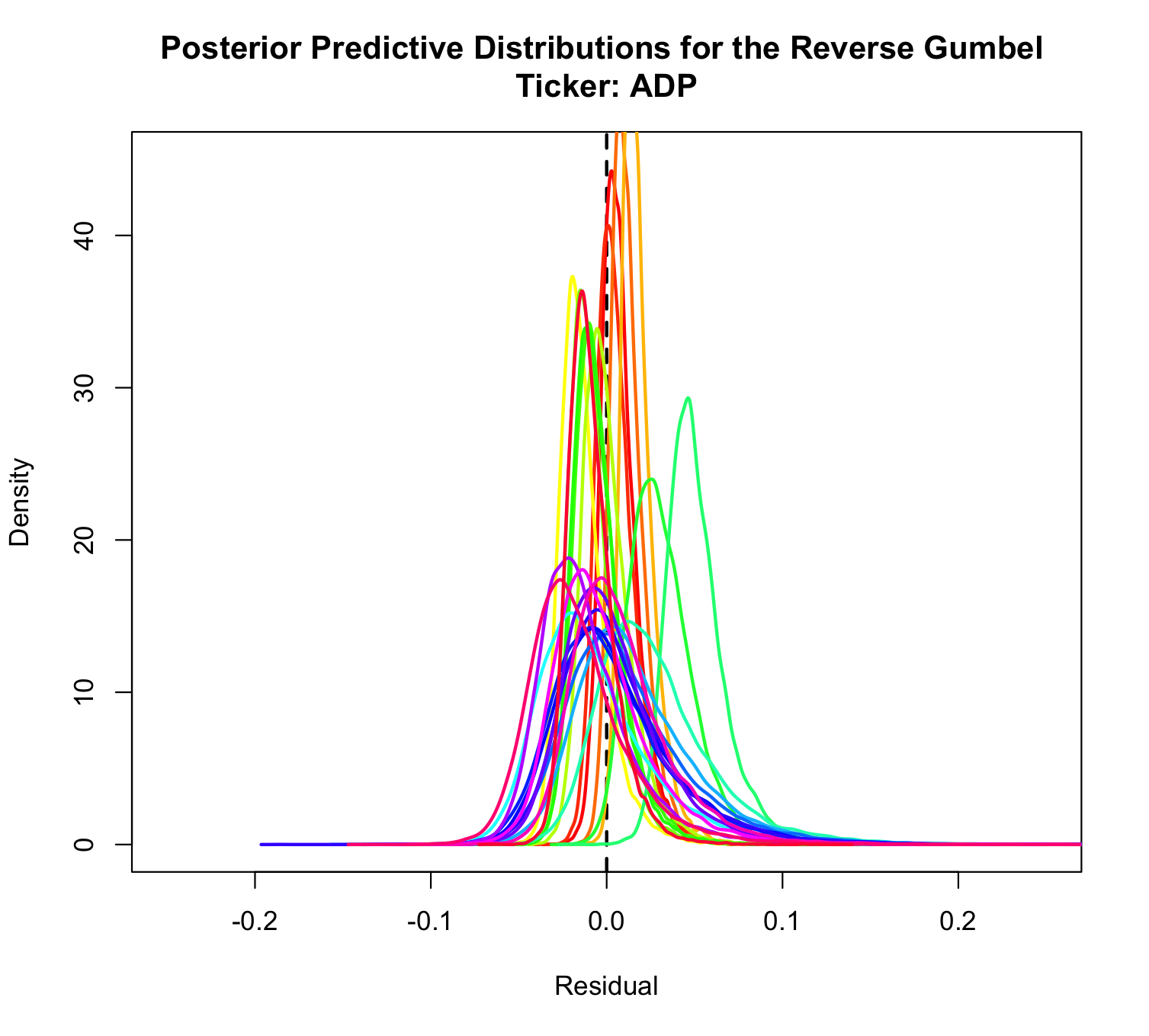}
    \caption{Posterior predictive distributions for ADP.}
    \label{fig:ADP}
\end{figure}

\begin{figure}[ht]
    \centering
    \includegraphics[width=0.33\textwidth]{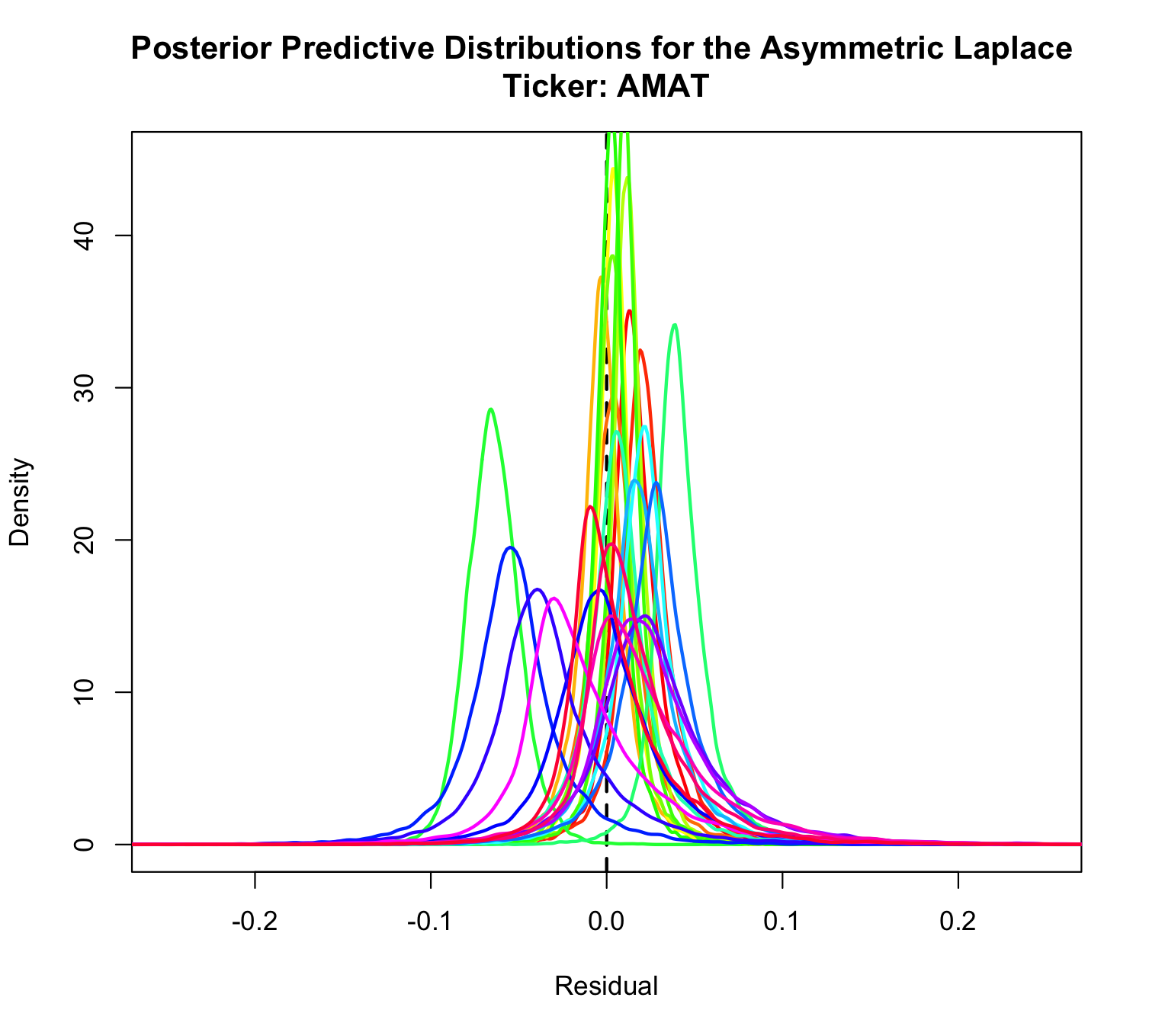}
    \includegraphics[width=0.33\textwidth]{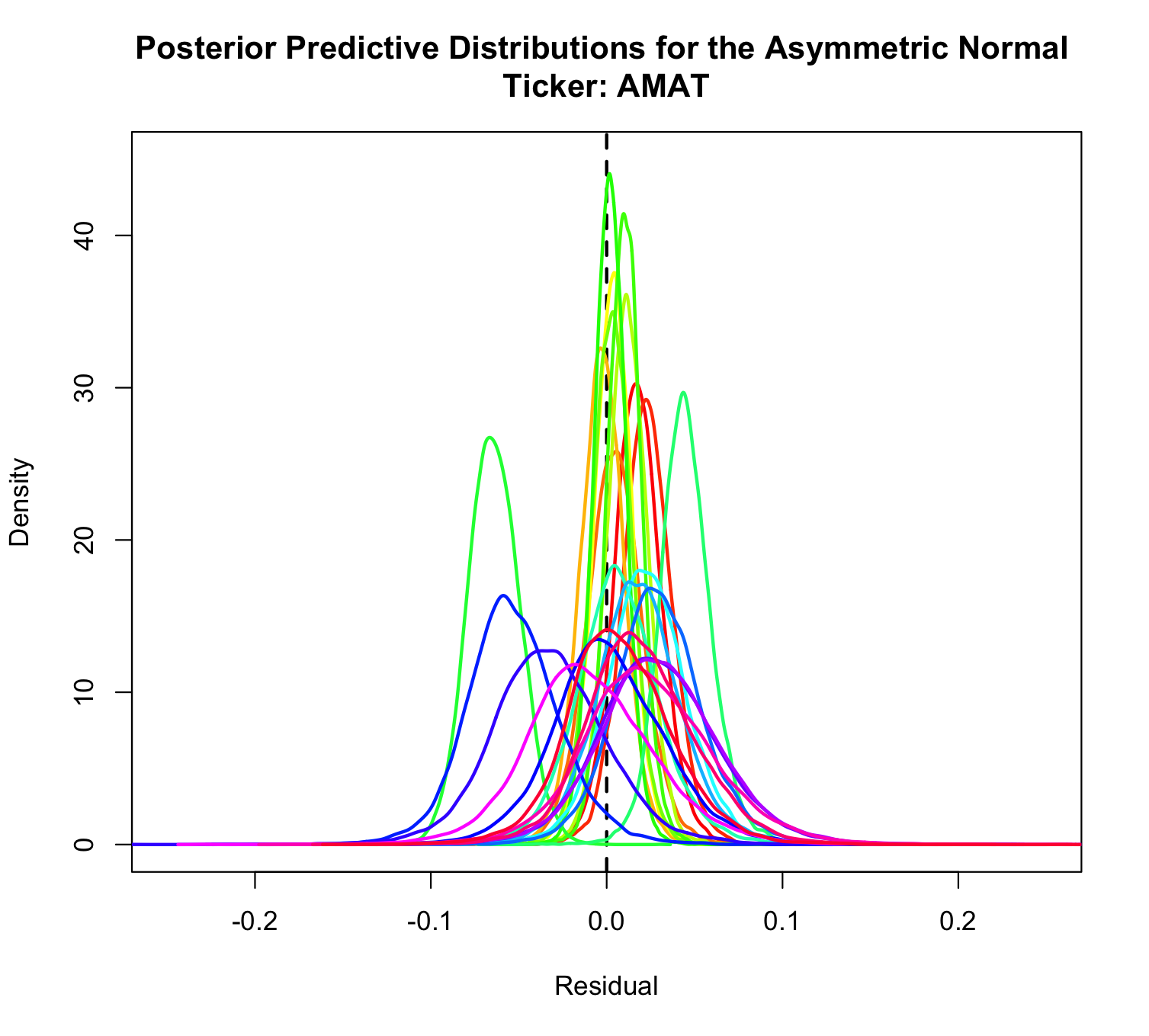}
    \includegraphics[width=0.33\textwidth]{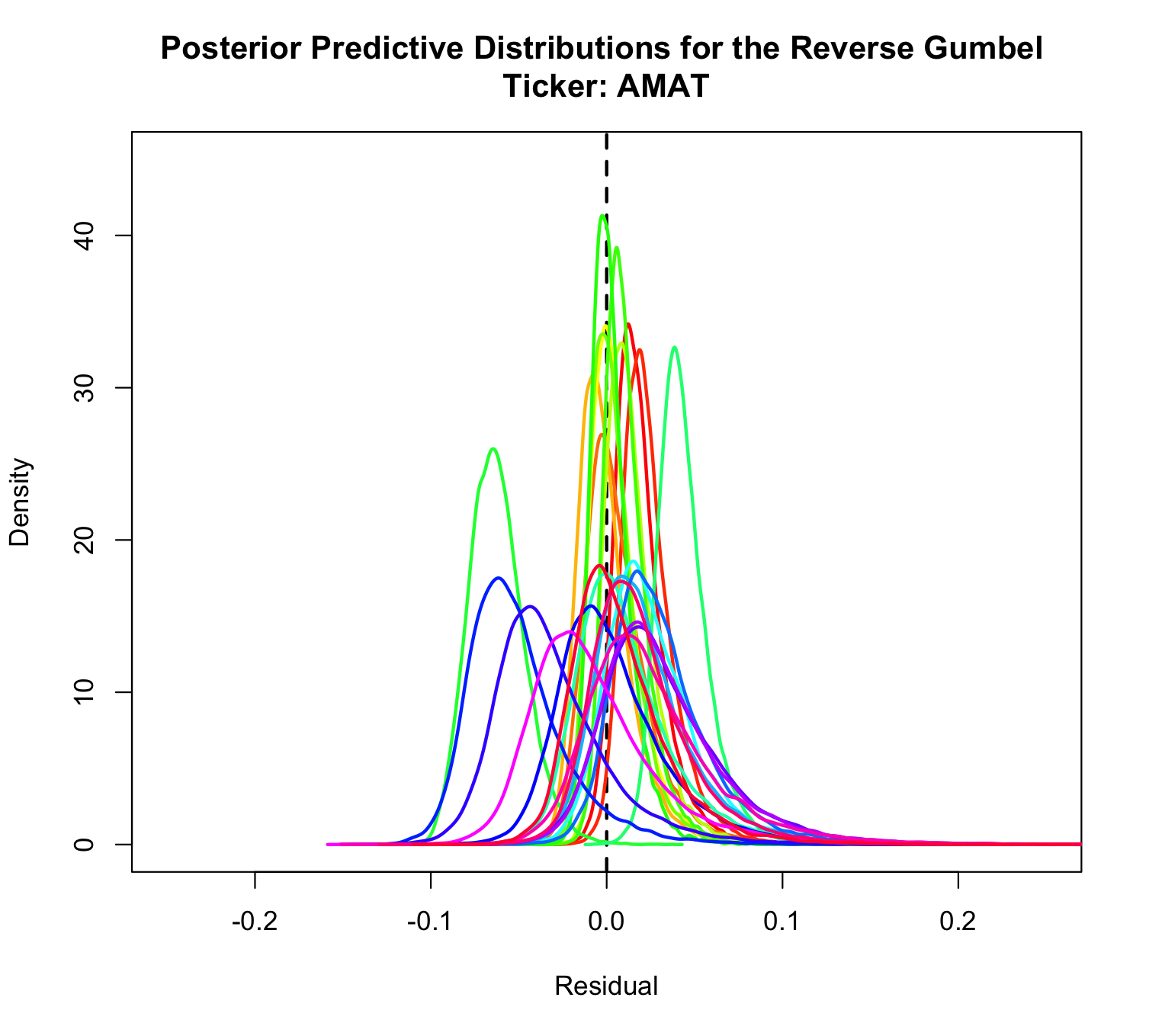}
    \caption{Posterior predictive distributions for AMAT.}
    \label{fig:AMAT}
\end{figure}

\begin{figure}[ht]
    \centering
    \includegraphics[width=0.33\textwidth]{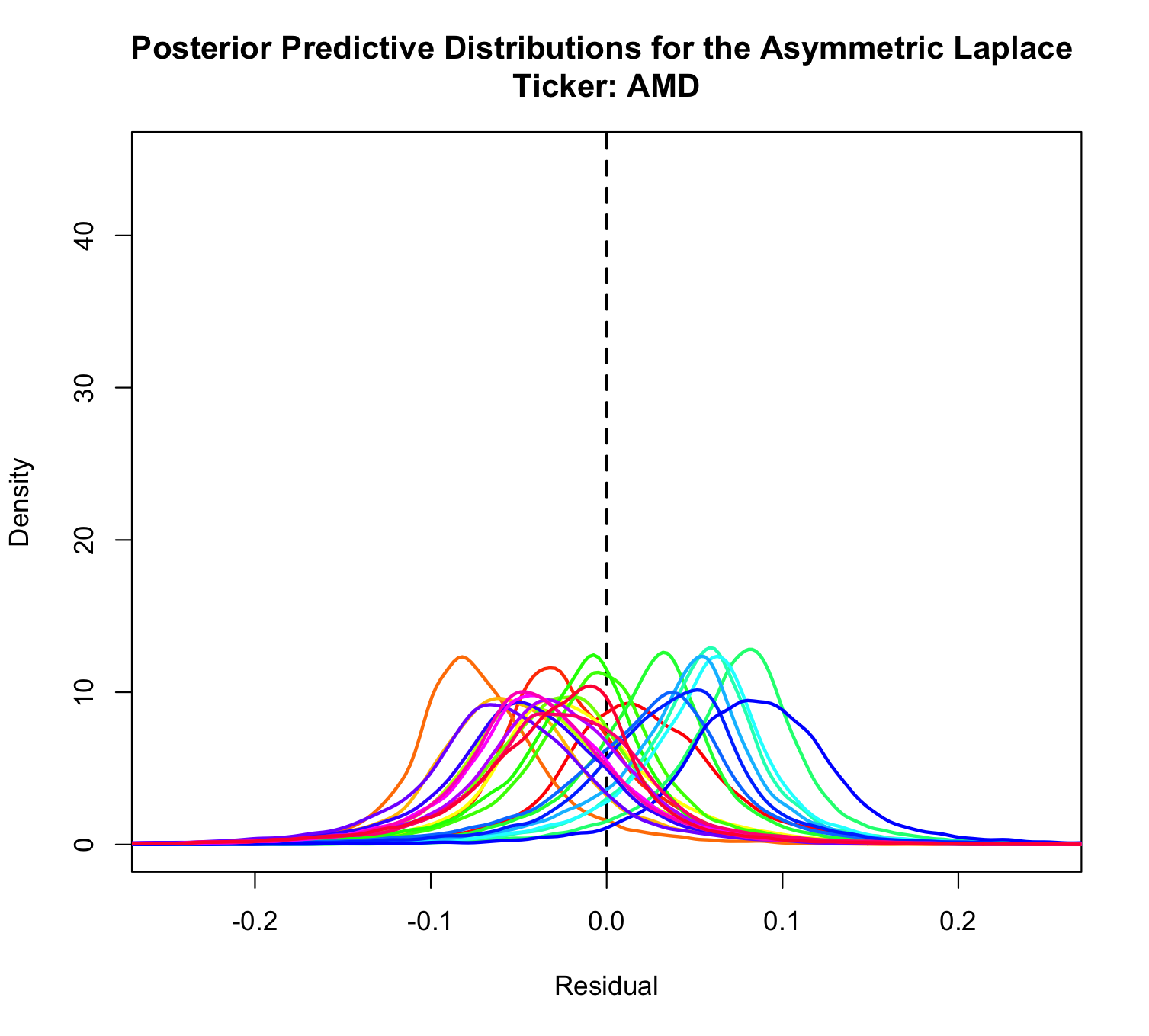}
    \includegraphics[width=0.33\textwidth]{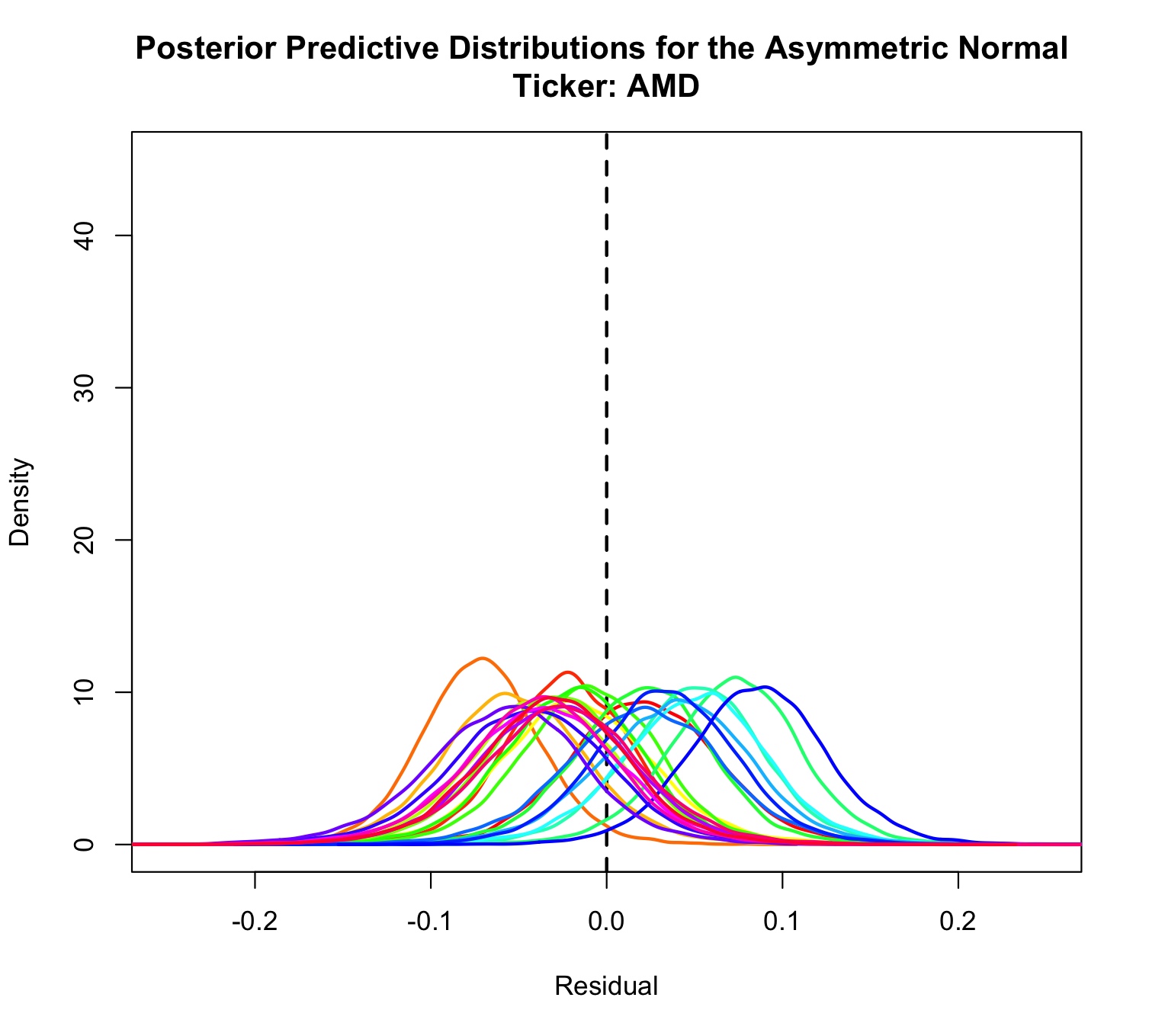}
    \includegraphics[width=0.33\textwidth]{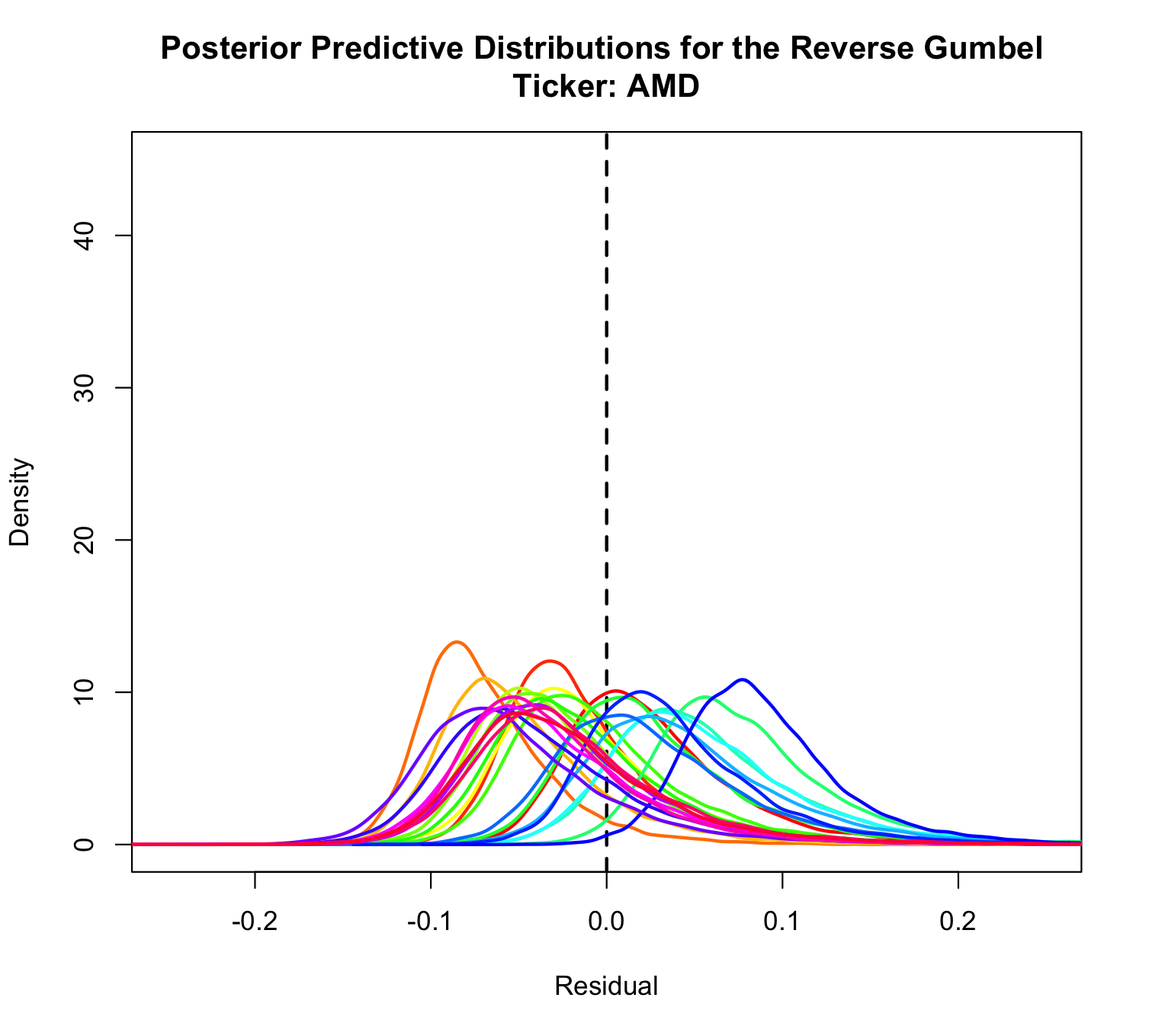}
    \caption{Posterior predictive distributions for AMD.}
    \label{fig:AMD}
\end{figure}

\begin{figure}[ht]
    \centering
    \includegraphics[width=0.33\textwidth]{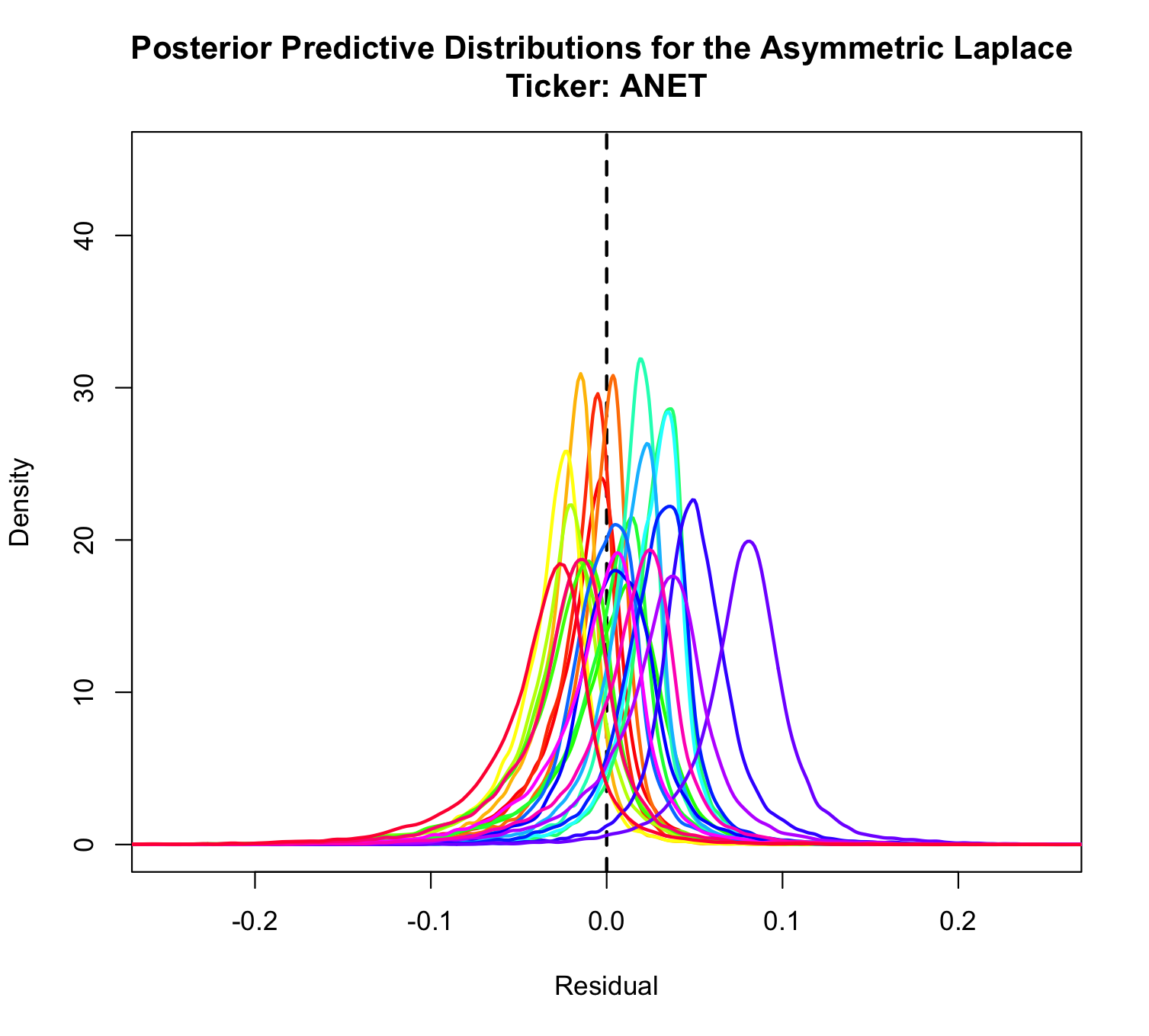}
    \includegraphics[width=0.33\textwidth]{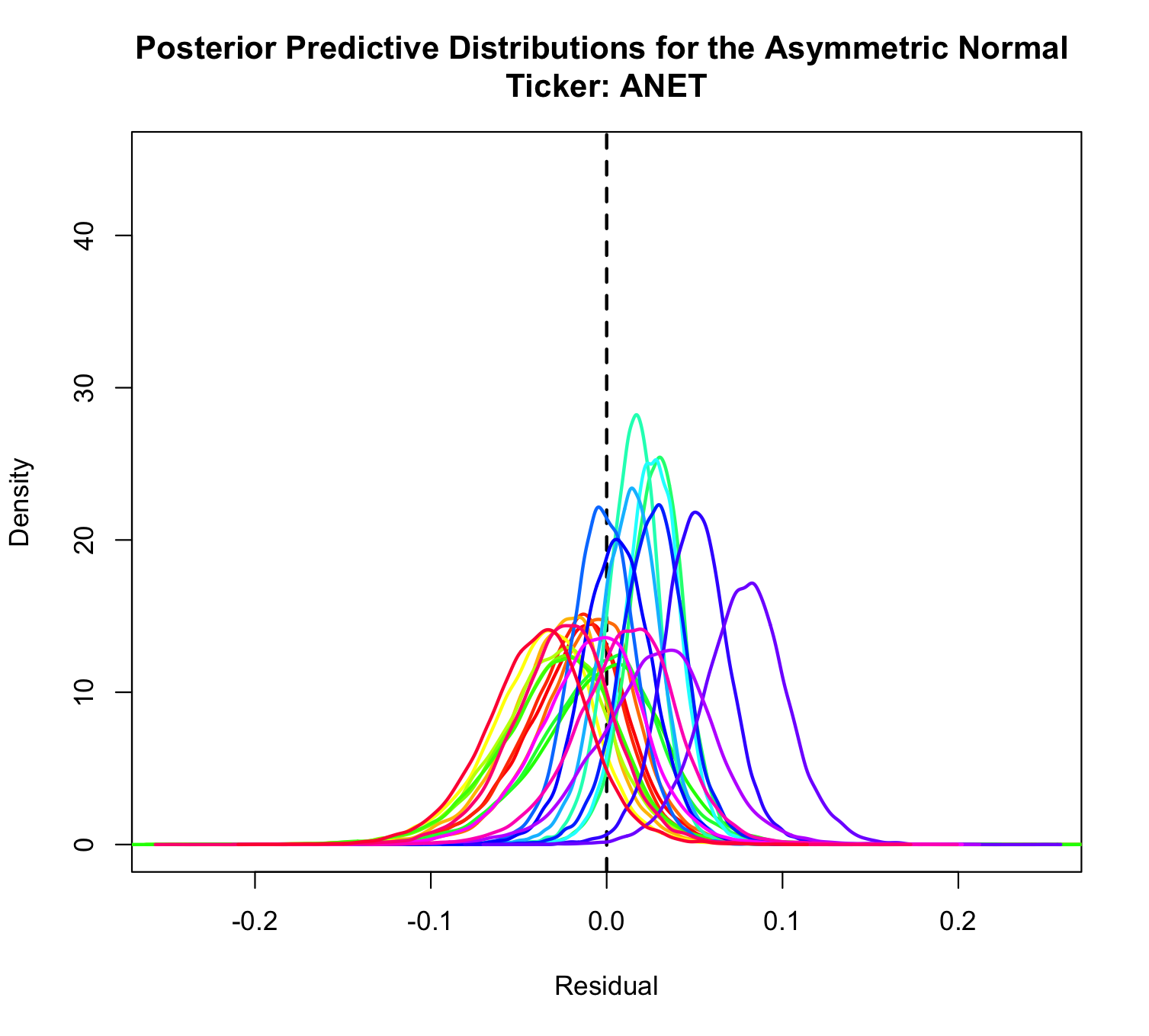}
    \includegraphics[width=0.33\textwidth]{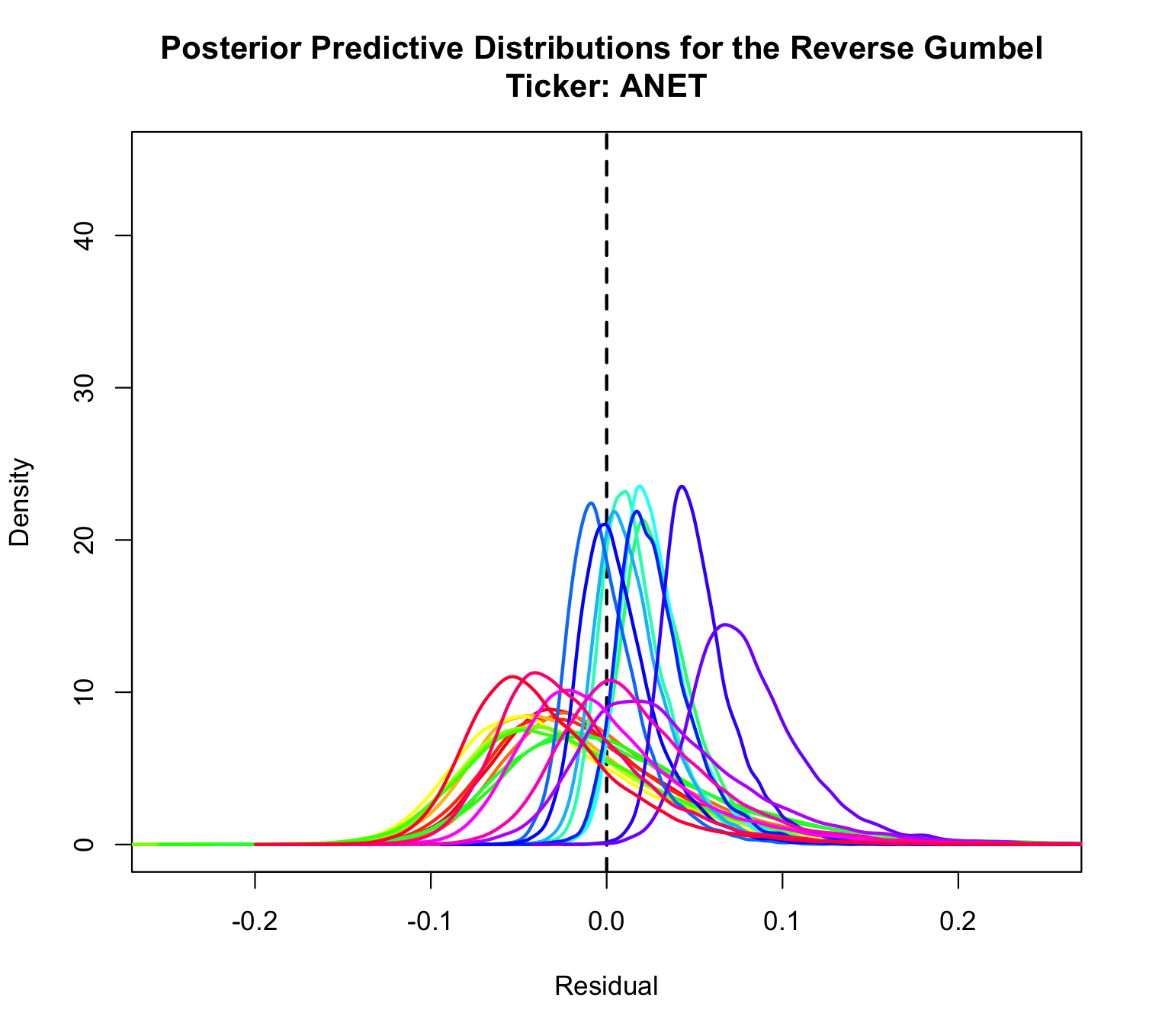}
    \caption{Posterior predictive distributions for ANET.}
    \label{fig:ANET}
\end{figure}

\begin{figure}[ht]
    \centering
    \includegraphics[width=0.33\textwidth]{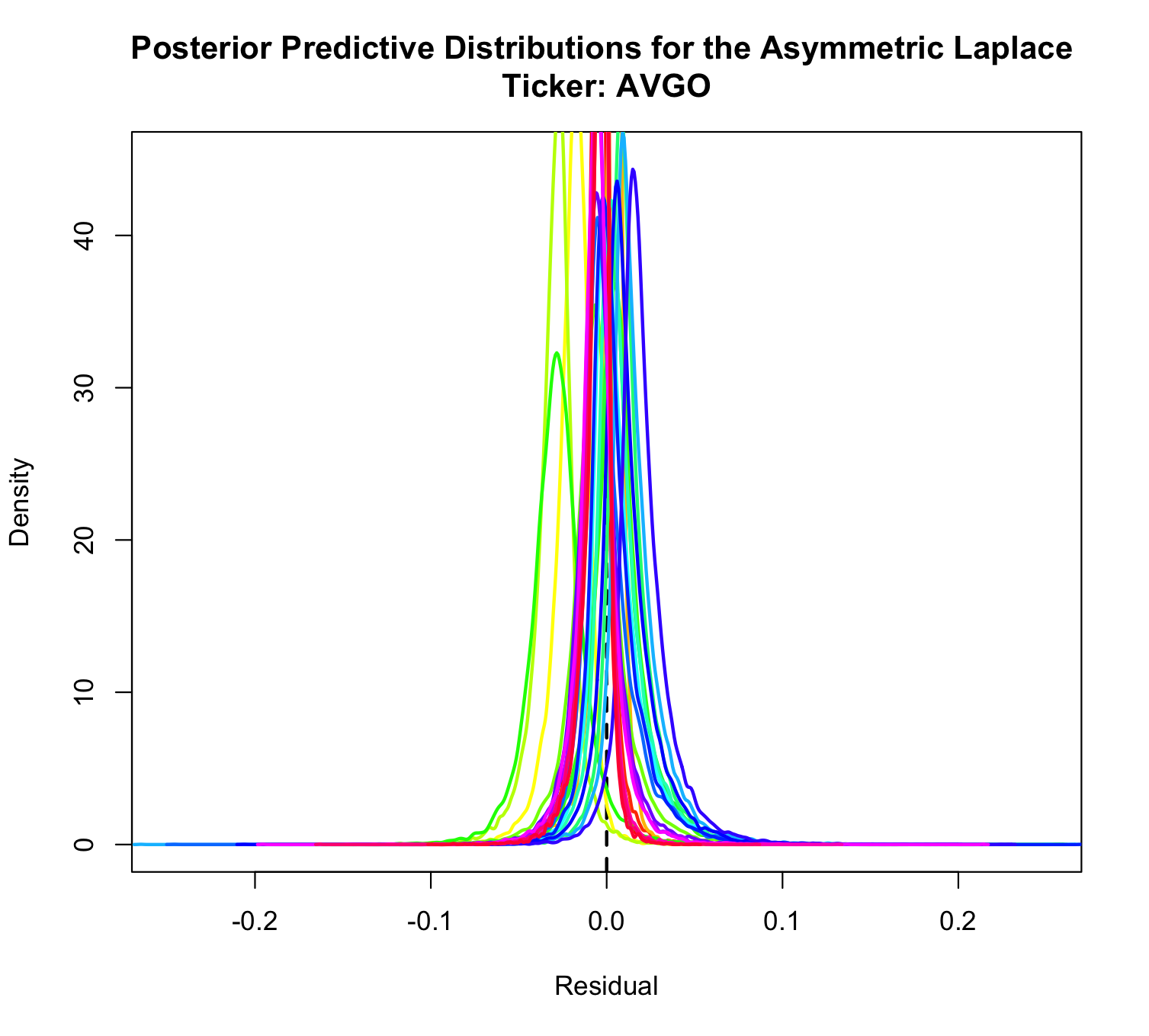}
    \includegraphics[width=0.33\textwidth]{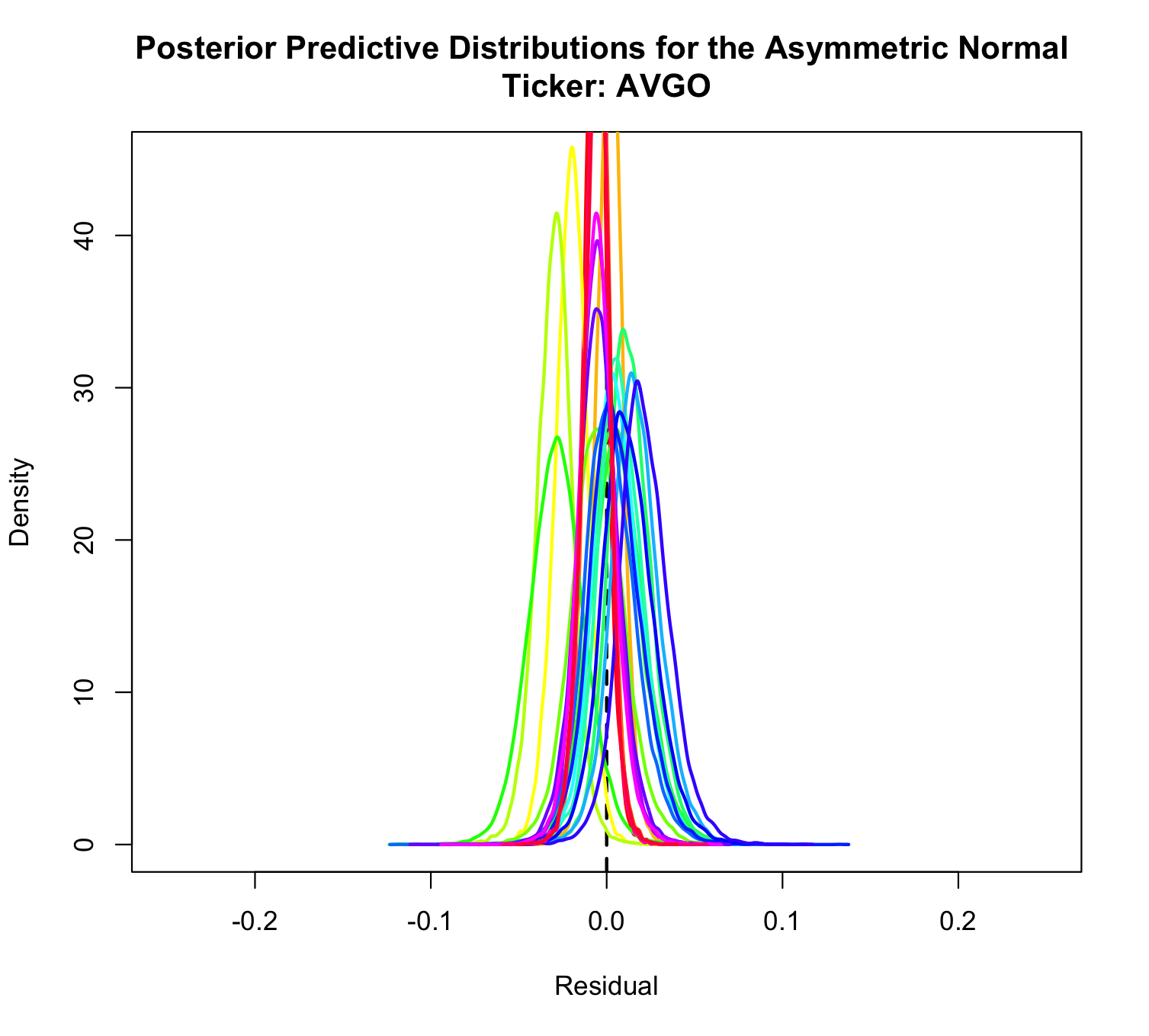}
    \includegraphics[width=0.33\textwidth]{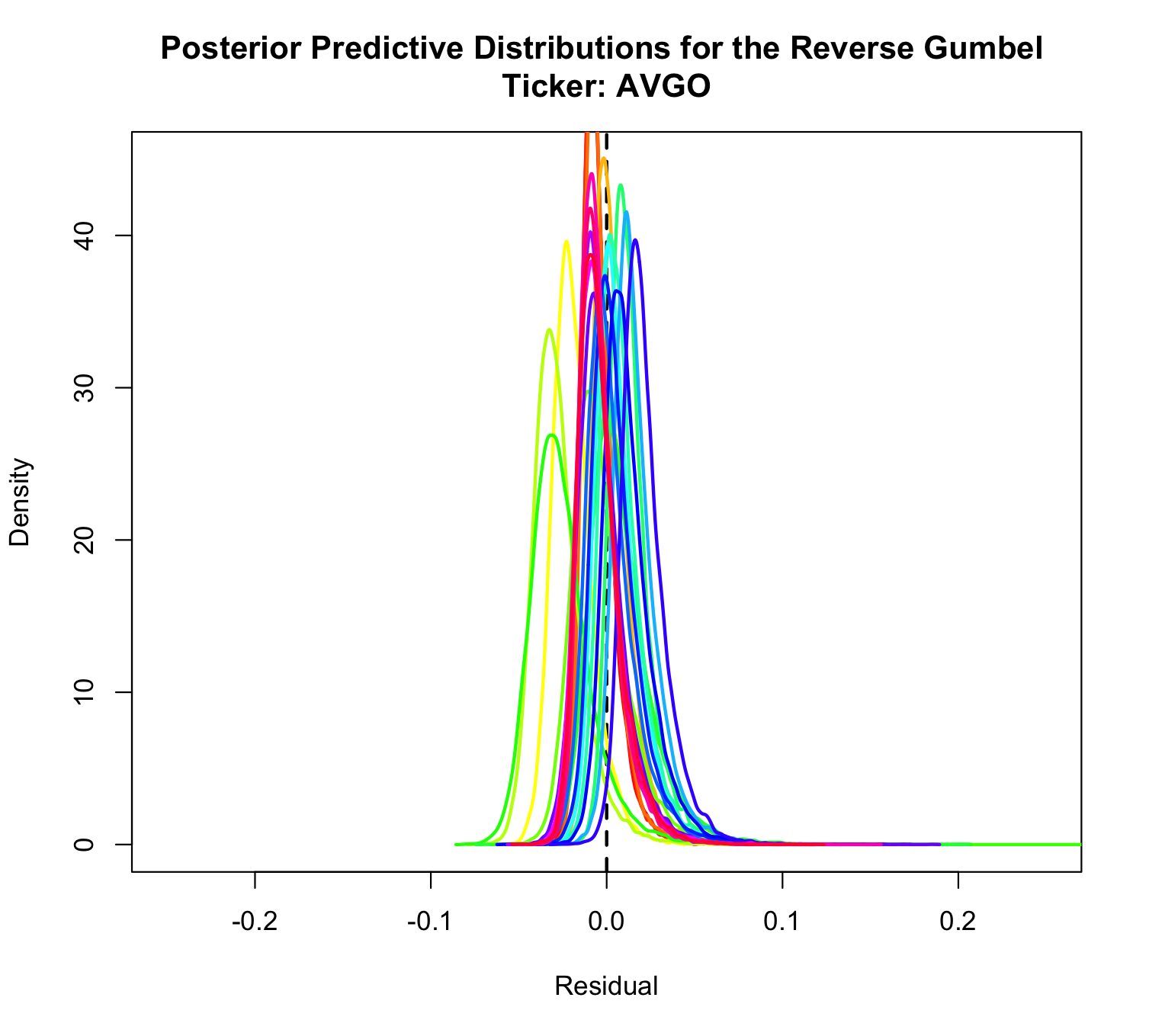}
    \caption{Posterior predictive distributions for AVGO.}
    \label{fig:AVGO}
\end{figure}

\begin{figure}[ht]
    \centering
    \includegraphics[width=0.33\textwidth]{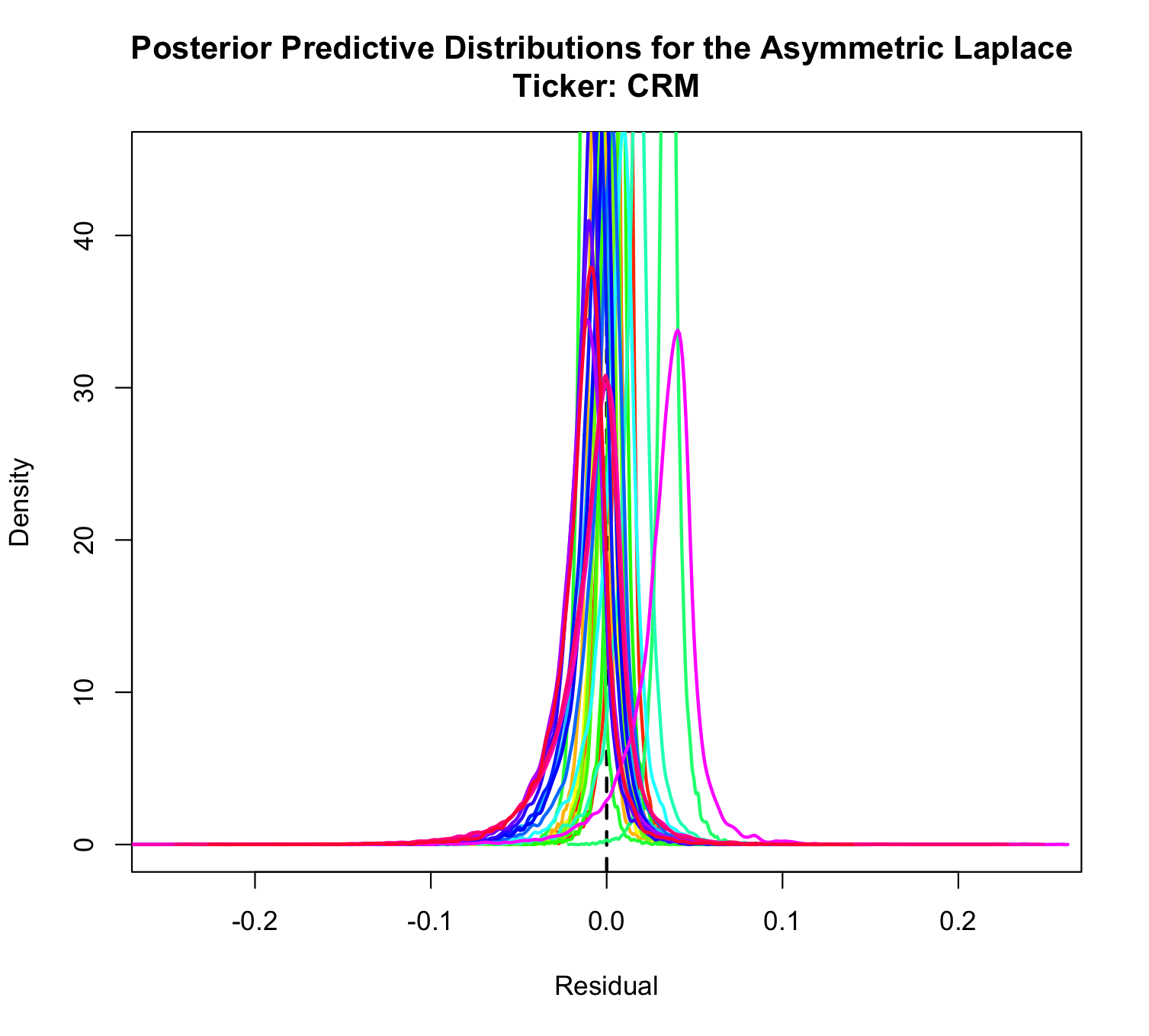}
    \includegraphics[width=0.33\textwidth]{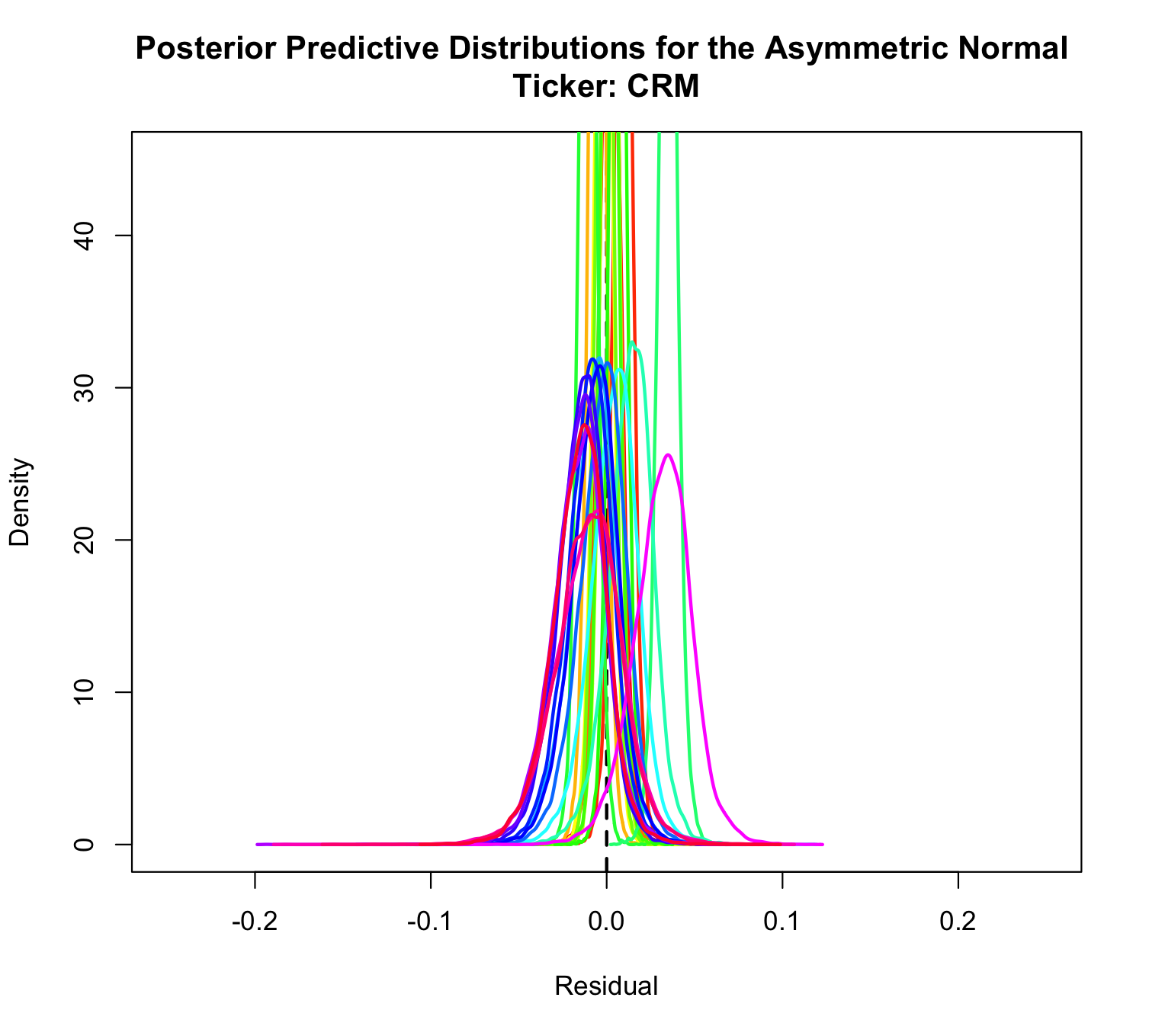}
    \includegraphics[width=0.33\textwidth]{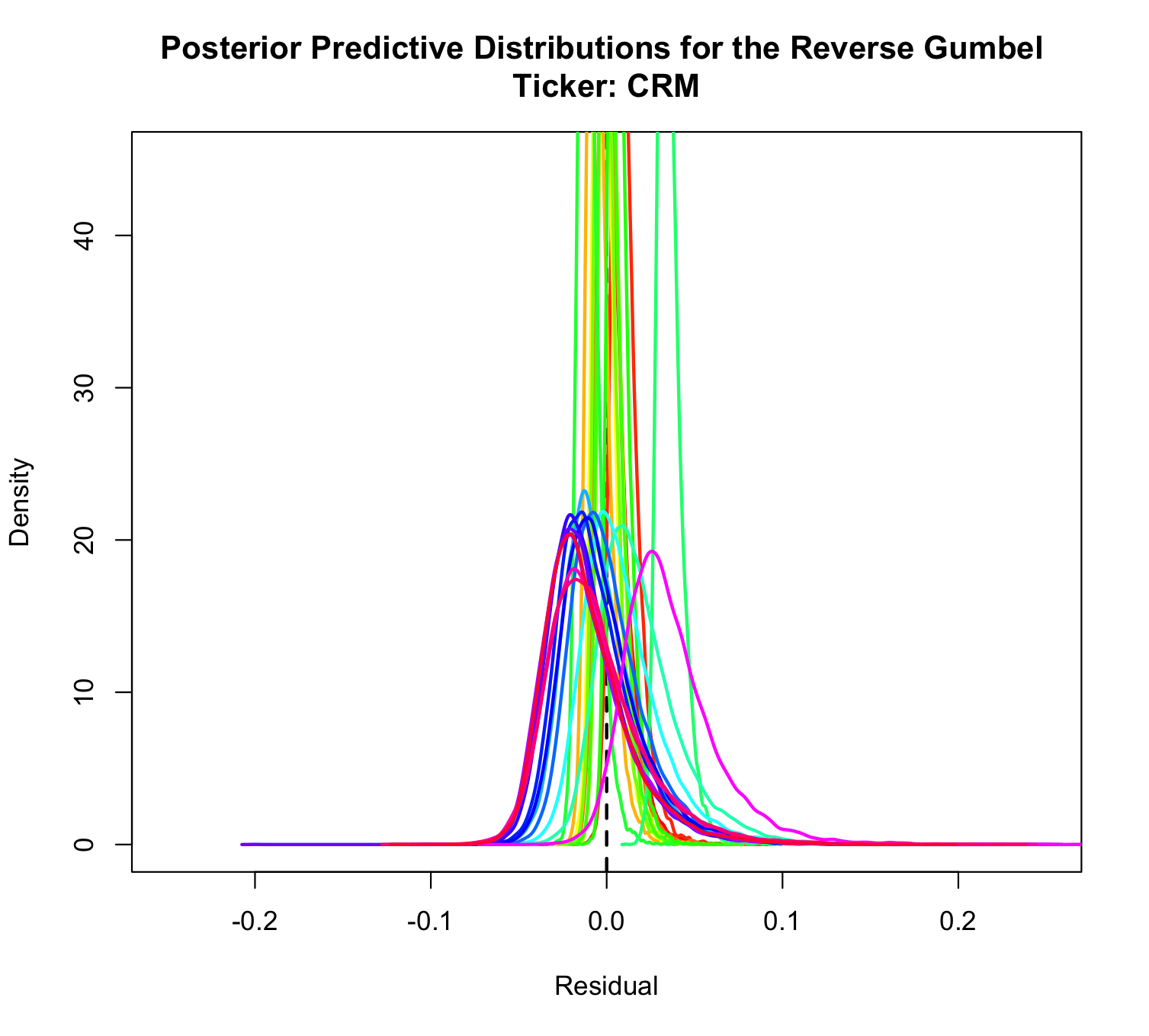}
    \caption{Posterior predictive distributions for CRM.}
    \label{fig:CRM}
\end{figure}

\begin{figure}[ht]
    \centering
    \includegraphics[width=0.33\textwidth]{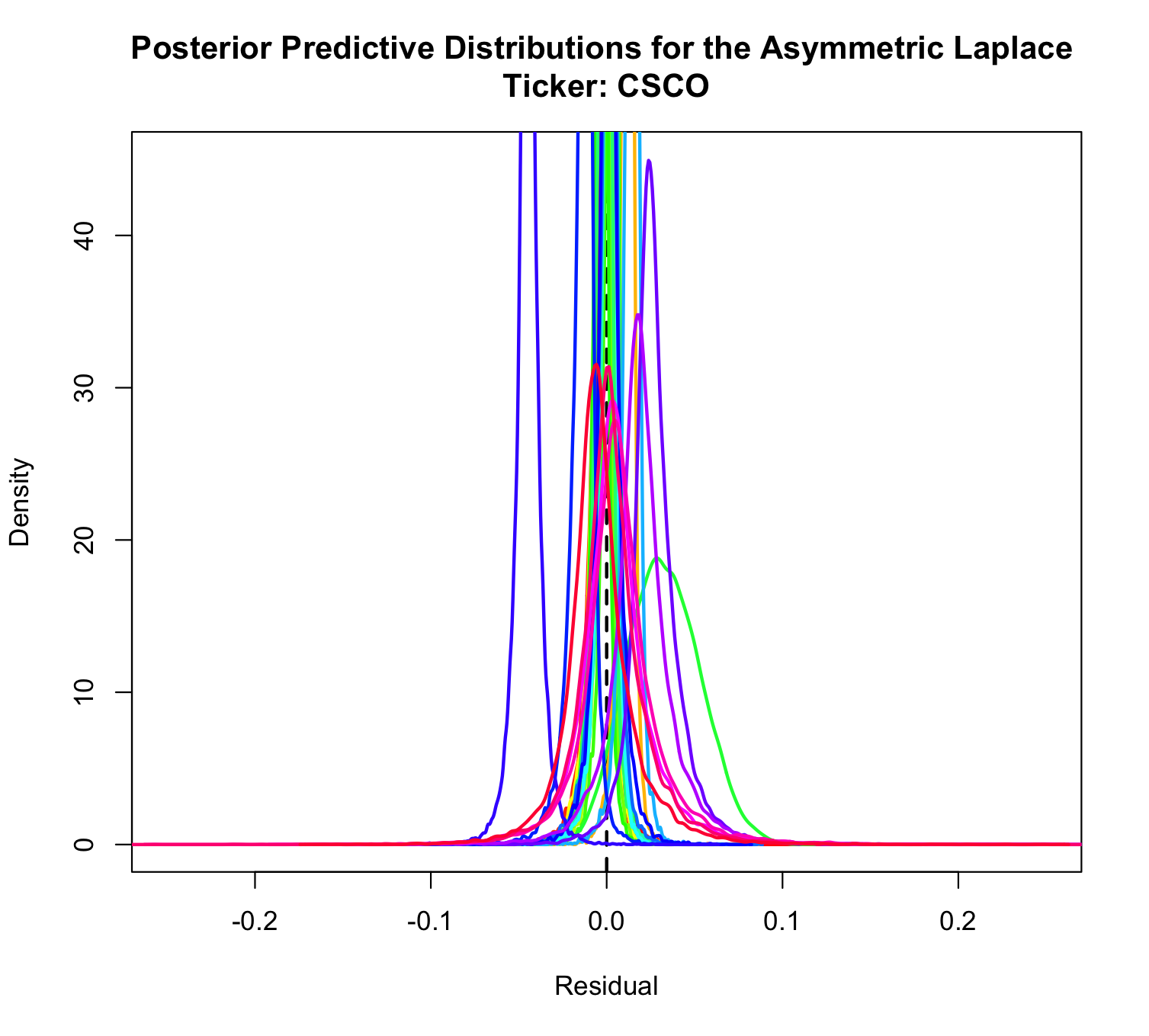}
    \includegraphics[width=0.33\textwidth]{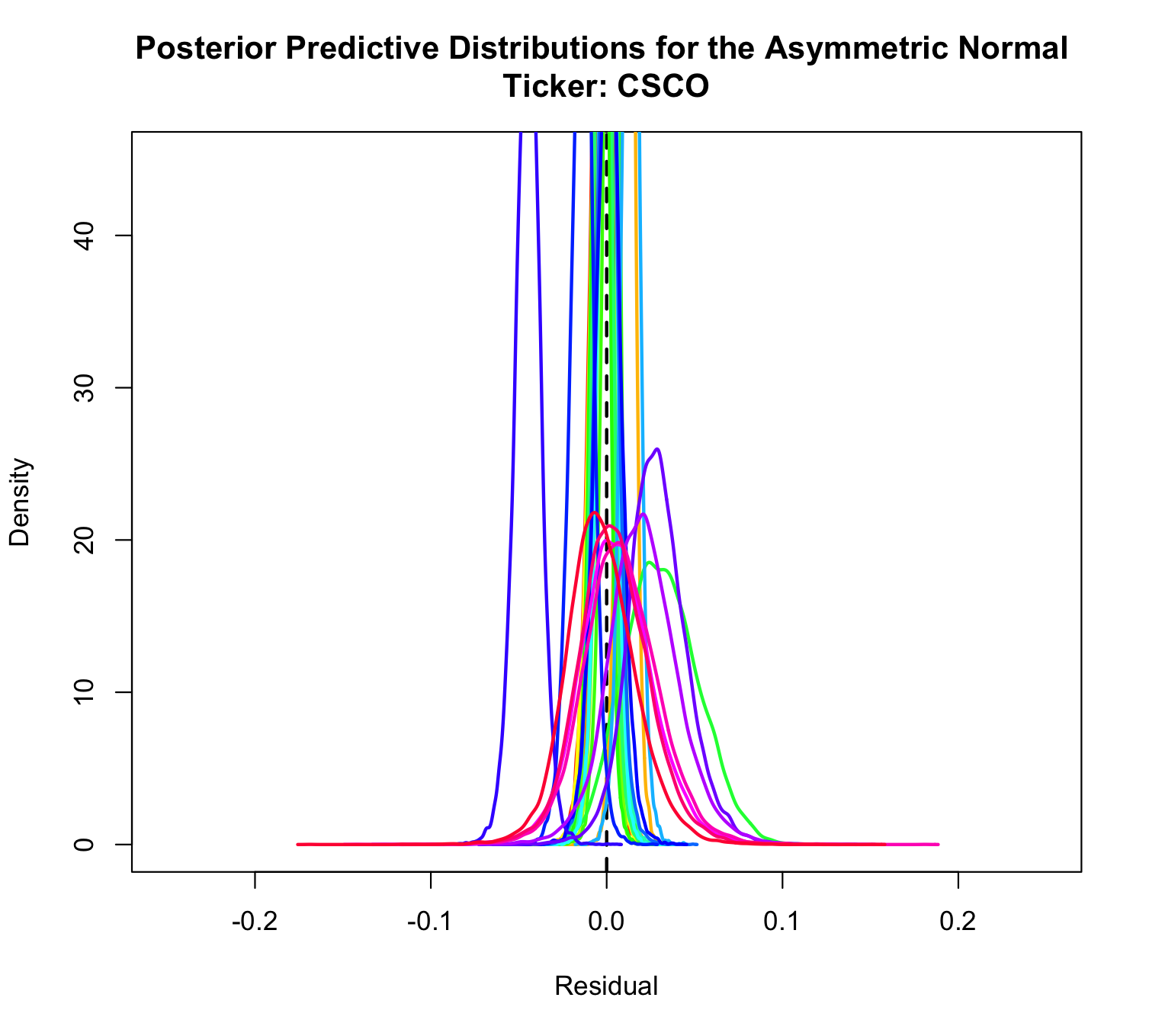}
    \includegraphics[width=0.33\textwidth]{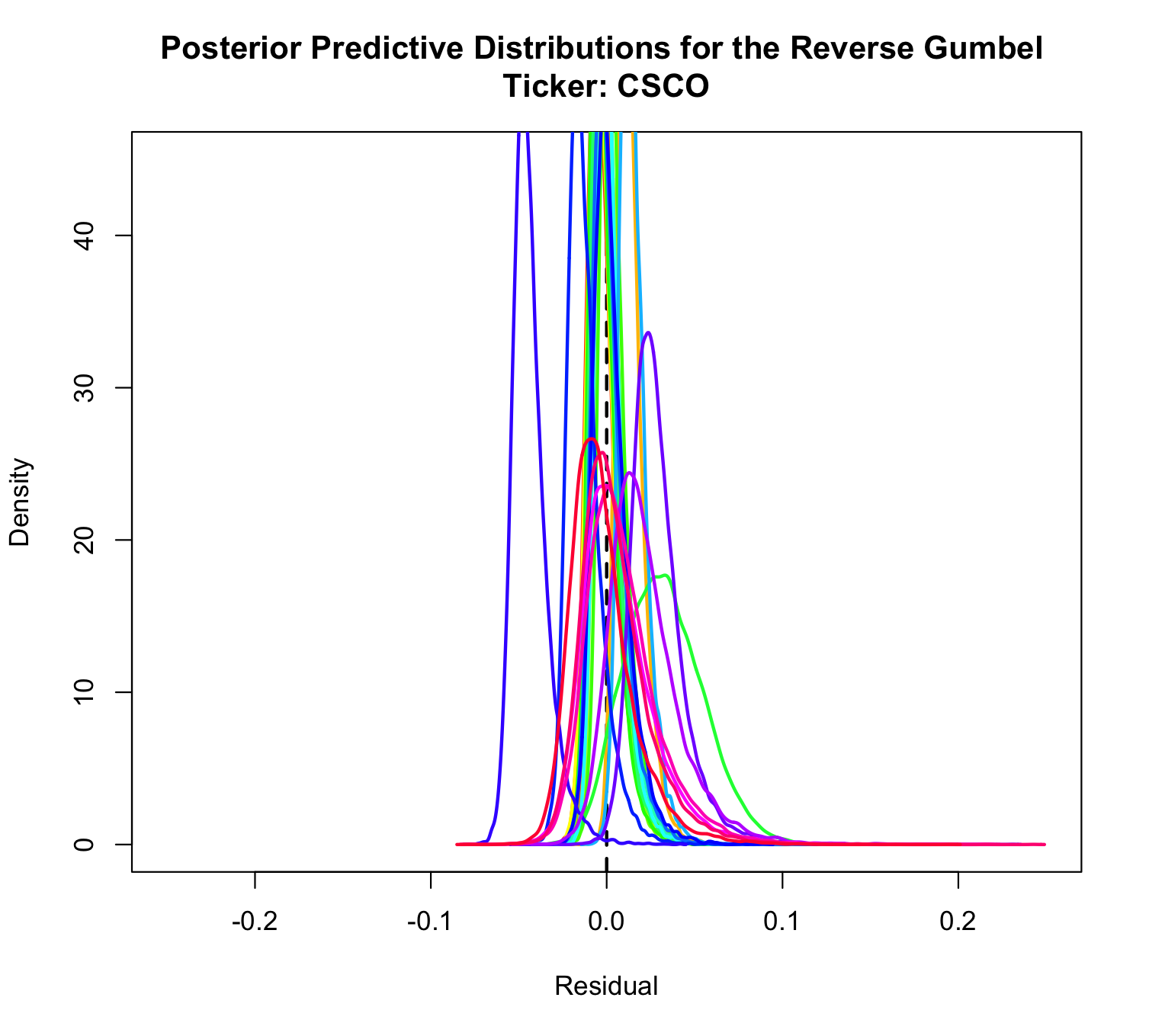}
    \caption{Posterior predictive distributions for CSCO.}
    \label{fig:CSCO}
\end{figure}

\begin{figure}[ht]
    \centering
    \includegraphics[width=0.33\textwidth]{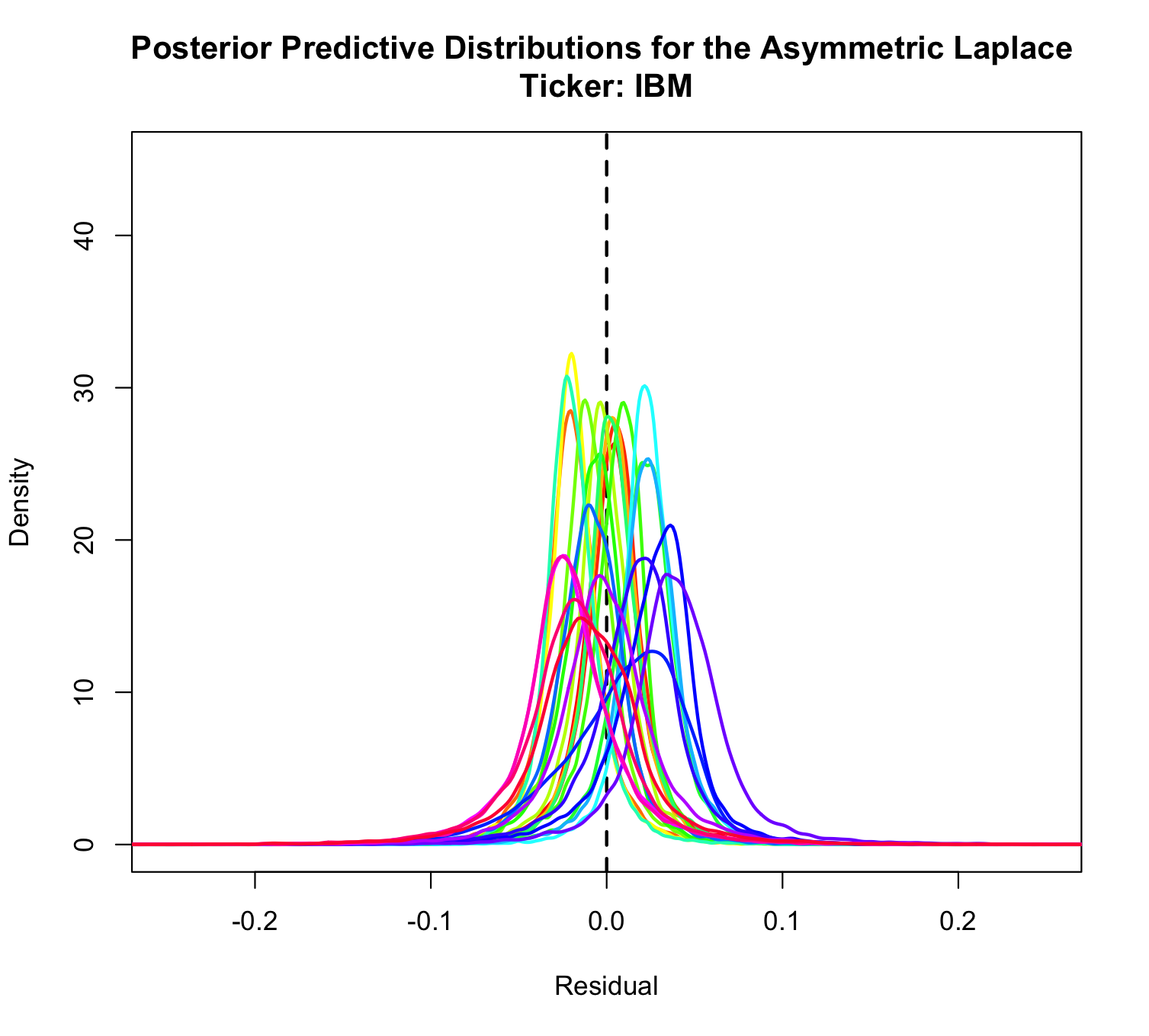}
    \includegraphics[width=0.33\textwidth]{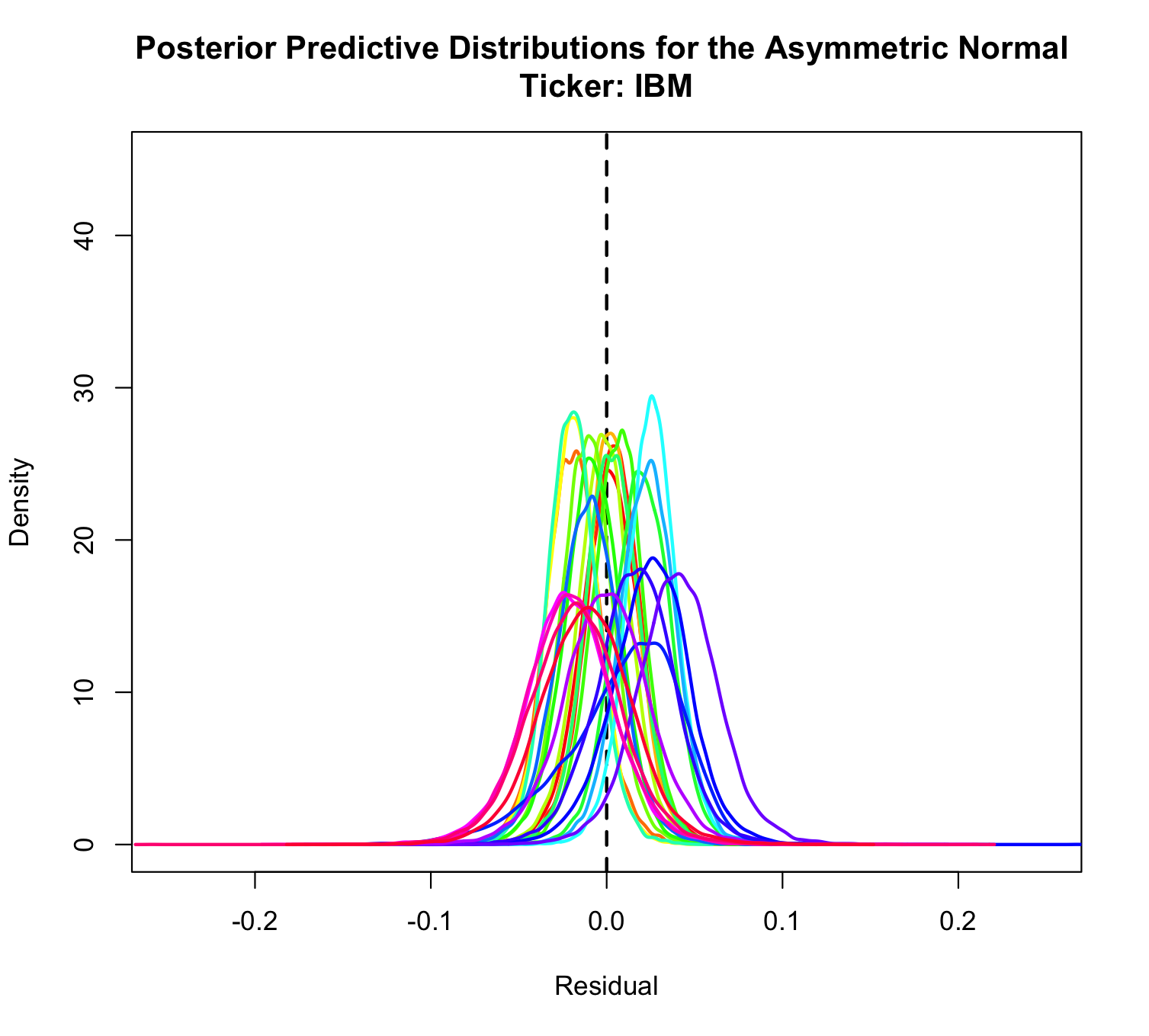}
    \includegraphics[width=0.33\textwidth]{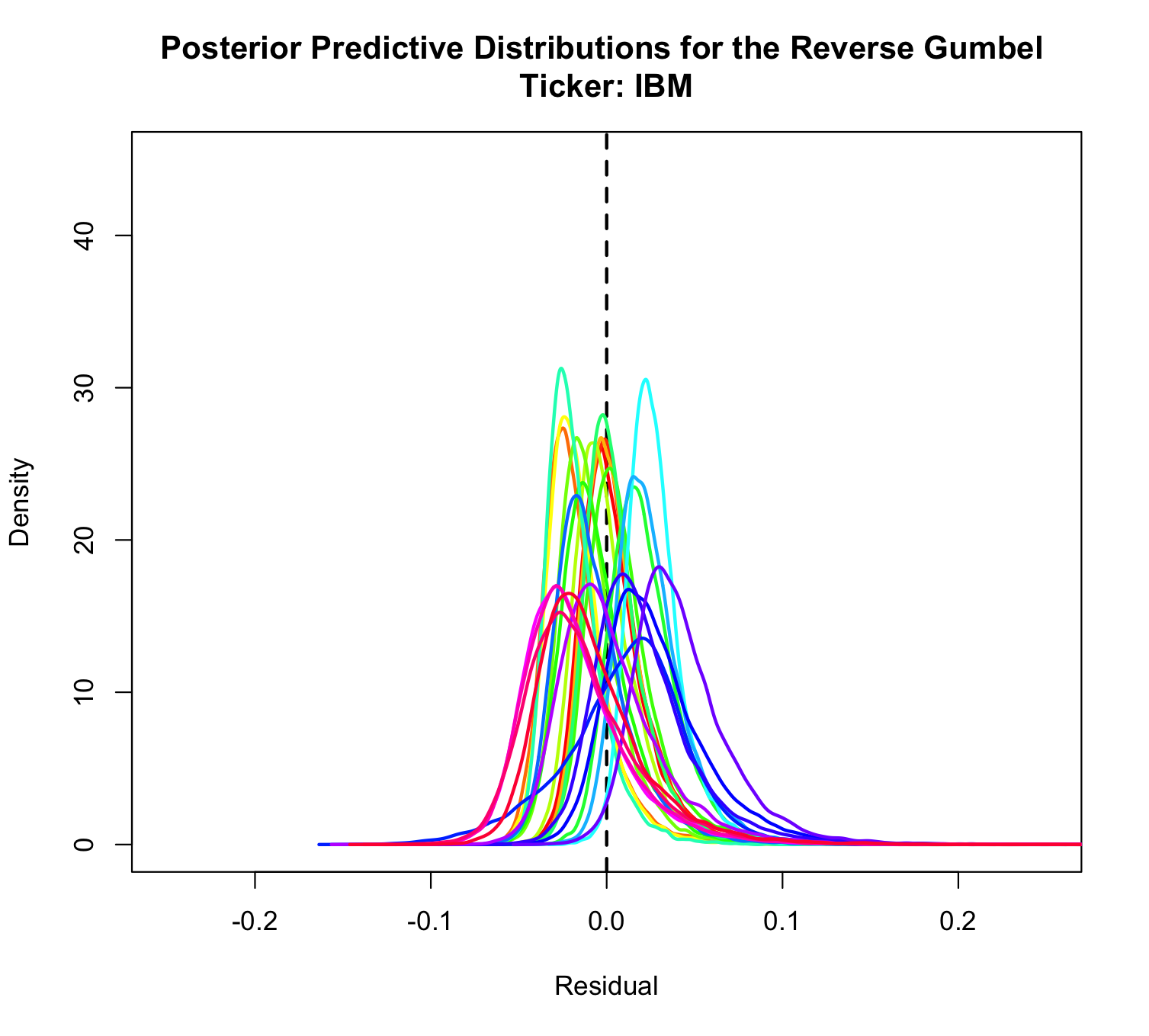}
    \caption{Posterior predictive distributions for IBM.}
    \label{fig:IBM}
\end{figure}

\begin{figure}[ht]
    \centering
    \includegraphics[width=0.33\textwidth]{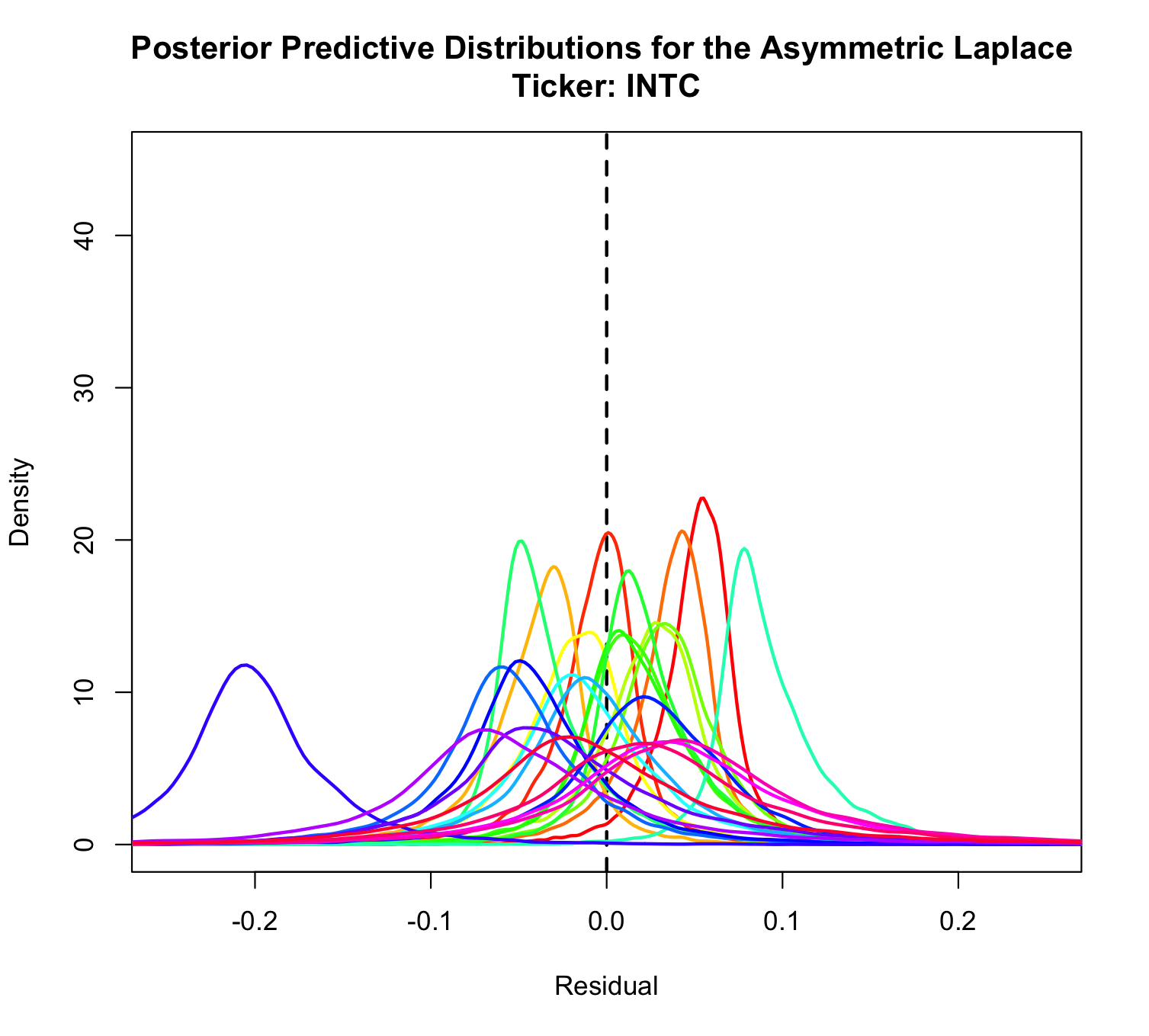}
    \includegraphics[width=0.33\textwidth]{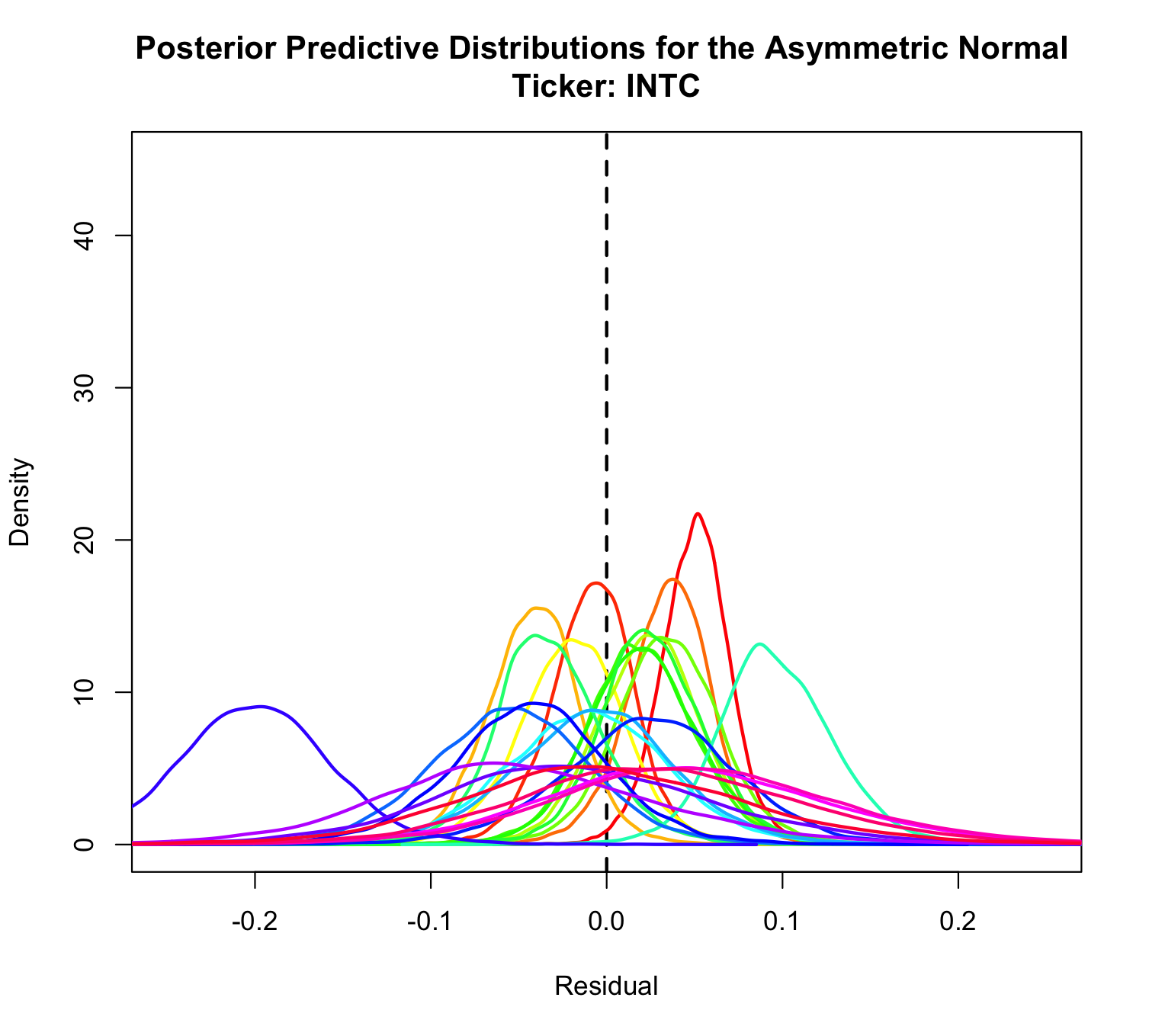}
    \includegraphics[width=0.33\textwidth]{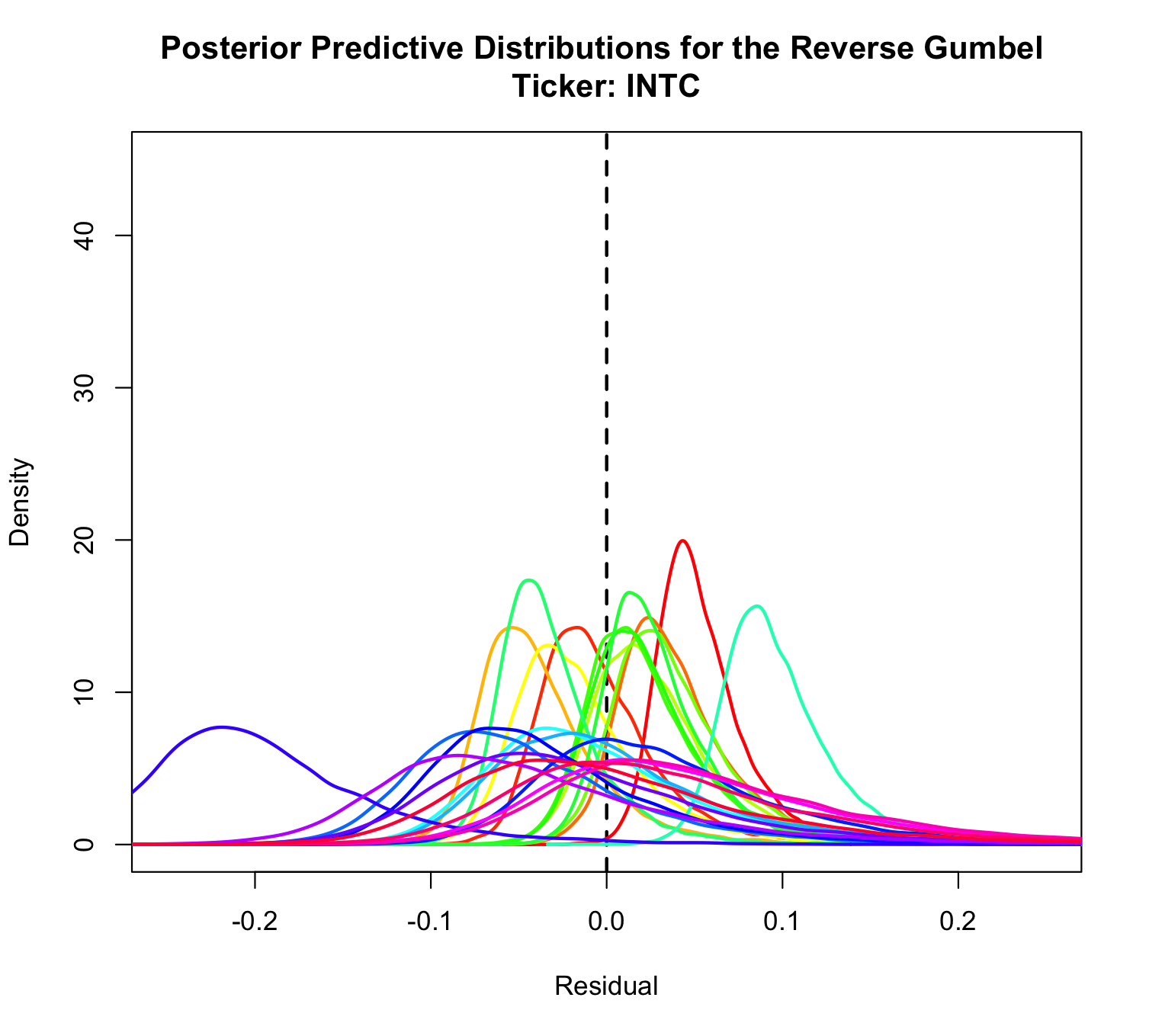}
    \caption{Posterior predictive distributions for INTC.}
    \label{fig:INTC}
\end{figure}

\begin{figure}[ht]
    \centering
    \includegraphics[width=0.33\textwidth]{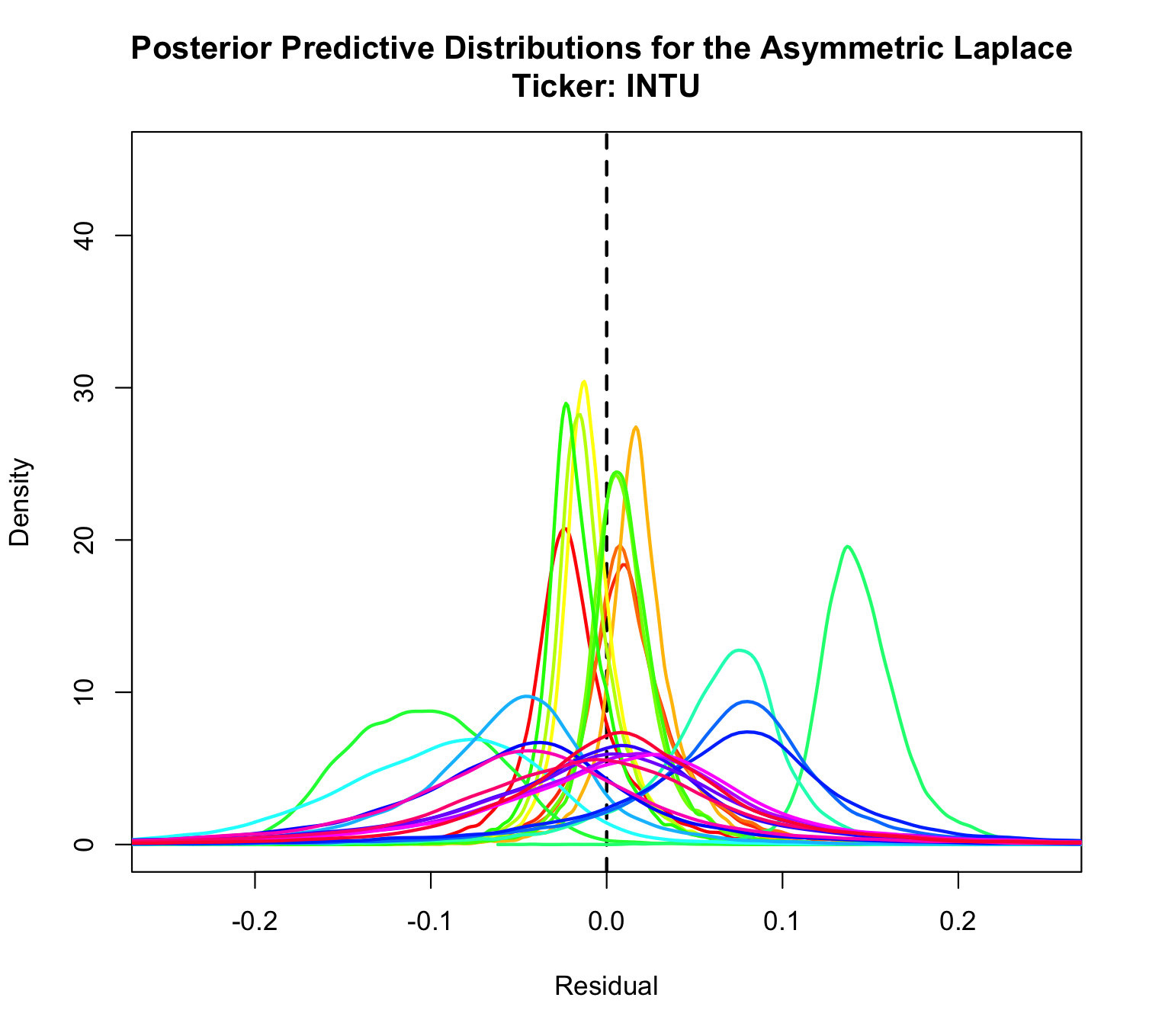}
    \includegraphics[width=0.33\textwidth]{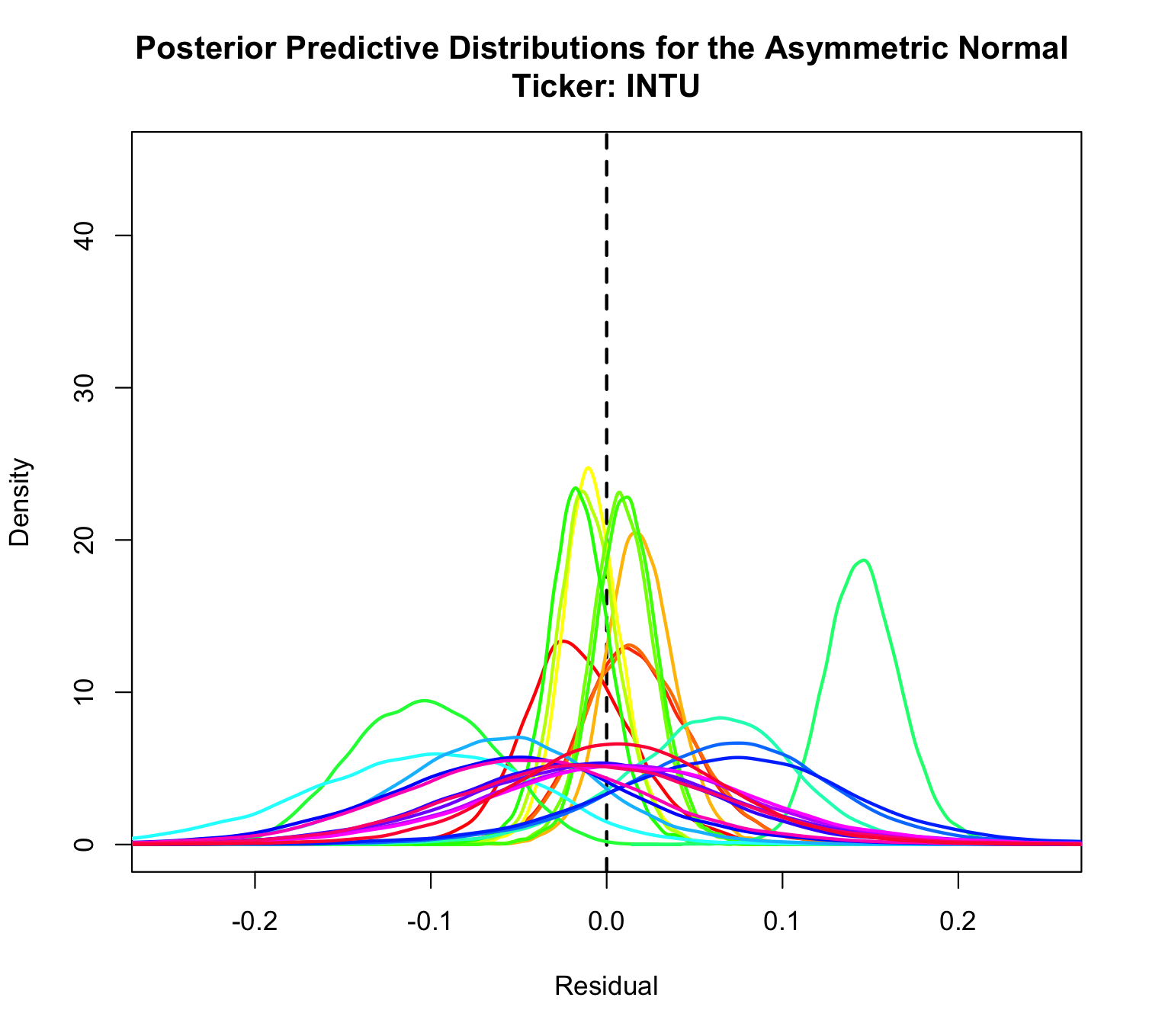}
    \includegraphics[width=0.33\textwidth]{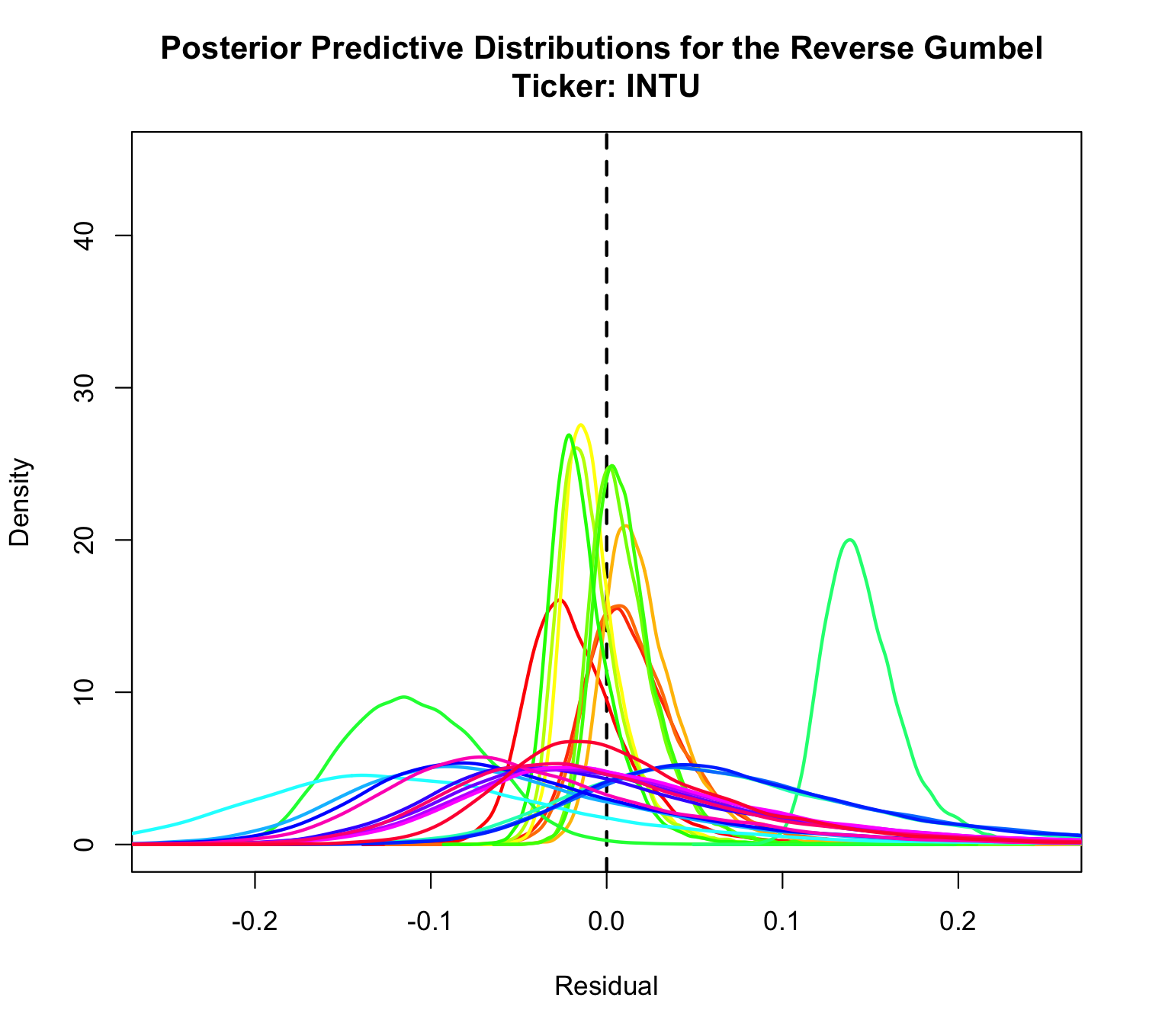}
    \caption{Posterior predictive distributions for INTU.}
    \label{fig:INTU}
\end{figure}

\begin{figure}[ht]
    \centering
    \includegraphics[width=0.33\textwidth]{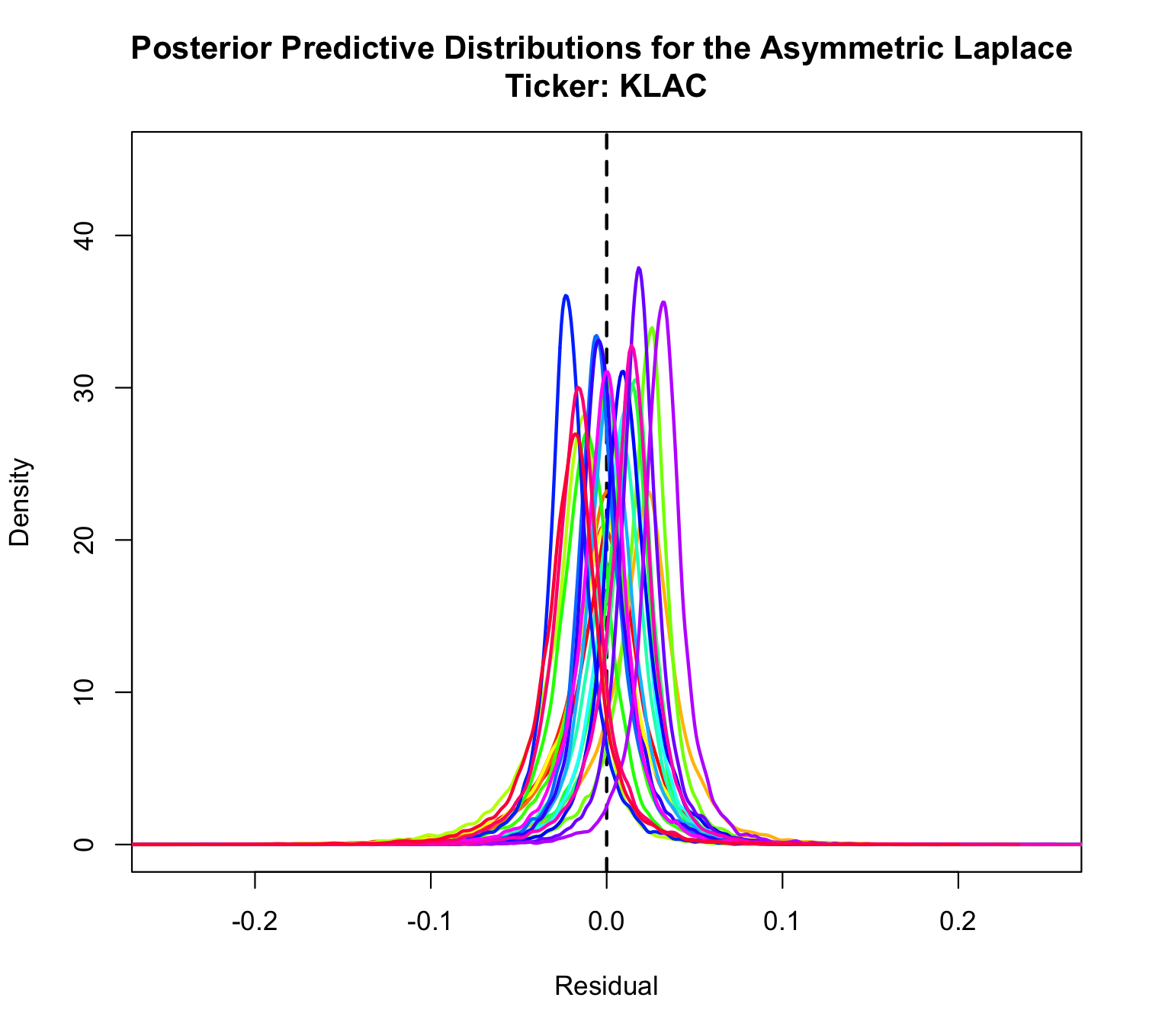}
    \includegraphics[width=0.33\textwidth]{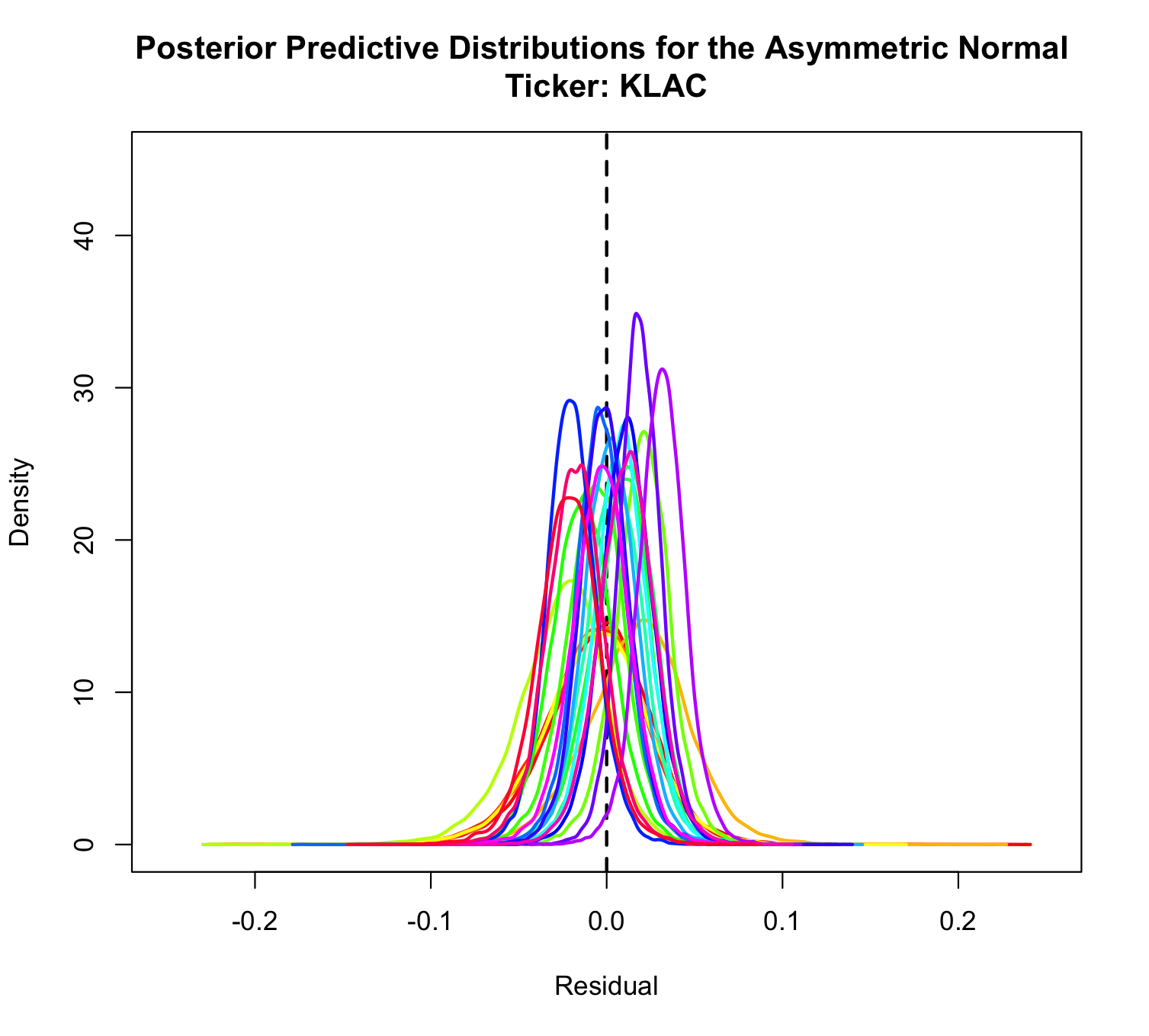}
    \includegraphics[width=0.33\textwidth]{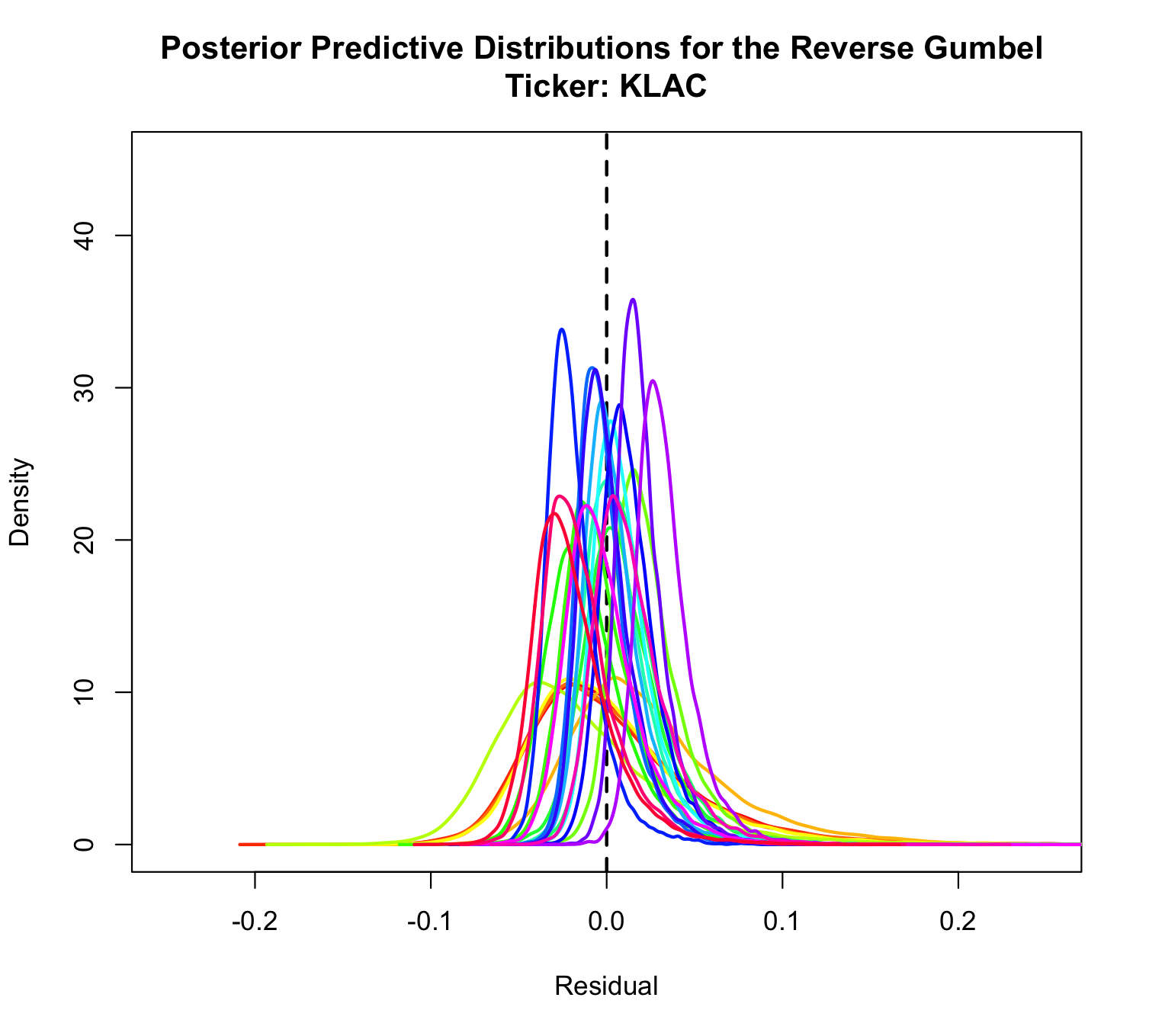}
    \caption{Posterior predictive distributions for KLAC.}
    \label{fig:KLAC}
\end{figure}

\begin{figure}[ht]
    \centering
    \includegraphics[width=0.33\textwidth]{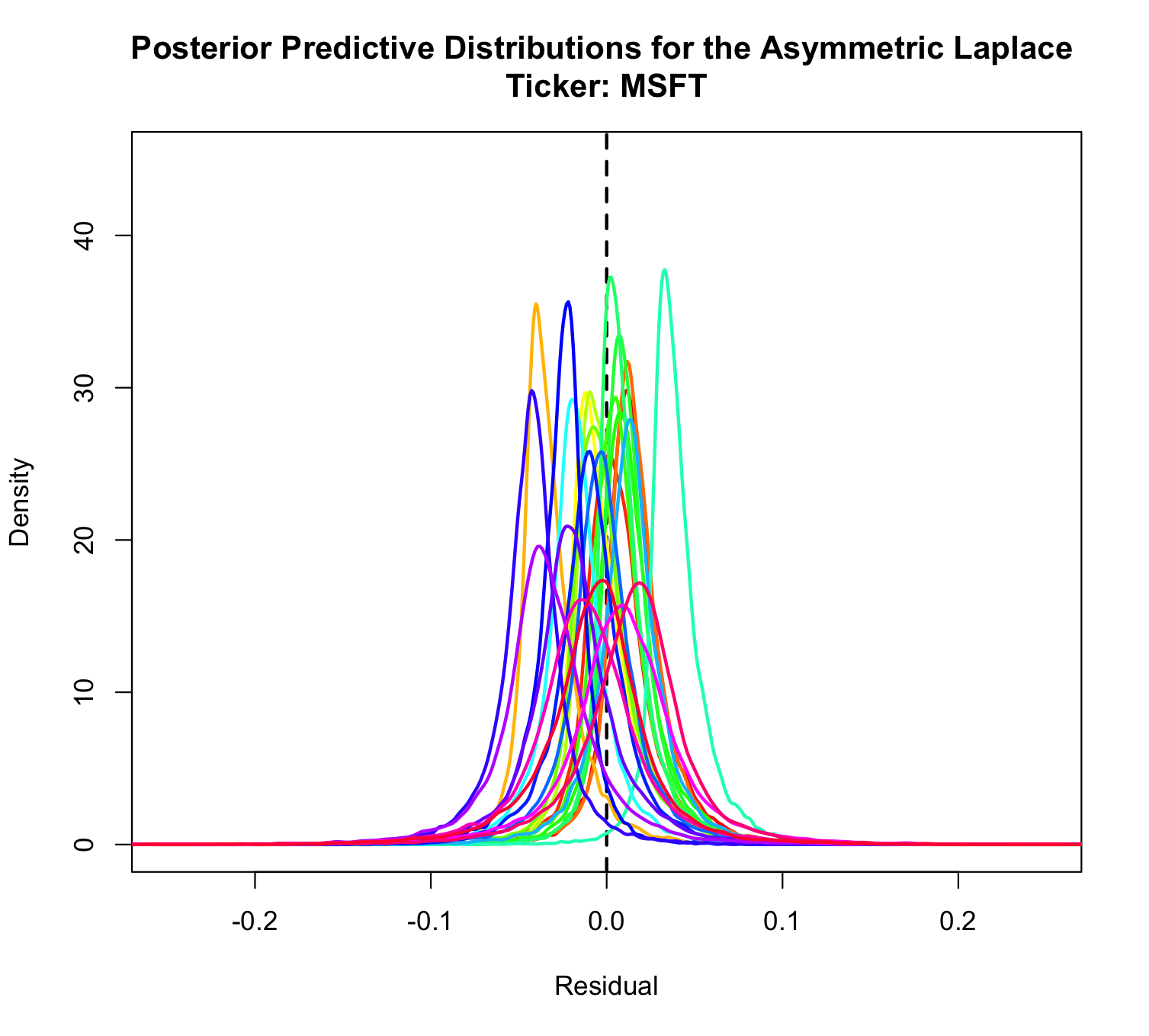}
    \includegraphics[width=0.33\textwidth]{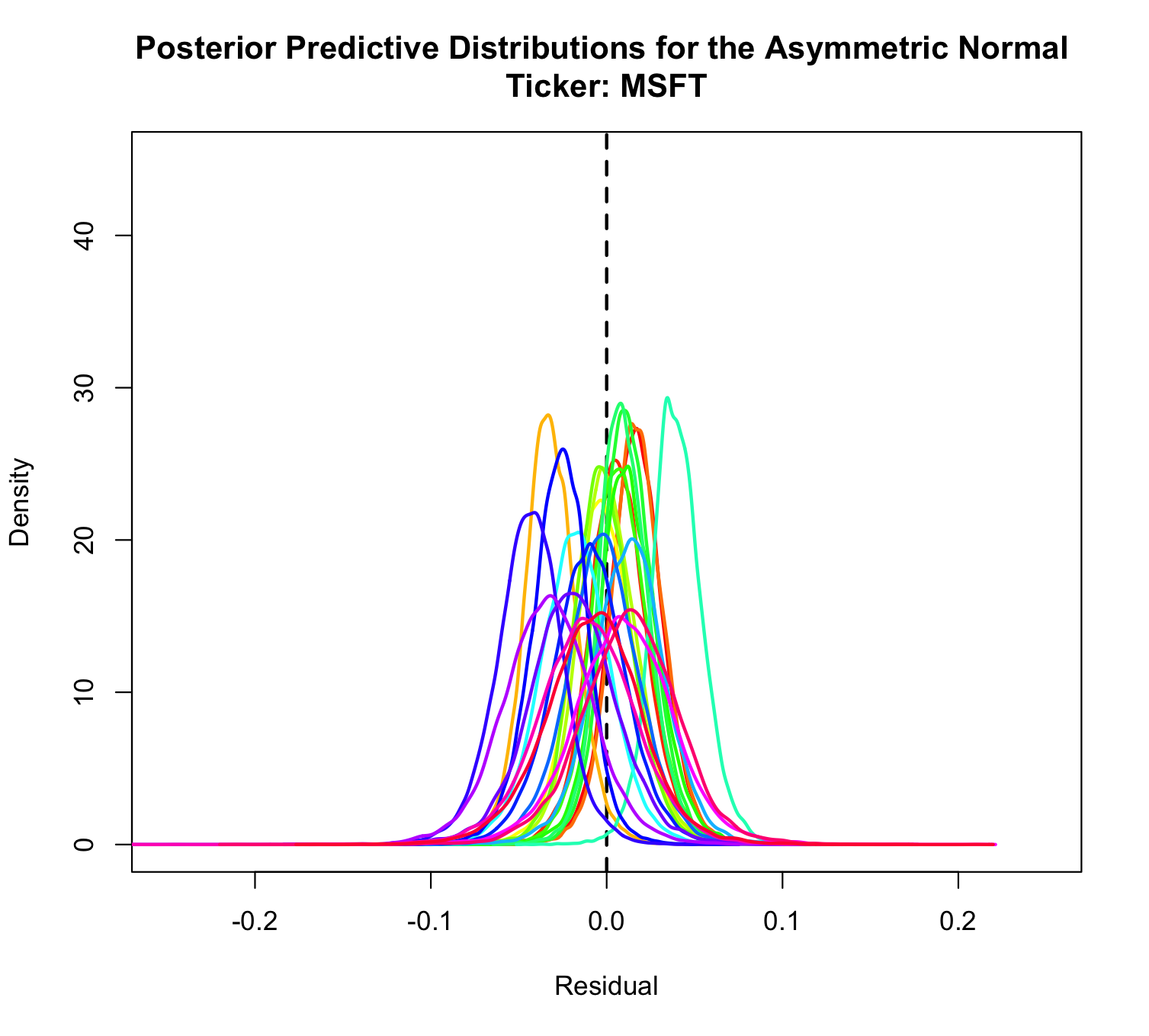}
    \includegraphics[width=0.33\textwidth]{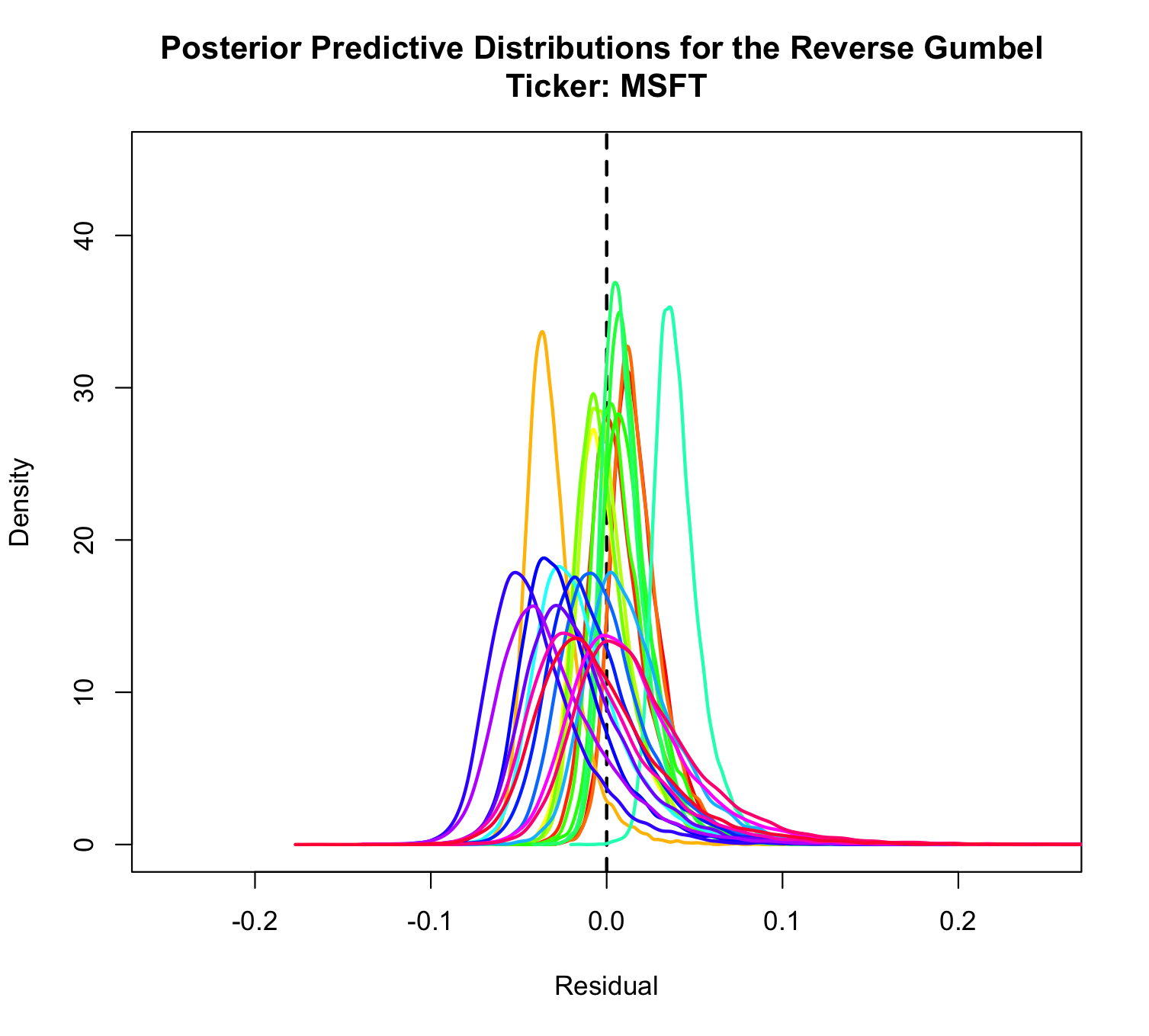}
    \caption{Posterior predictive distributions for MSFT.}
    \label{fig:MSFT}
\end{figure}

\begin{figure}[ht]
    \centering
    \includegraphics[width=0.33\textwidth]{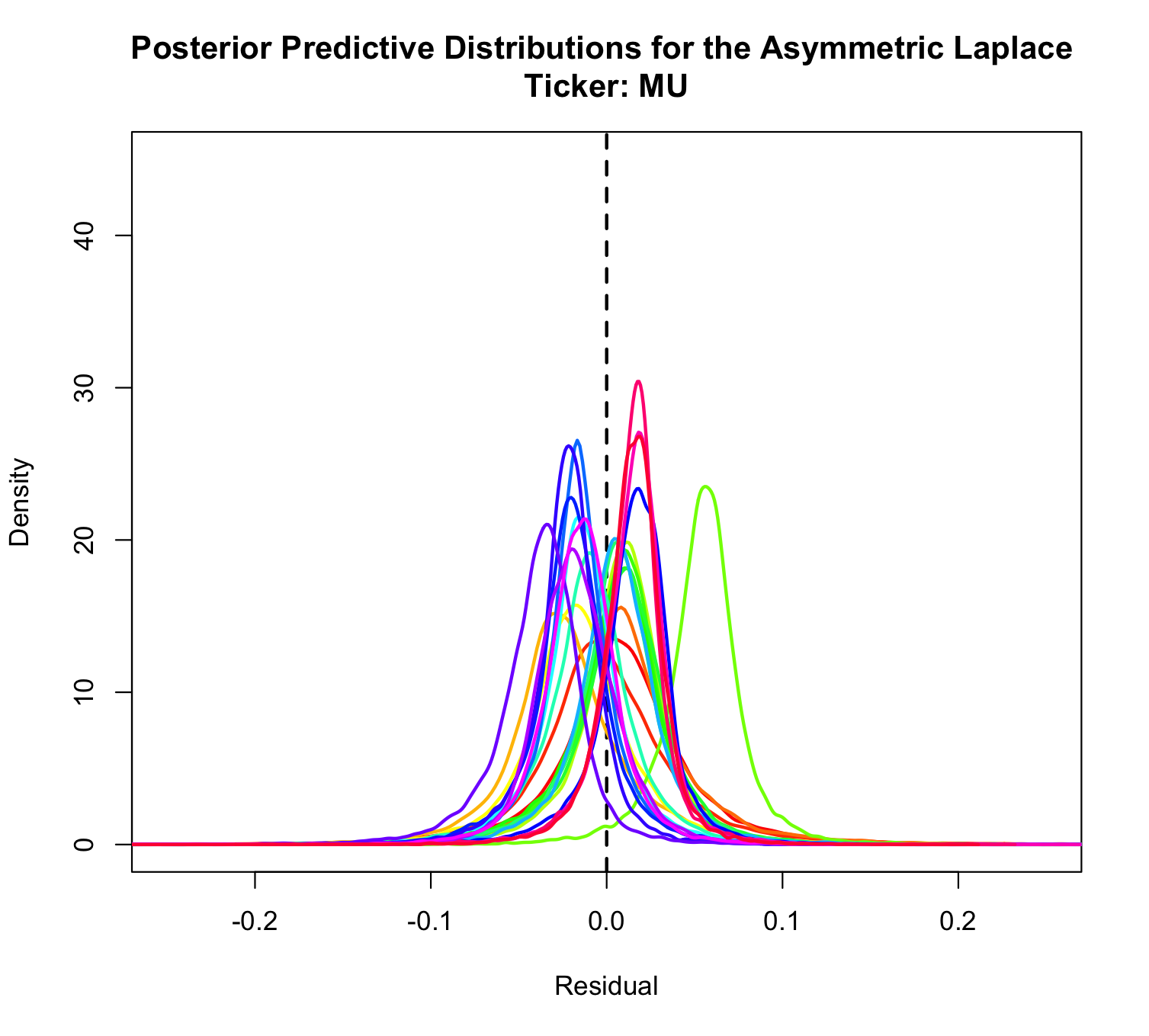}
    \includegraphics[width=0.33\textwidth]{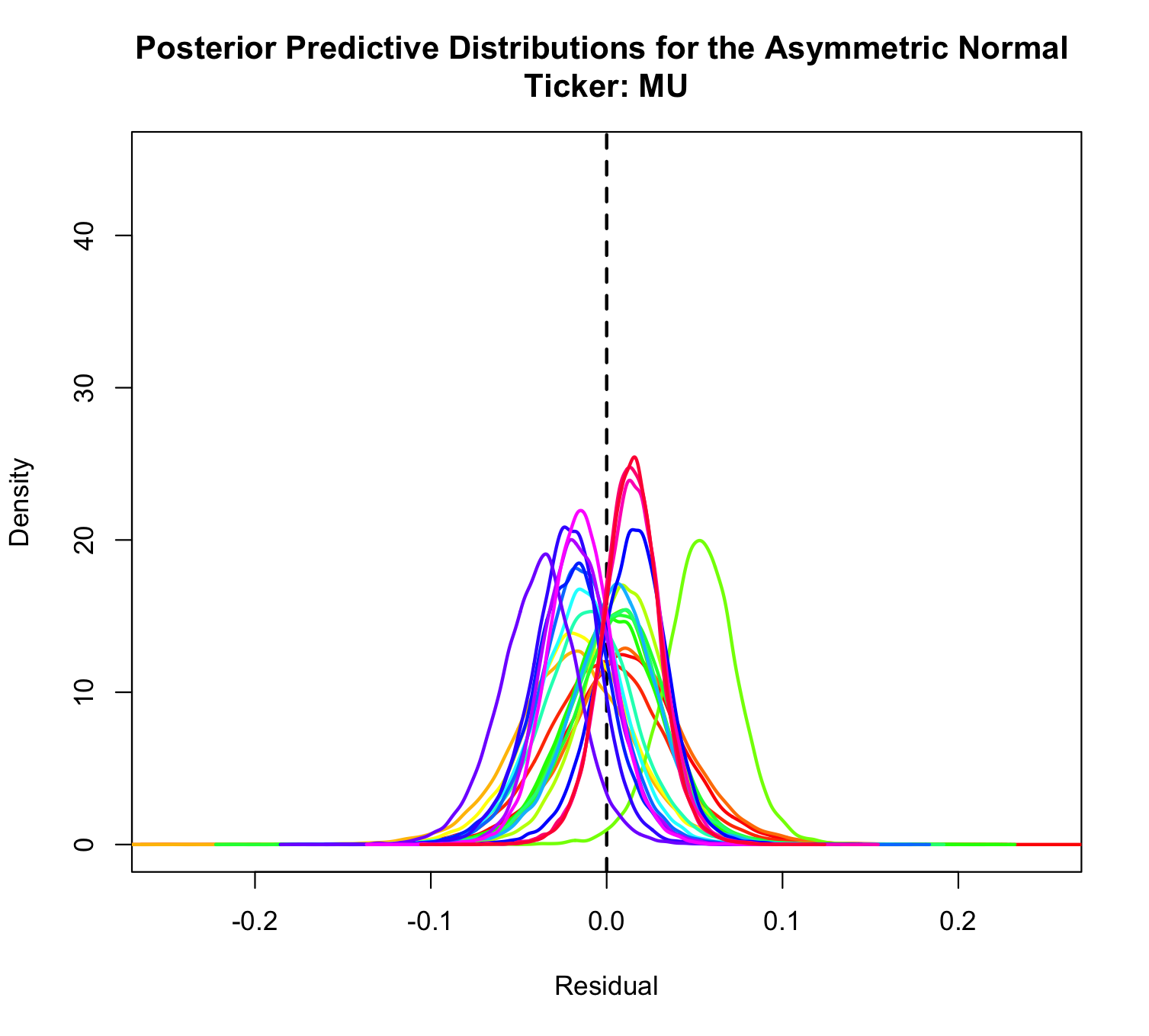}
    \includegraphics[width=0.33\textwidth]{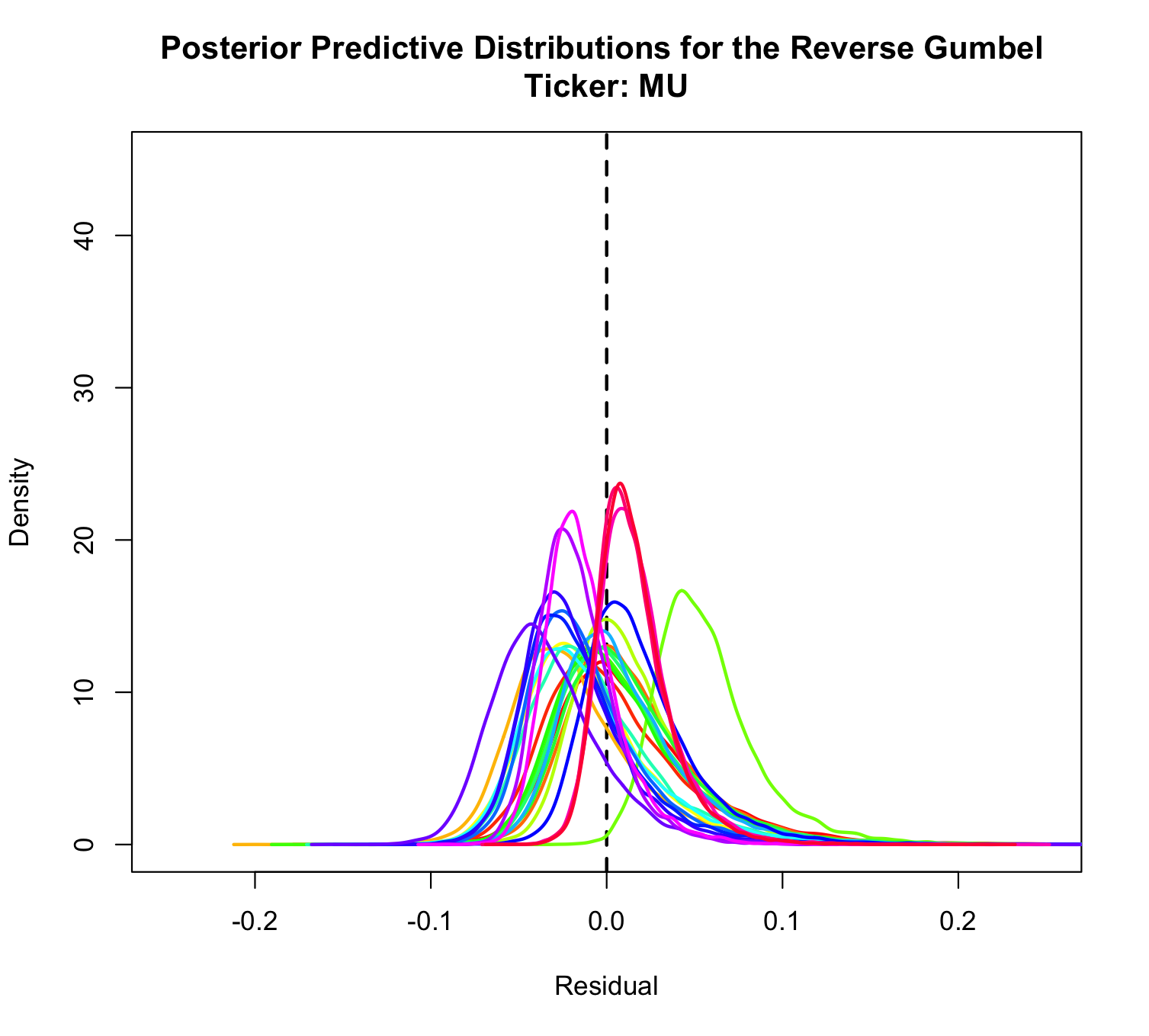}
    \caption{Posterior predictive distributions for MU.}
    \label{fig:MU}
\end{figure}

\begin{figure}[ht]
    \centering
    \includegraphics[width=0.33\textwidth]{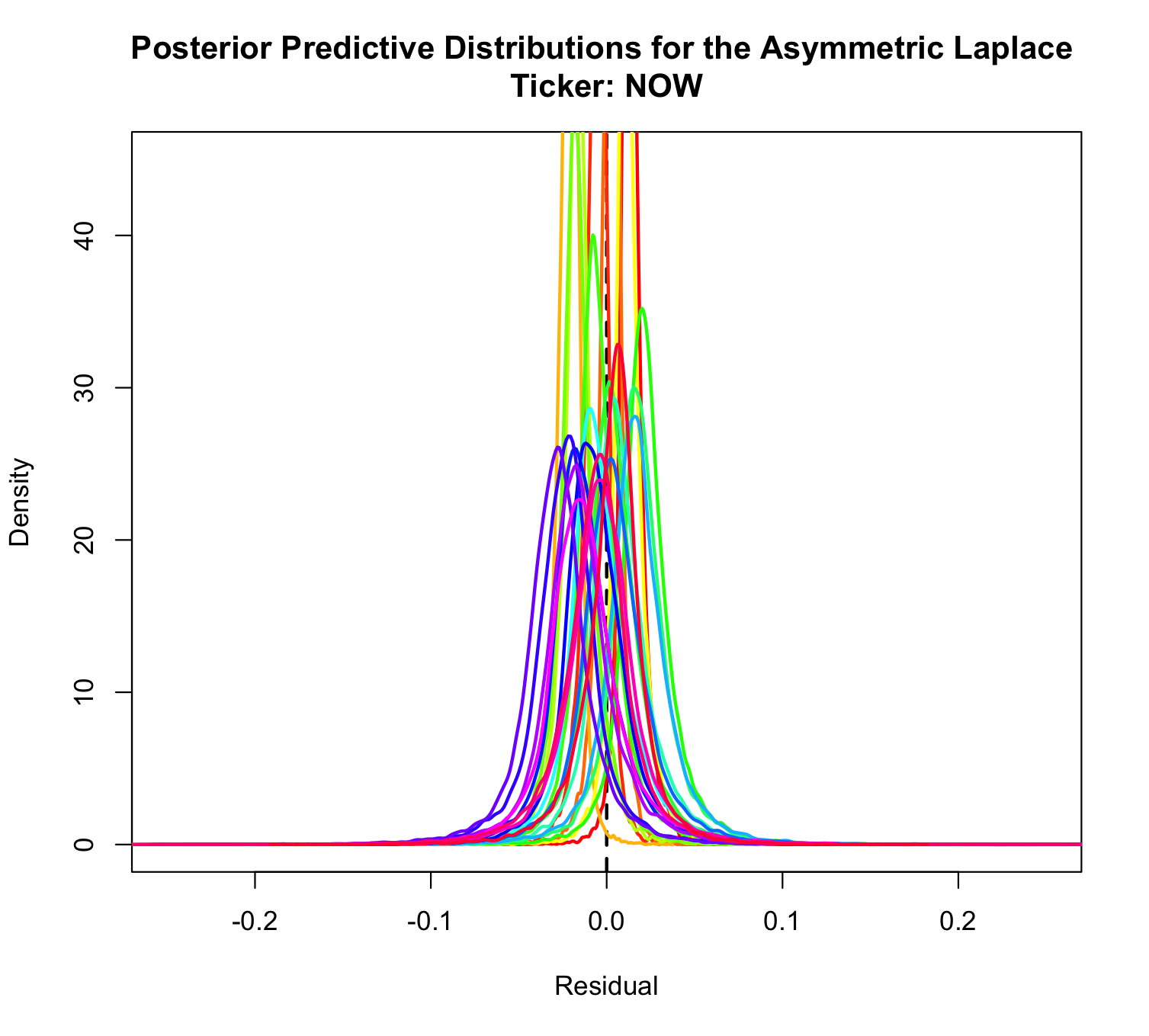}
    \includegraphics[width=0.33\textwidth]{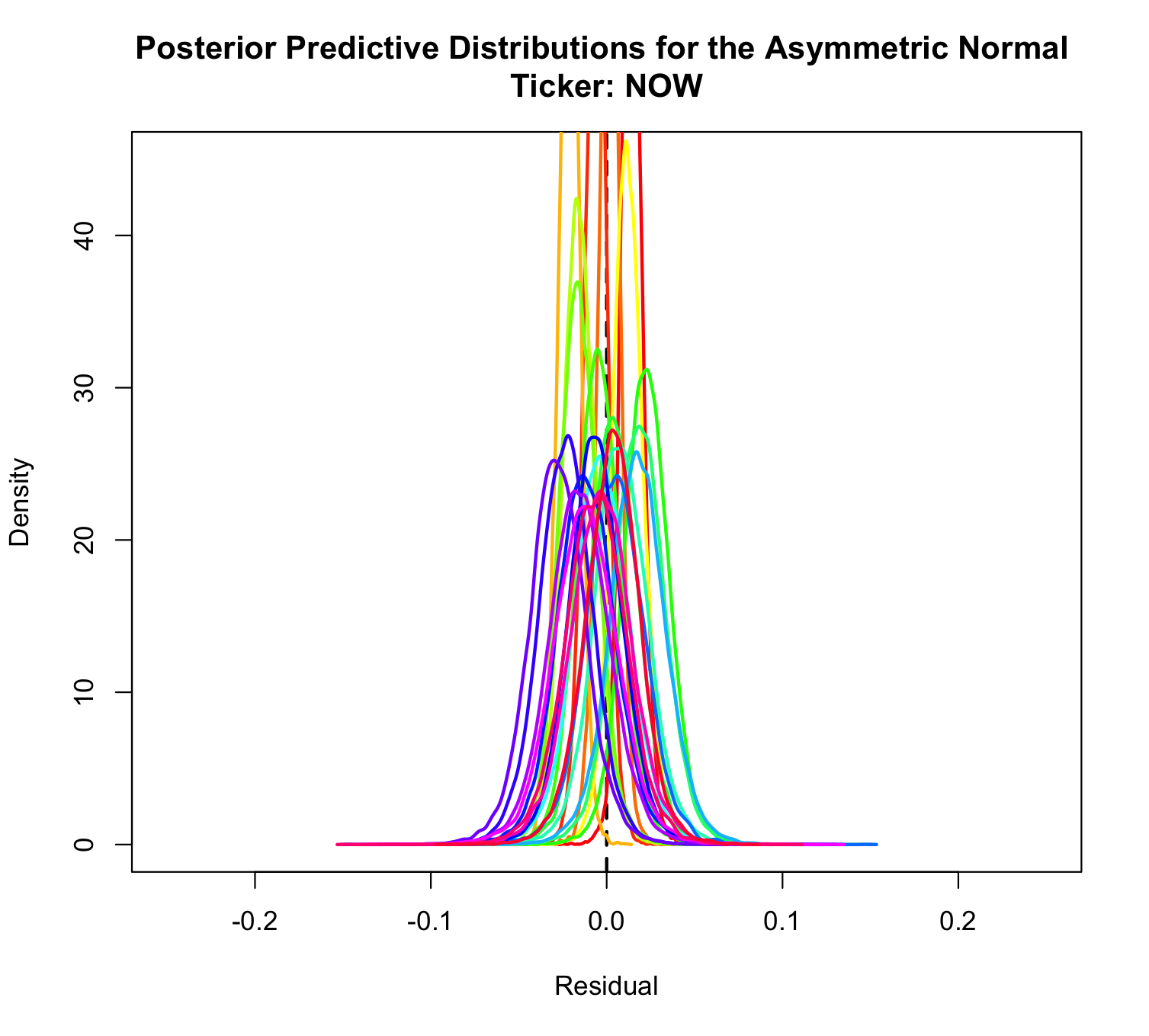}
    \includegraphics[width=0.33\textwidth]{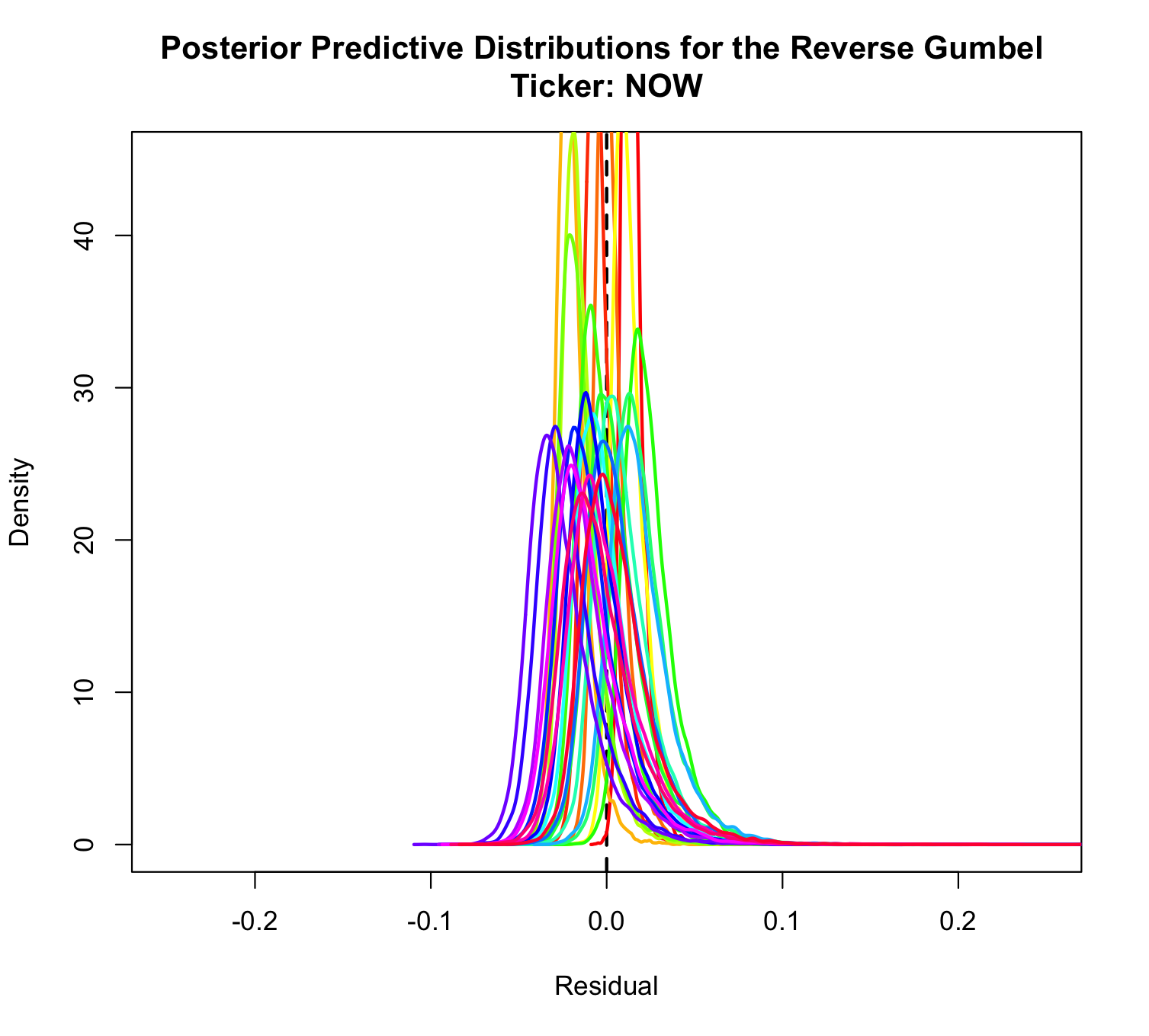}
    \caption{Posterior predictive distributions for NOW.}
    \label{fig:NOW}
\end{figure}

\begin{figure}[ht]
    \centering
    \includegraphics[width=0.33\textwidth]{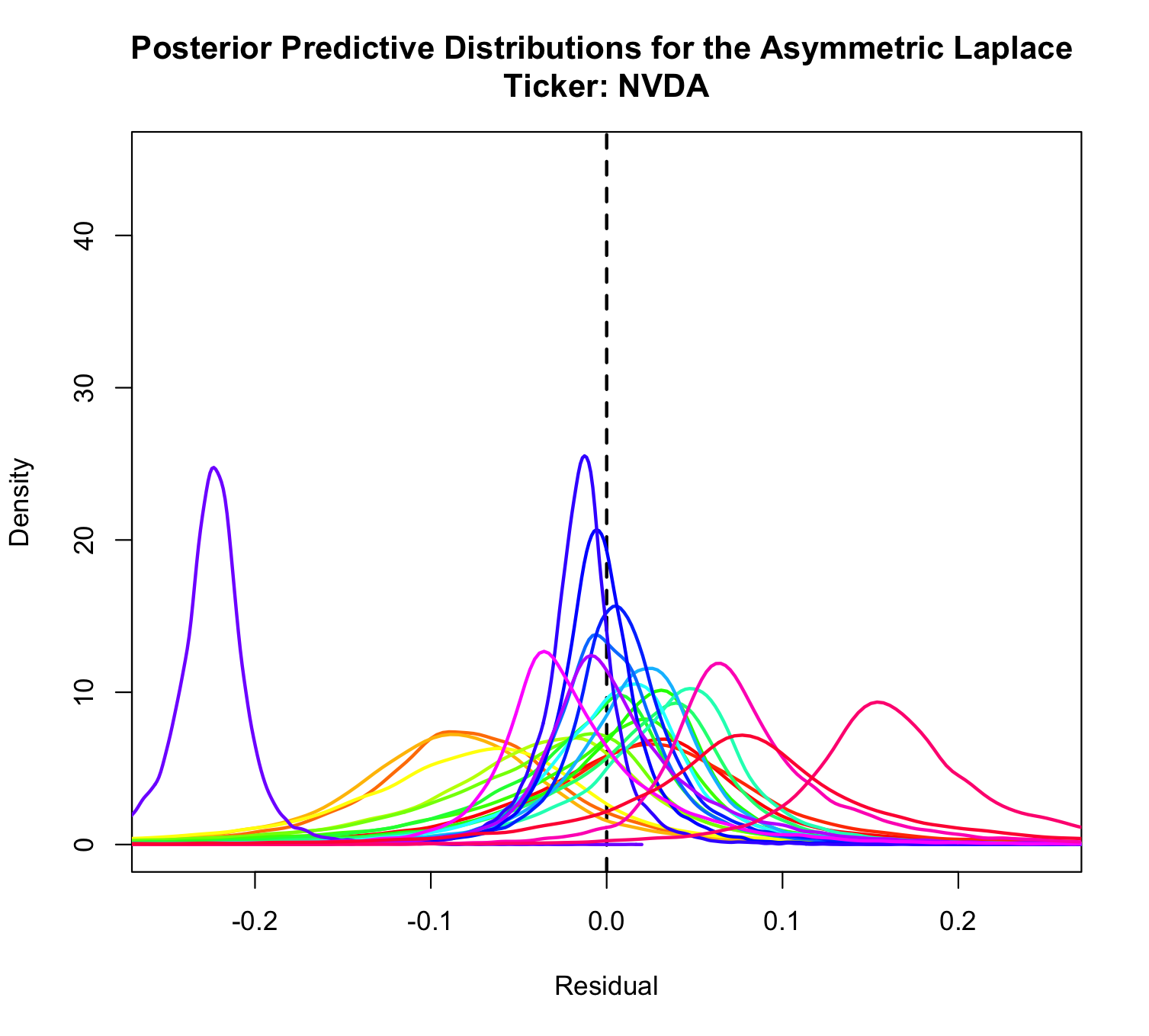}
    \includegraphics[width=0.33\textwidth]{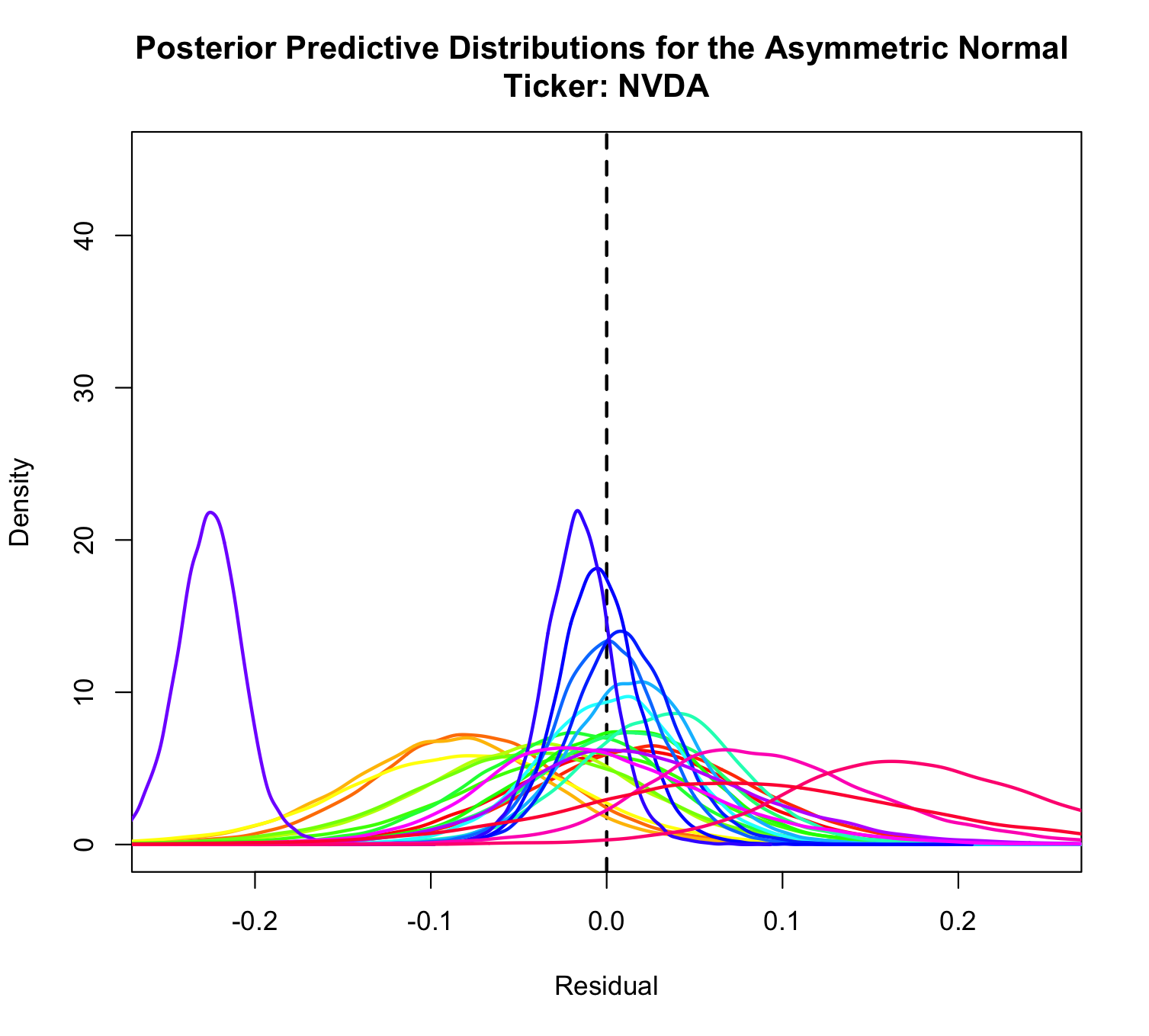}
    \includegraphics[width=0.33\textwidth]{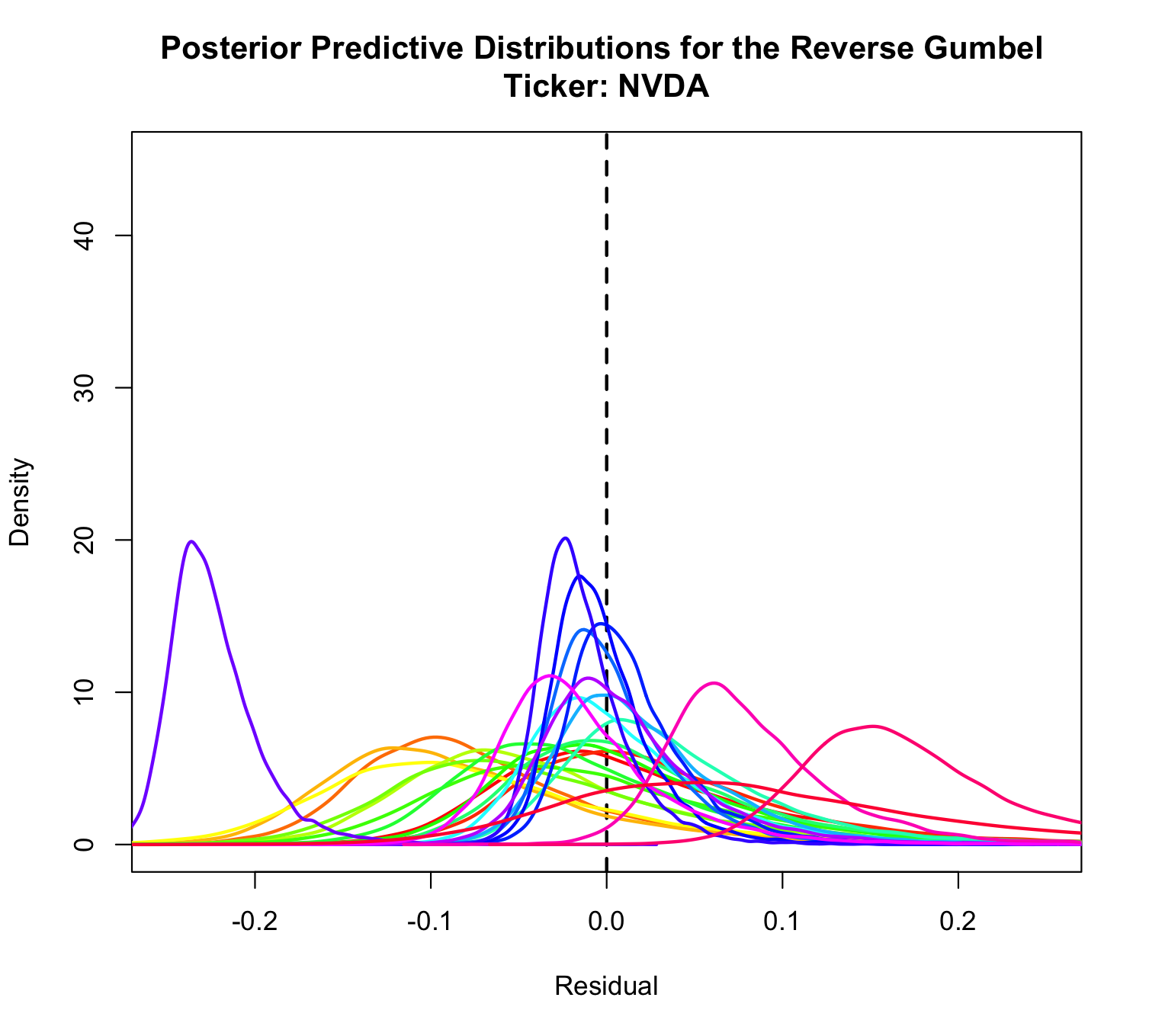}
    \caption{Posterior predictive distributions for NVDA.}
    \label{fig:NVDA}
\end{figure}

\begin{figure}[ht]
    \centering
    \includegraphics[width=0.33\textwidth]{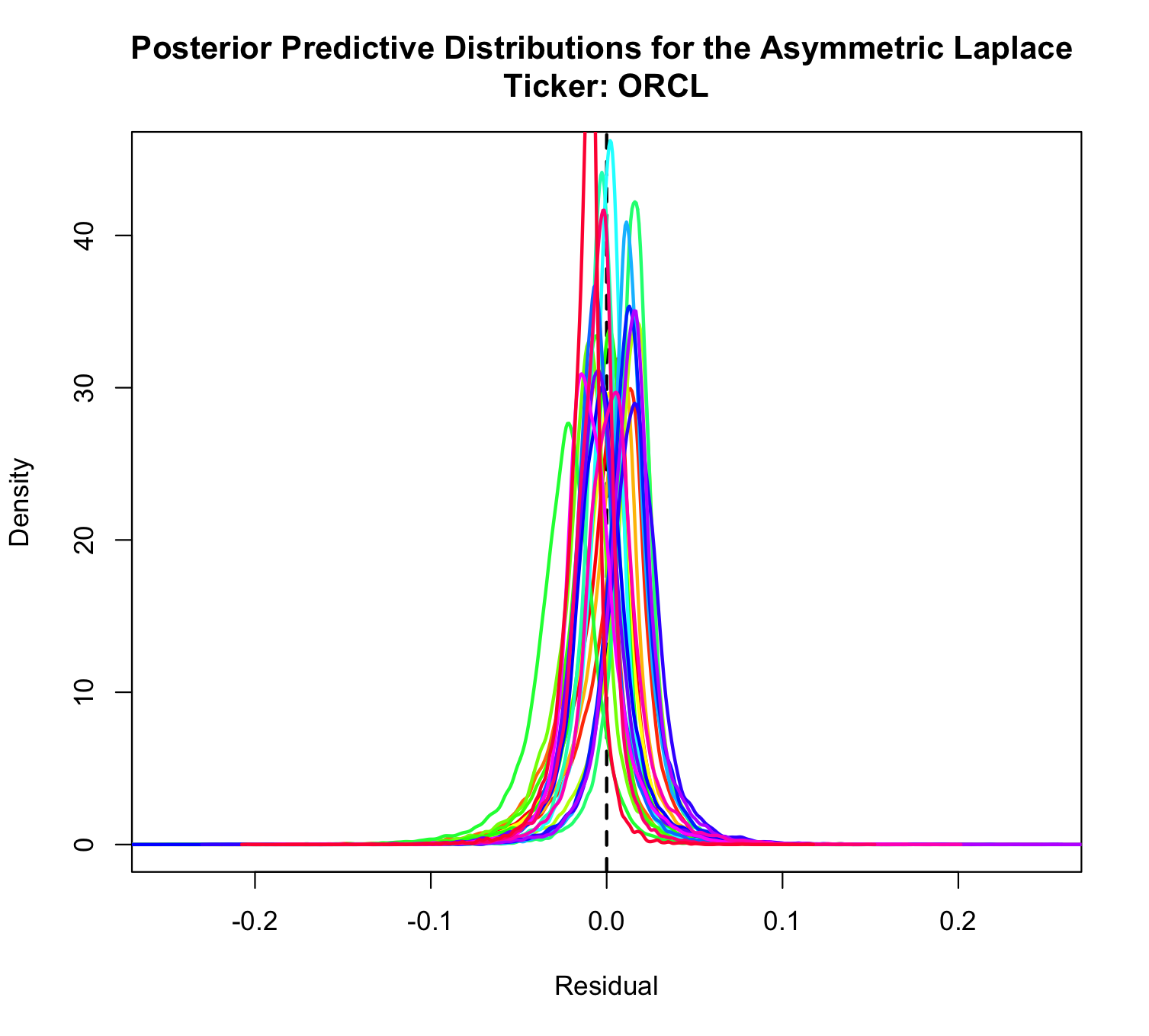}
    \includegraphics[width=0.33\textwidth]{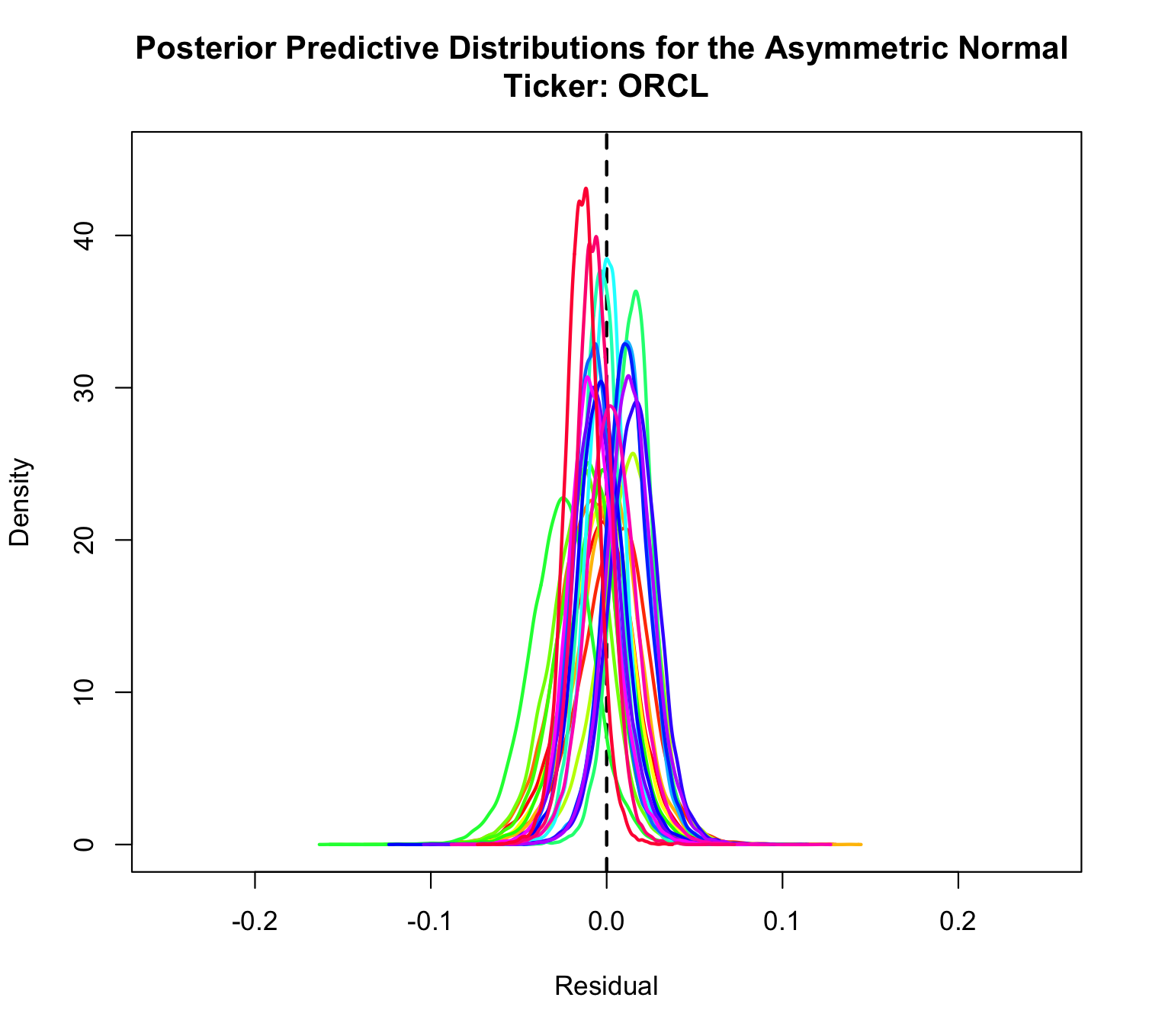}
    \includegraphics[width=0.33\textwidth]{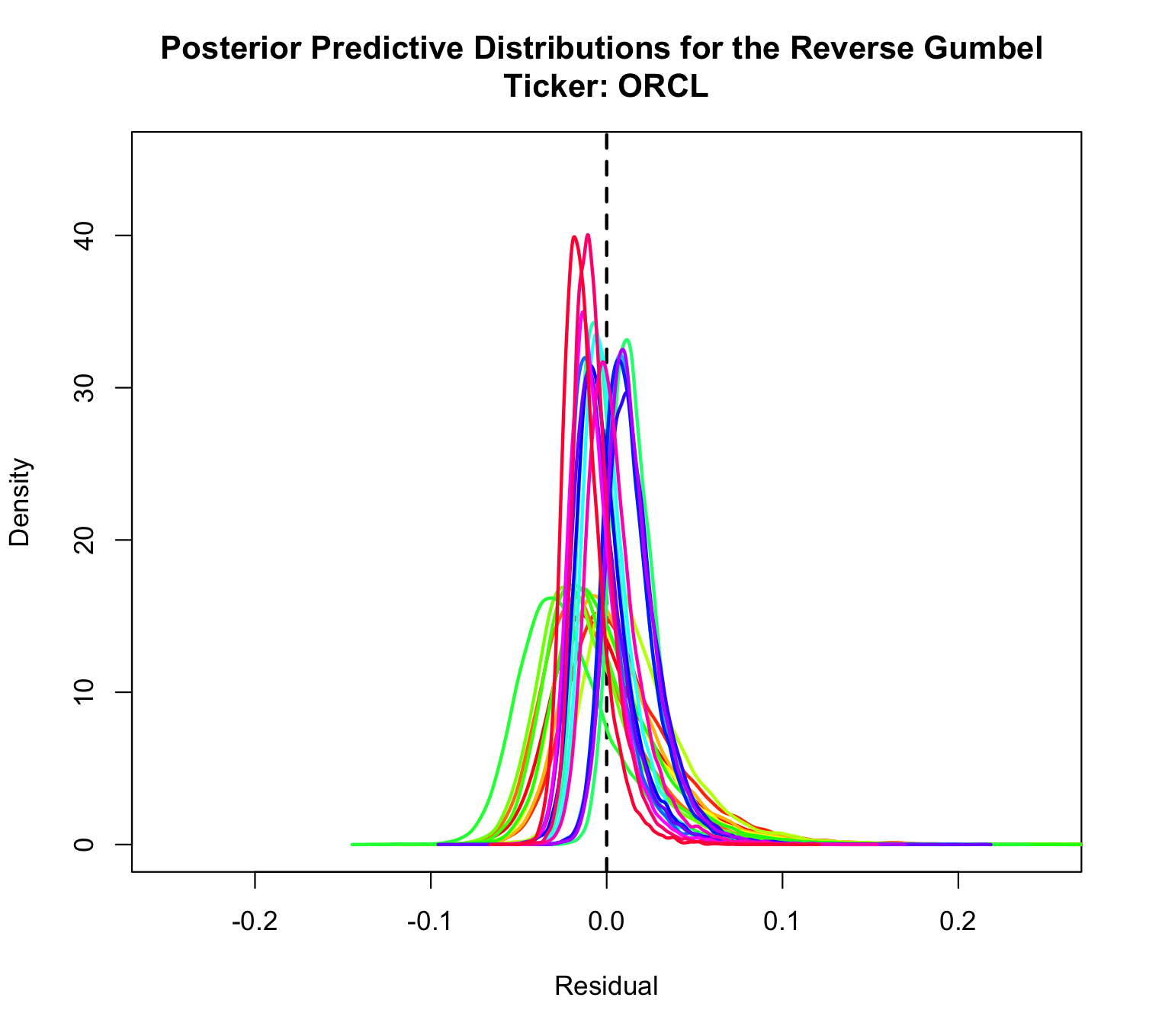}
    \caption{Posterior predictive distributions for ORCL.}
    \label{fig:ORCL}
\end{figure}

\begin{figure}[ht]
    \centering
    \includegraphics[width=0.33\textwidth]{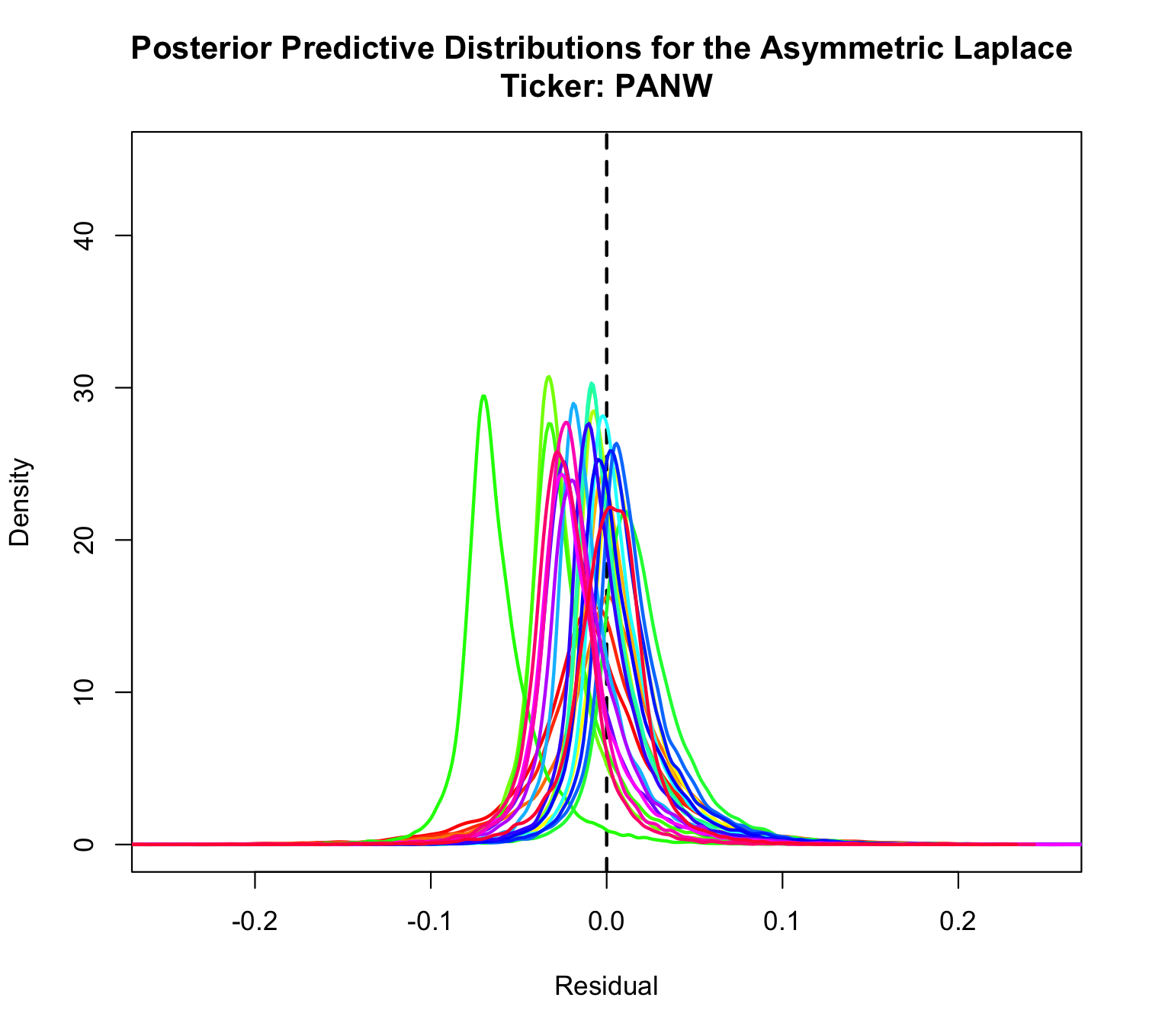}
    \includegraphics[width=0.33\textwidth]{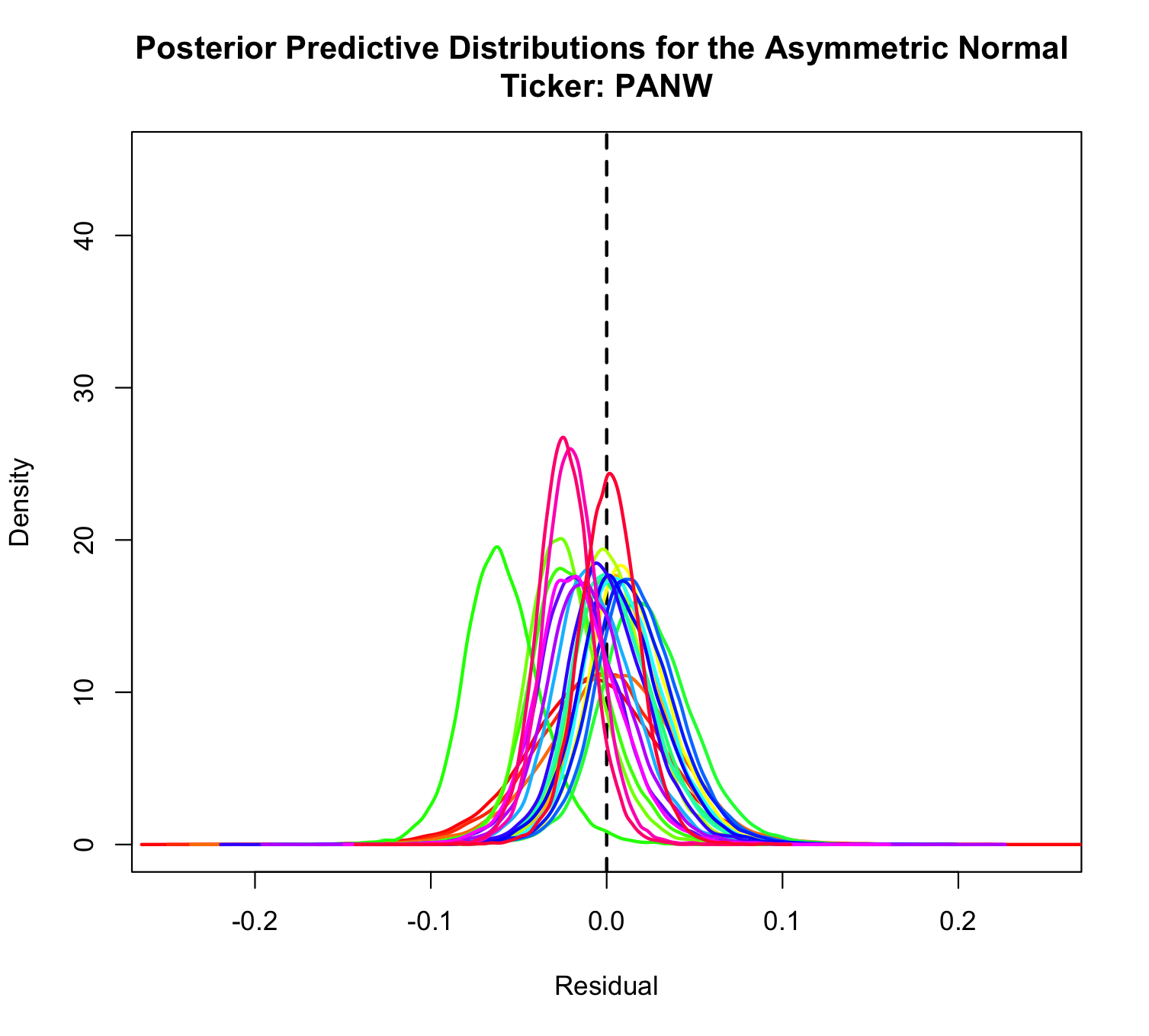}
    \includegraphics[width=0.33\textwidth]{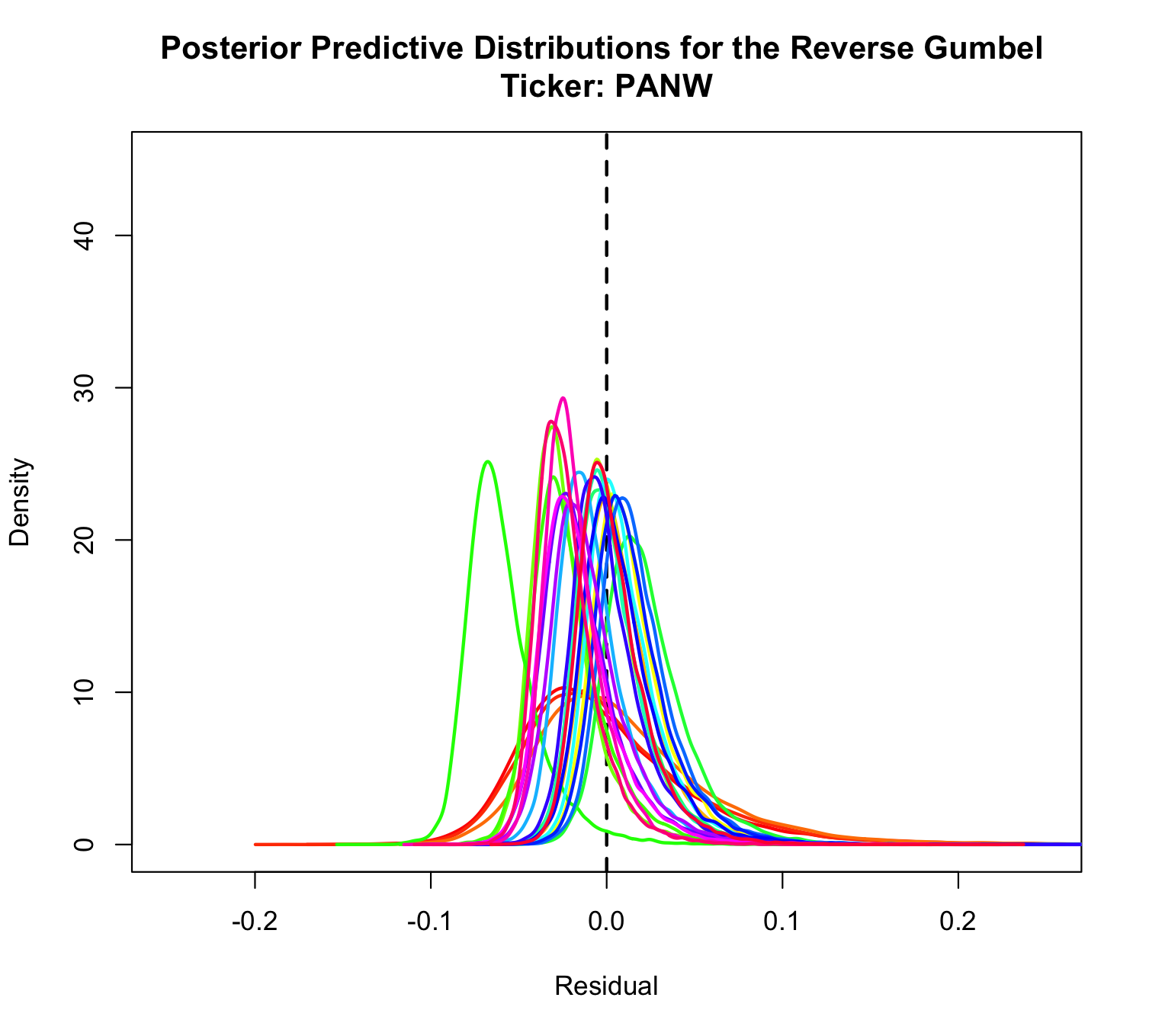}
    \caption{Posterior predictive distributions for PANW.}
    \label{fig:PANW}
\end{figure}

\begin{figure}[ht]
    \centering
    \includegraphics[width=0.33\textwidth]{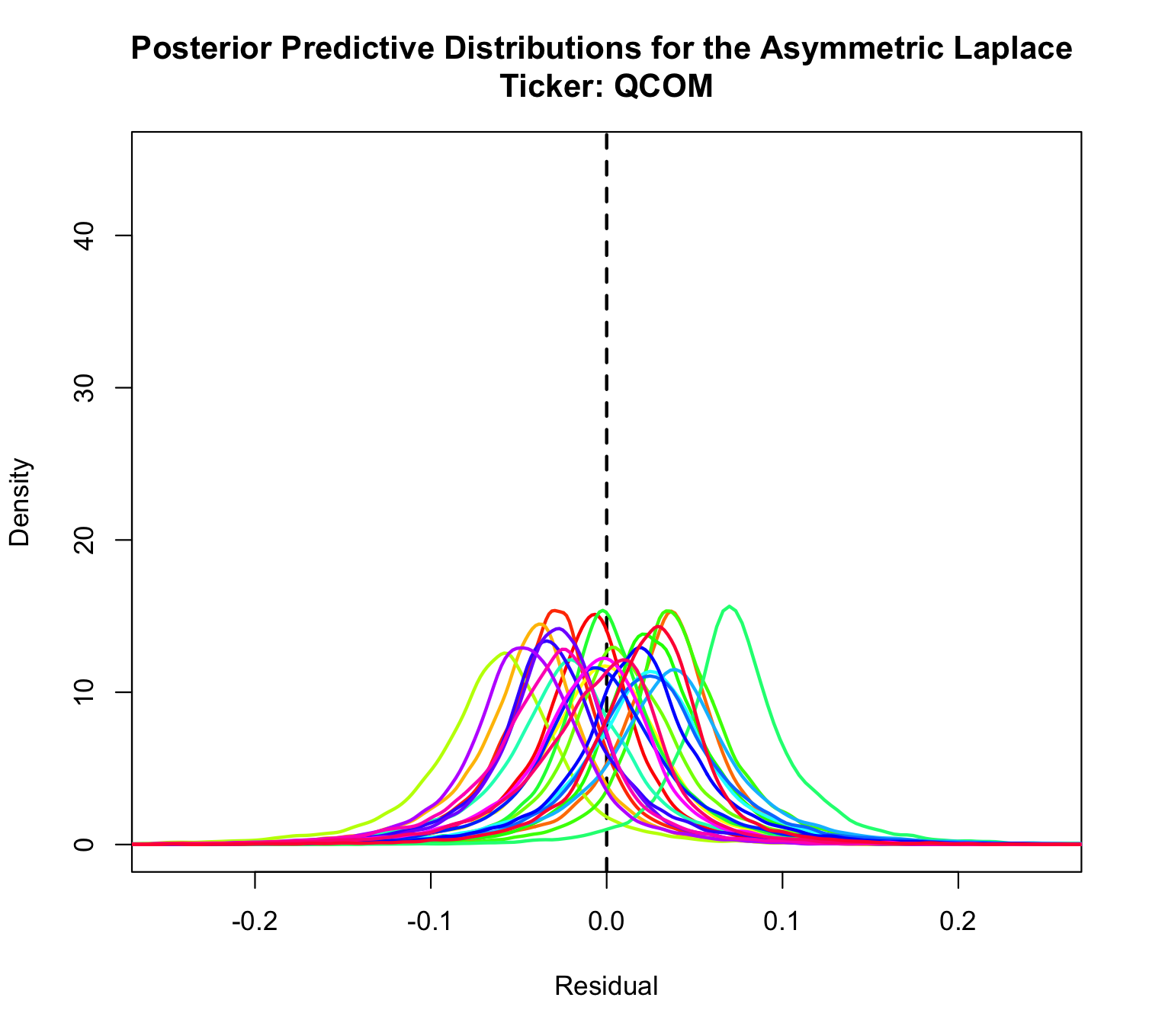}
    \includegraphics[width=0.33\textwidth]{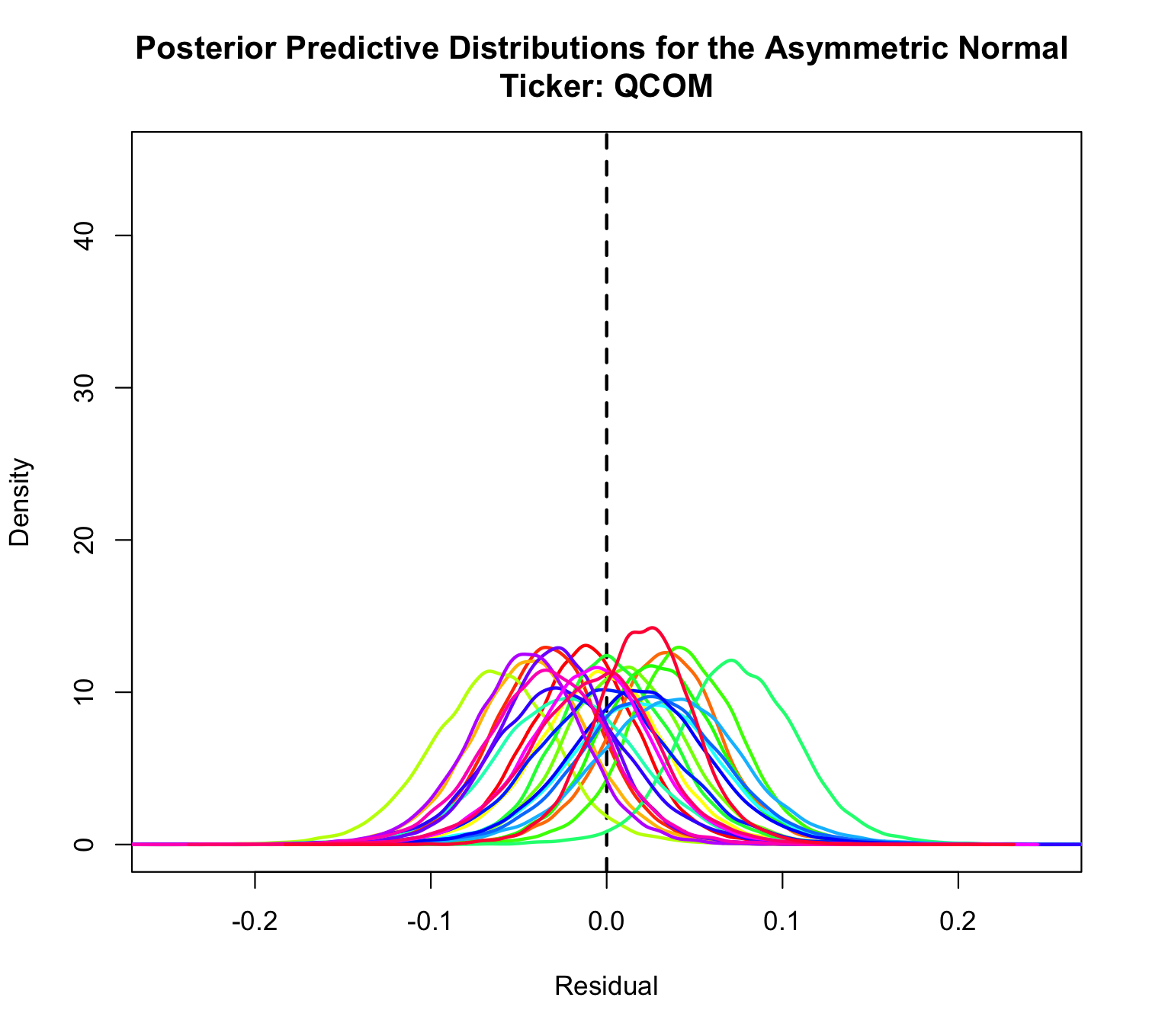}
    \includegraphics[width=0.33\textwidth]{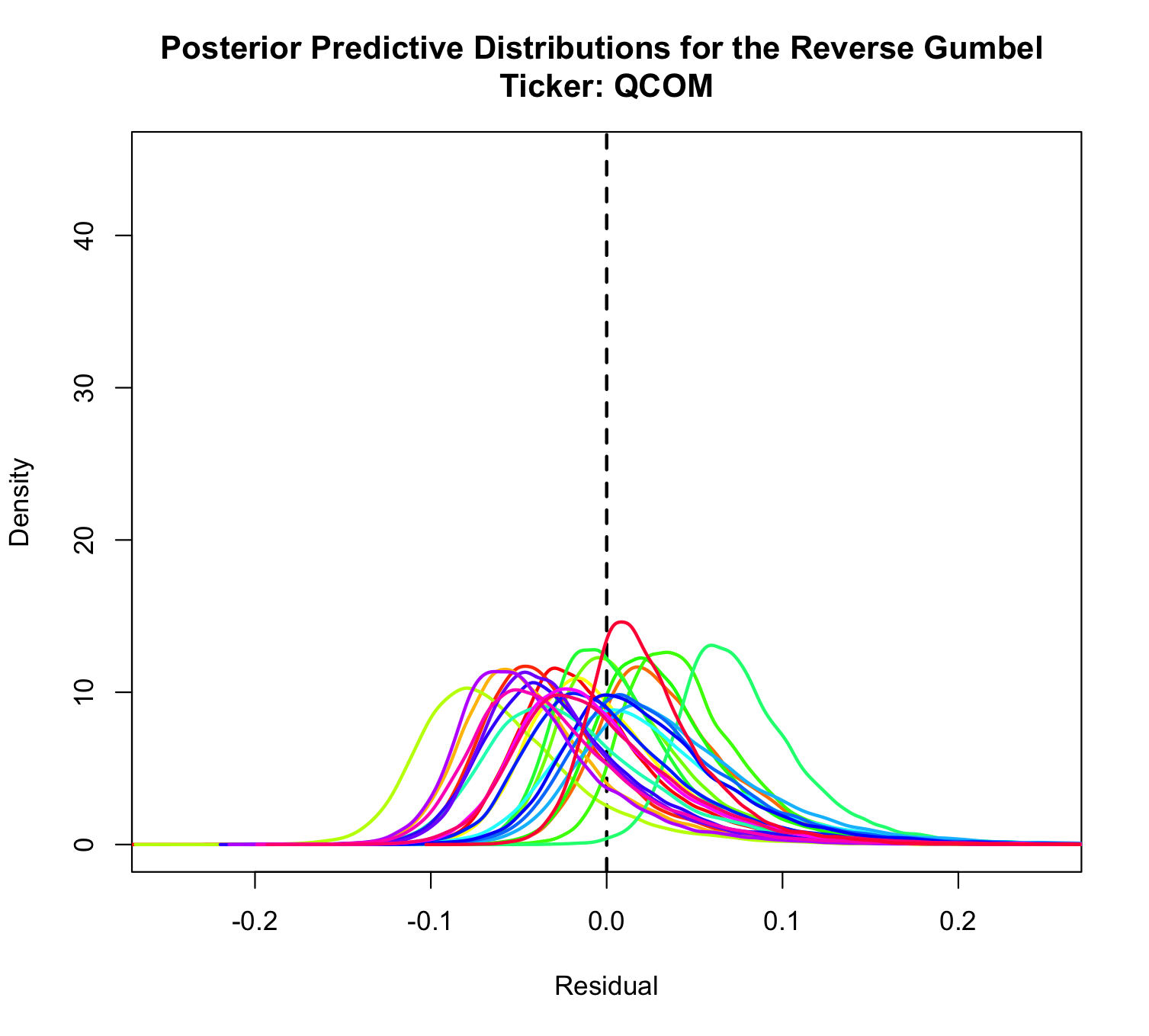}
    \caption{Posterior predictive distributions for QCOM.}
    \label{fig:QCOM}
\end{figure}

\begin{figure}[ht]
    \centering
    \includegraphics[width=0.33\textwidth]{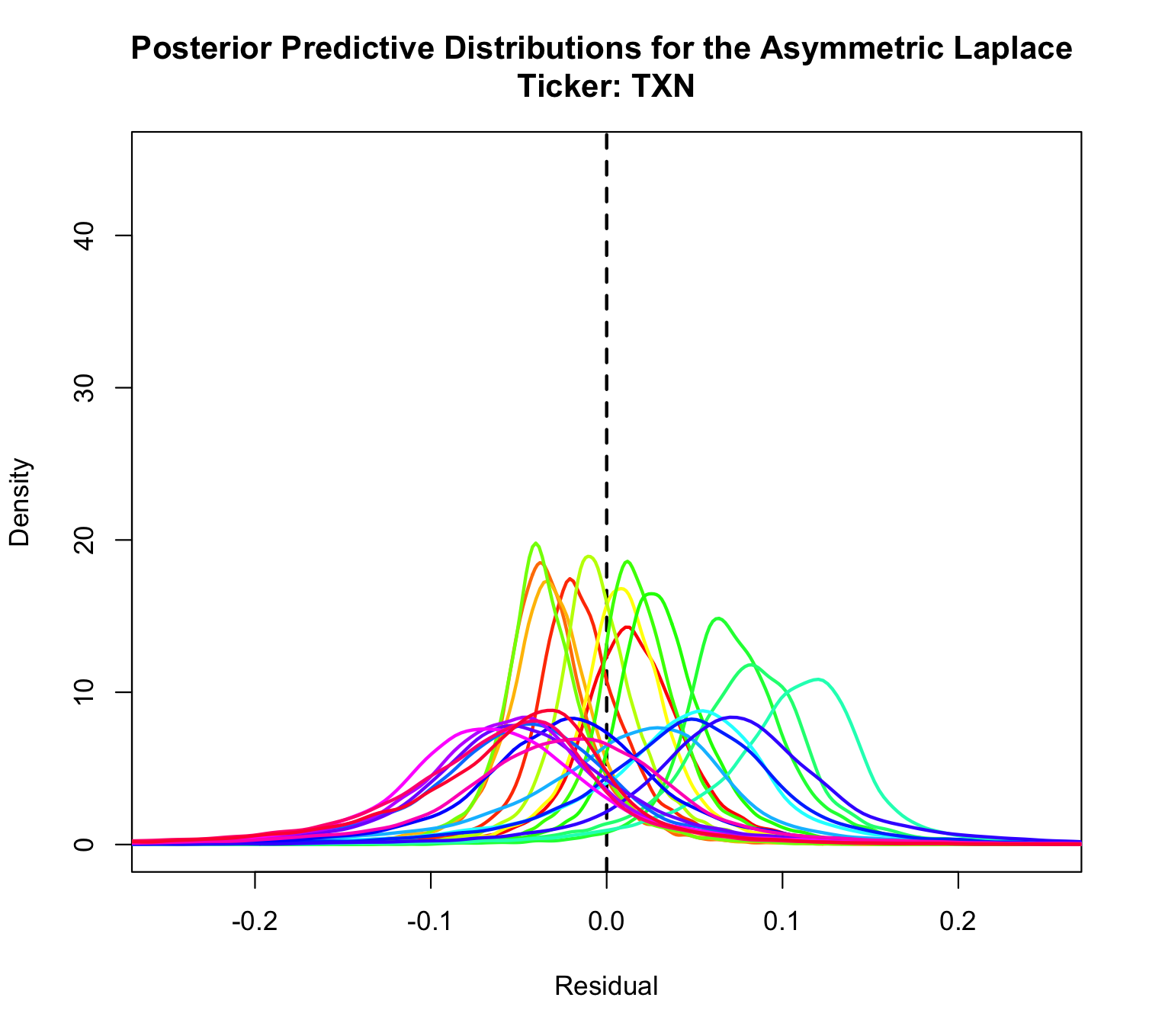}
    \includegraphics[width=0.33\textwidth]{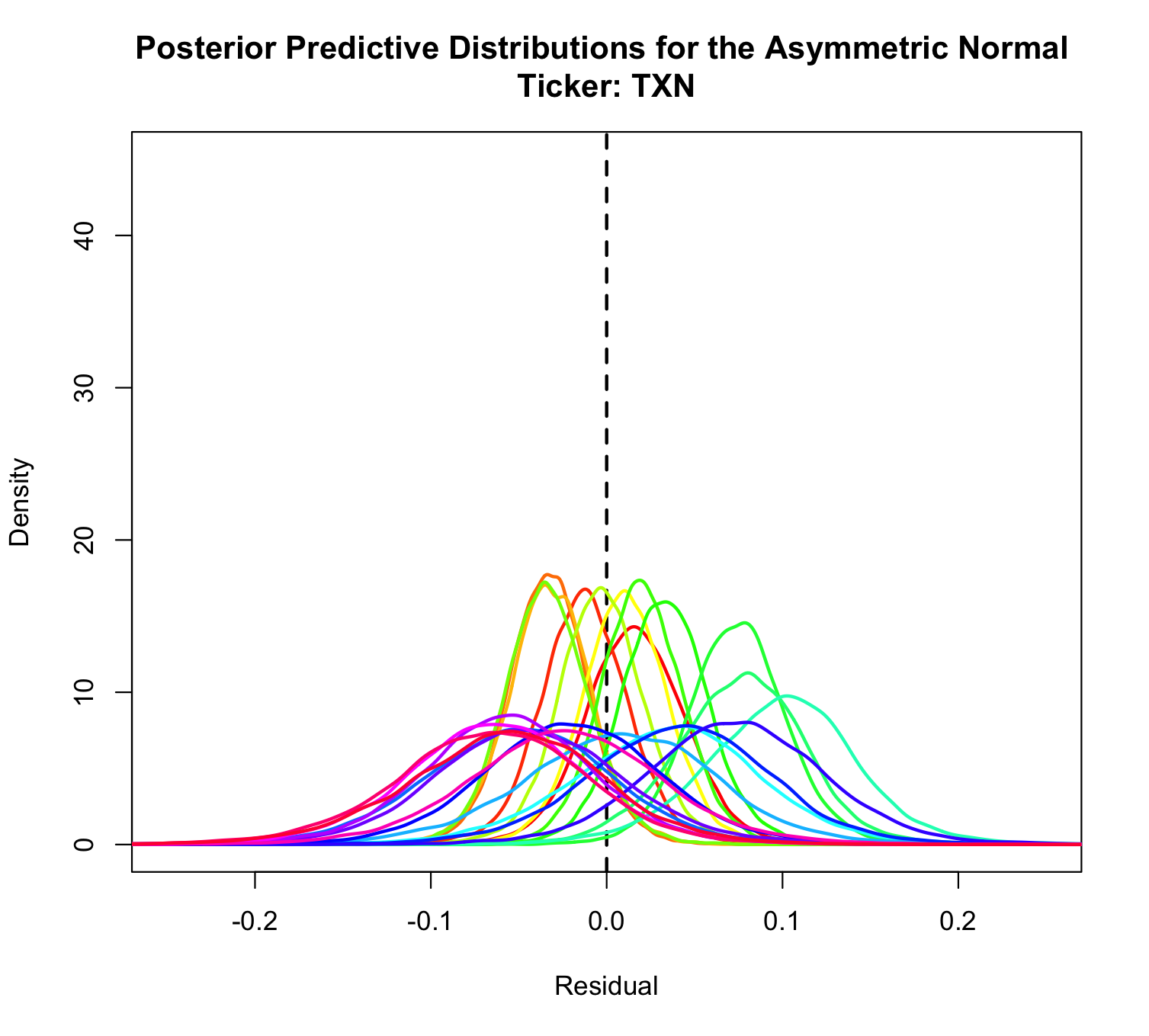}
    \includegraphics[width=0.33\textwidth]{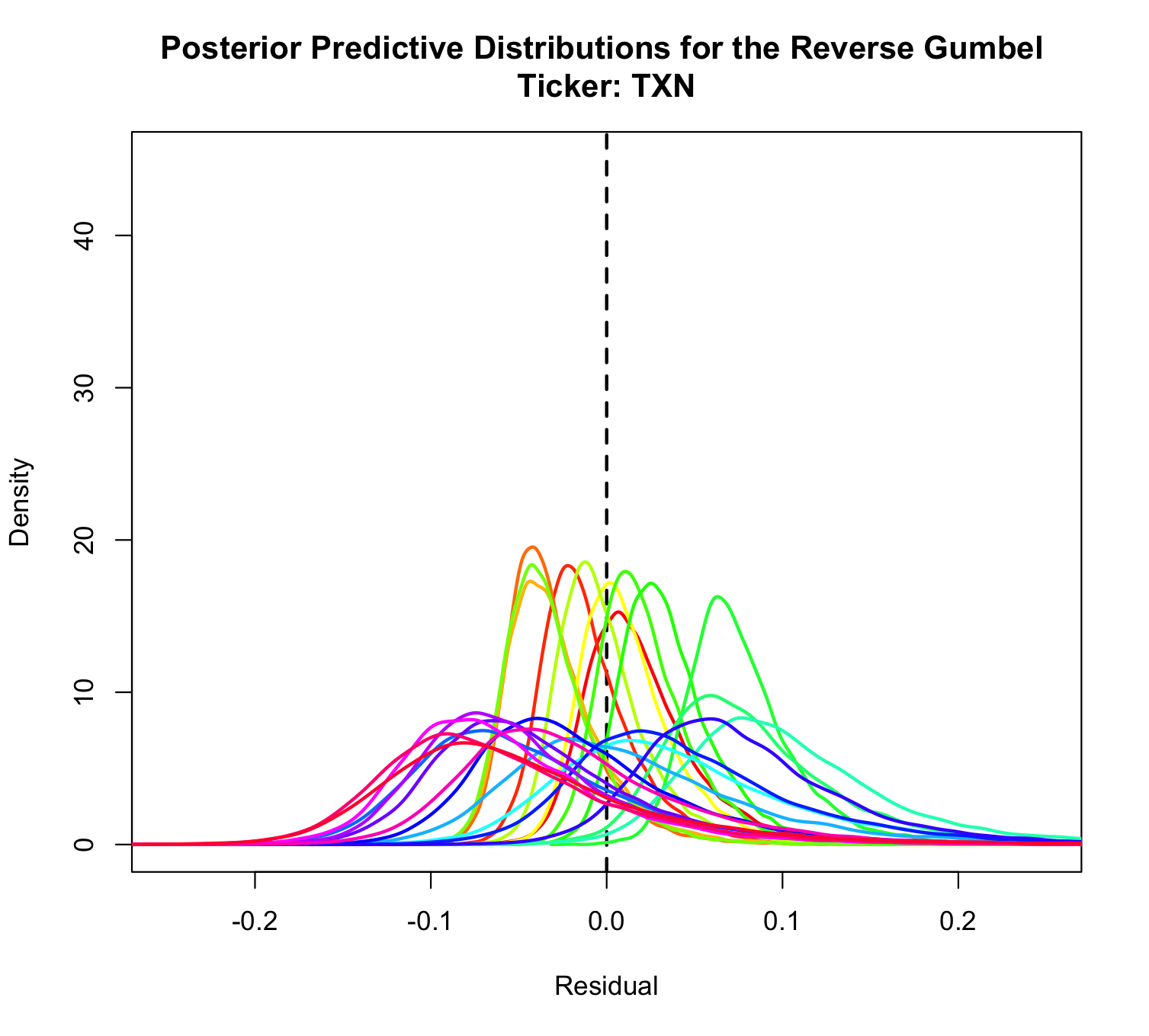}
    \caption{Posterior predictive distributions for TXN.}
    \label{fig:TXN}
\end{figure}

\end{document}